\documentclass[useAMS,usenatbib]{mn2e}

\pdfoutput=1

\usepackage[letterpaper,total={17.8cm,24.0cm},centering]{geometry} 
\usepackage{multirow}
\usepackage{amssymb}
\usepackage[utf8]{inputenc}

\usepackage{tikz}
\usetikzlibrary{fit}
\usetikzlibrary{decorations.pathmorphing}
\usetikzlibrary{backgrounds}
\usetikzlibrary{positioning}
\usetikzlibrary{shapes}
\usetikzlibrary{trees}

\tikzstyle{unit}=[circle,
		  thick,
		  minimum size=0.4cm,
		  draw=black]

\tikzstyle{visible}=[unit,
		    fill=gray!30]

\tikzstyle{hidden}=[unit,
		    fill=white]
		    
\tikzstyle{textunit}=[unit,
                    draw=white]

\tikzstyle{spacer}=[opacity=0]

\tikzstyle{layer}=[rectangle,
		   fill=white,
		   inner sep=0.15cm,
		   draw=black,
		   rounded corners=2mm,
		   minimum width=3cm]

\tikzstyle{wide}=[minimum width=5cm]

\tikzstyle{background}=[rectangle,
		   fill=gray!10,
		   inner sep=0.1cm,
		   draw=black,
		   rounded corners=2mm]

\tikzstyle{every node}=[font=\small]

\usepackage{amsmath}
\usepackage{url}
\usepackage{subcaption}
\usepackage{color,soul}

\renewcommand{\v}[1]{\mathbf{#1}}
\newcommand{\m}[1]{\mathbf{#1}}
\newcommand{\gztwo}{Galaxy~Zoo~2}

\title[Convnets for galaxy morphology prediction]{Rotation-invariant convolutional neural networks for galaxy morphology prediction}
\author[Dieleman, Willett \& Dambre]{Sander Dieleman$^{1}$\thanks{E-mail:
sander.dieleman@ugent.be (SD), willett@physics.umn.edu (KWW)}, Kyle W. Willett$^{2}$\footnotemark[1] and Joni Dambre$^{1}$\\
$^{1}$Electronics and Information Systems department, Ghent University, Sint-Pietersnieuwstraat 41, 9000 Ghent, Belgium\\
$^{2}$School of Physics and Astronomy, University of Minnesota, 116 Church St SE, Minneapolis, MN 55455, USA}
\begin{document}

\date{Accepted 20 March 2015}

\pagerange{\pageref{firstpage}--\pageref{lastpage}} \pubyear{2014}

\maketitle

\label{firstpage}

\begin{abstract}
Measuring the morphological parameters of galaxies is a key requirement for studying their formation and evolution. Surveys such as the Sloan Digital Sky Survey (SDSS) have resulted in the availability of very large collections of images, which have permitted population-wide analyses of galaxy morphology. Morphological analysis has traditionally been carried out mostly via visual inspection by trained experts, which is time-consuming and does not scale to large ($\gtrsim10^4$) numbers of images.

Although attempts have been made to build automated classification systems, these have not been able to achieve the desired level of accuracy. The Galaxy~Zoo project successfully applied a crowdsourcing strategy, inviting online users to classify images by answering a series of questions. Unfortunately, even this approach does not scale well enough to keep up with the increasing availability of galaxy images.

We present a deep neural network model for galaxy morphology classification which exploits translational and rotational symmetry. It was developed in the context of the \emph{Galaxy Challenge}, an international competition to build the best model for morphology classification based on annotated images from the Galaxy~Zoo project. 

For images with high agreement among the Galaxy~Zoo participants, our model is able to reproduce their consensus with near-perfect accuracy ($> 99\%$) for most questions. Confident model predictions are highly accurate, which makes the model suitable for filtering large collections of images and forwarding challenging images to experts for manual annotation. This approach greatly reduces the experts' workload without affecting accuracy. The application of these algorithms to larger sets of training data will be critical for analysing results from future surveys such as the LSST.
\end{abstract}


\section{Introduction}
Galaxies exhibit a wide variety of shapes, colours and sizes. These properties are indicative of their age, formation conditions, and interactions with other galaxies over the course of many Gyr. Studies of galaxy formation and evolution use morphology to probe the physical processes that give rise to them. In particular, large, all-sky surveys of galaxies are critical for disentangling the complicated relationships between parameters such as halo mass, metallicity, environment, age, and morphology; deeper surveys probe the changes in morphology starting at high redshifts and taking place over timescales of billions of years. 

Such studies require both the observation of large numbers of galaxies and accurate classification of their morphologies. Large-scale surveys such as the Sloan Digital Sky Survey (SDSS)\footnote{\url{http://www.sdss.org/}} have resulted in the availability of image data for millions of celestial objects. However, manually inspecting all these images to annotate them with morphological information is impractical for either individual astronomers or small teams.

Attempts to build automated classification systems for galaxy morphologies have historically had difficulties in reaching the levels of reliability required for scientific analysis \citep{citeulike:9525101}. The Galaxy~Zoo project\footnote{\url{http://www.galaxyzoo.org/}} was conceived to accelerate this task through the method of crowdsourcing. The original goal of the project was to obtain reliable morphological classifications for $\sim900,000$ galaxies by allowing members of the public to contribute classifications via a web platform. The project was much more successful than anticipated, with the entire catalog being annotated within a timespan of several months (originally projected to take years). Since its original inception, several iterations of the project with different sets of images and more detailed classification taxonomies have followed.

Two recent sets of developments since the launch of Galaxy~Zoo have made an automated approach more feasible: first, the large strides in the fields of image classification and computer vision in general, primarily through the use of \emph{deep neural networks} \citep{Krizhevsky2012,razavian2014cnn}. Although neural networks have existed for several decades \citep{mcculloch1943logical,fukushima1980neocognitron}, they have recently returned to the forefront of machine learning research. A significant increase in available computing power, along with new techniques such as rectified linear units \citep{NAI10} and dropout regularization \citep{Hinton2012,JMLR:v15:srivastava14a}, have made it possible to build more powerful neural network models (see Section~\ref{sec:deep-learning} for descriptions of these techniques).

Secondly, large sets of reliably annotated images of galaxies are now available as a consequence of the success of Galaxy~Zoo. Such data can be used to train machine learning models and increase the accuracy of their morphological classifications. Deep neural networks in particular tend to scale very well as the number of available training examples increases. Nevertheless it is also possible to train deep neural networks on more modestly sized datasets using techniques such as regularization, data augmentation, parameter sharing and model averaging, which we discuss in Section \ref{sec:overfitting} and following.

An automated approach is also becoming indispensable: modern telescopes continue to collect more and more images every day. Future telescopes will vastly increase the number of galaxy images that can be morphologically classified, including multi-wavelength imaging, deeper fields, synoptic observing, and true all-sky coverage. As a result, the crowdsourcing approach cannot be expected to scale indefinitely with the growing amount of data. Supplementing both expert and crowdsourced catalogues with automated classifications is a logical and necessary next step.

In this paper, we propose a convolutional neural network model for galaxy morphology classification that is specifically tailored to the properties of images of galaxies. It efficiently exploits both translational and rotational symmetry in the images, and autonomously learns several levels of increasingly abstract representations of the images that are suitable for classification. The model was developed in the context of the \emph{Galaxy Challenge}\footnote{\url{https://www.kaggle.com/c/galaxy-zoo-the-galaxy-challenge}}, an international competition to build the best model for automatic galaxy morphology classification based on a set of annotated images from the \gztwo~project. This model finished in first place out of 326 participants\footnote{The model was independently developed by SD for the Kaggle competition. KWW co-designed and administered the competition, but shared no data or code with any participants until after the closing date.}. The model can efficiently and automatically annotate catalogs of images with morphology information, enabling quantitative studies of galaxy morphology on an unprecedented scale.

The rest of this paper is structured as follows: we introduce the Galaxy~Zoo project in Section~\ref{sec:galaxy-zoo} and Section~\ref{sec:galaxy-challenge} explains the set up of the Galaxy Challenge. We discuss related work in Section~\ref{sec:related-work}. Convolutional neural networks are described in Section~\ref{sec:background}, and our method to incorporate rotation invariance in these models is described in Section~\ref{sec:rotation-invariance}. We provide a complete overview of our modelling approach in Section~\ref{sec:approach} and report results in Section~\ref{sec:results}. We analyse the model in Section~\ref{sec:analysis}. Finally, we draw conclusions in Section~\ref{sec:conclusion}.

\section{Galaxy~Zoo}
\label{sec:galaxy-zoo}

\begin{figure*}
\centering
\includegraphics[width=0.90\textwidth]{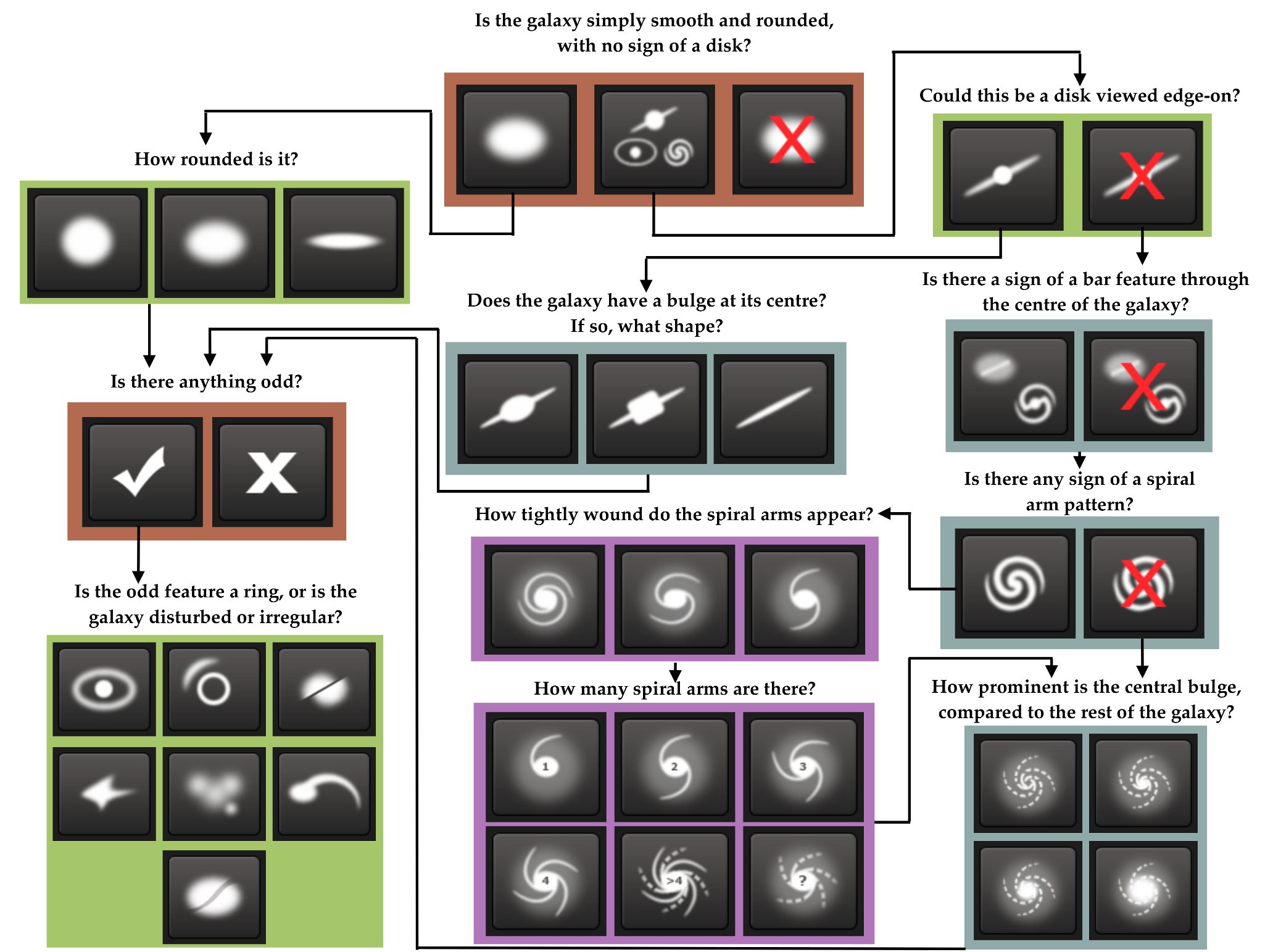}
\caption{The \gztwo~decision tree. Reproduced from Figure~1 in \citet{willett2013galaxy}.}
\label{fig:decision-tree}
\end{figure*}

Galaxy~Zoo is an online crowdsourcing project where users are asked to describe the morphology of galaxies based on colour images \citep{lin08,lin11}. Our model and analysis uses data from the \gztwo~iteration of the project, which uses colour images from the SDSS and a more detailed classification scheme than the original project \citep{willett2013galaxy}. Participants are asked various questions such as `How rounded is the galaxy?' and `Does it have a central bulge?', with the users' answers determining which question will be asked next. The questions form a decision tree (Figure~\ref{fig:decision-tree}) which is designed to encompass all points in the traditional Hubble tuning fork as well as a range of more irregular morphologies. The classification scheme has 11 questions and 37 answers in total (Table~\ref{tab:questions}).

\begin{table}
\scriptsize
\centering
 \begin{tabular}{llllr}
    \noalign{\hrule height 1pt}
    & \textbf{question} & & \textbf{answers} & \textbf{next} \\
    \hline
    \multirow{3}{*}{Q1}  & \multirow{3}{4cm}{Is the galaxy simply smooth and rounded, with no sign of a disk?}          & A1.1  & smooth & Q7 \\
                         &                                                                                              & A1.2  & features or disk & Q2 \\
                         &                                                                                              & A1.3  & star or artifact & end \\
    \hline
    \multirow{2}{*}{Q2}  & \multirow{2}{4cm}{Could this be a disk viewed edge-on?}                                      & A2.1  & yes              & Q9 \\
                         &                                                                                              & A2.2  & no               & Q3 \\
    \hline
    \multirow{2}{*}{Q3}  & \multirow{2}{4cm}{Is there a sign of a bar feature through the centre of the galaxy?}        & A3.1  & yes              & Q4 \\
                         &                                                                                              & A3.2  & no               & Q4 \\
    \hline
    \multirow{2}{*}{Q4}  & \multirow{2}{4cm}{Is there any sign of a spiral arm pattern?}                                & A4.1  & yes              & Q10 \\
                         &                                                                                              & A4.2  & no               & Q5 \\
    \hline
    \multirow{4}{*}{Q5}  & \multirow{4}{4cm}{How prominent is the central bulge, compared with the rest of the galaxy?} & A5.1  & no bulge         & Q6 \\
                         &                                                                                              & A5.2  & just noticeable  & Q6 \\
			 &                                                                                              & A5.3  & obvious          & Q6 \\
			 &                                                                                              & A5.4  & dominant         & Q6 \\
    \hline
    \multirow{2}{*}{Q6}  & \multirow{2}{4cm}{Is there anything odd?}                                                    & A6.1  & yes              & Q8 \\
                         &                                                                                              & A6.2  & no               & end \\
    \hline
    \multirow{3}{*}{Q7}  & \multirow{3}{4cm}{How rounded is it?}                                                        & A7.1  & completely round & Q6 \\
                         &                                                                                              & A7.2  & in between       & Q6 \\
                         &                                                                                              & A7.3  & cigar-shaped     & Q6 \\
    \hline
    \multirow{7}{*}{Q8}  & \multirow{7}{4cm}{Is the odd feature a ring, or is the galaxy disturbed or irregular?}       & A8.1  & ring             & end \\
                         &                                                                                              & A8.2  & lens or arc      & end \\
                         &                                                                                              & A8.3  & disturbed        & end \\
                         &                                                                                              & A8.4  & irregular        & end \\
                         &                                                                                              & A8.5  & other            & end \\
                         &                                                                                              & A8.6  & merger           & end \\
                         &                                                                                              & A8.7  & dust lane        & end \\
    \hline
    \multirow{3}{*}{Q9}  & \multirow{3}{4cm}{Does the galaxy have a bulge at its centre? If so, what shape?}            & A9.1  & rounded          & Q6 \\
                         &                                                                                              & A9.2  & boxy             & Q6 \\
                         &                                                                                              & A9.3  & no bulge         & Q6 \\
    \hline
    \multirow{3}{*}{Q10} & \multirow{3}{4cm}{How tightly wound do the spiral arms appear?}                              & A10.1 & tight            & Q11 \\
                         &                                                                                              & A10.2 & medium           & Q11 \\
                         &                                                                                              & A10.3 & loose            & Q11 \\
    \hline
    \multirow{6}{*}{Q11} & \multirow{6}{4cm}{How many spiral arms are there?}                                           & A11.1 & 1                & Q5 \\
                         &                                                                                              & A11.2 & 2                & Q5 \\
                         &                                                                                              & A11.3 & 3                & Q5 \\
                         &                                                                                              & A11.4 & 4                & Q5 \\
                         &                                                                                              & A11.5 & more than four   & Q5 \\
                         &                                                                                              & A11.6 & can't tell       & Q5 \\
    \noalign{\hrule height 1pt}
 \end{tabular} 
\caption{All questions that can be asked about an image, with the corresponding answers that participants can choose from. Question~1 is the only one that is asked of every image. The final column indicates the next question to be asked when a particular answer is given. Reproduced from Table~2 in \citet{willett2013galaxy}.}
 \label{tab:questions}
\end{table}

Because of the structure of the decision tree, each individual participant answered only a subset of the questions for each classification.
When many participants have classified the same image, their answers are aggregated into a set of weighted vote fractions for the entire decision tree.  These vote fractions are used to estimate confidence levels for each answer, and are indicative of the difficulty users experienced in classifying the image.
More than half a million people have contributed classifications to Galaxy~Zoo, with each image independently classified by 40 to 50 people\footnote{Note that the vote fractions are post-processed to increase their reliability, for example by weighting users based on their consistency with the majority, and by compensating for classification bias induced by different image apparent magnitudes and sizes \citep{willett2013galaxy}.}.

Data from the Galaxy~Zoo projects have already been used in a wide variety of studies on galaxy structure, formation, and evolution \citep{skibba2009galaxy,bamford2009galaxy,schawinski2009galaxy,lintott2009galaxy,darg2010galaxy,masters2010galaxy,masters2011galaxy,sim13,mel14,willett15}. Comparisons of Galaxy~Zoo morphologies to smaller samples from both experts and automated classifications show high levels of agreement, testifying to the accuracy of the crowdsourced annotations \citep{bamford2009galaxy,willett2013galaxy}.

\section{The Galaxy Challenge}
\label{sec:galaxy-challenge}

Our model was developed in the context of the Galaxy Challenge, an online international competition organized by Galaxy~Zoo, sponsored by Winton Capital, and hosted on the Kaggle platform for data prediction contests. It was held from December~20$^{\rm th}$, 2013 to April~4$^{\rm th}$, 2014. The goal of the competition was to build a model that could predict galaxy morphology from images like the ones that were used in the \gztwo~project.

Images of galaxies and morphological data for the competition were taken from the \gztwo~main spectroscopic sample. Galaxies were selected to cover the full observed range of morphology, colour, and size, since the goal was to develop a general algorithm that could be applied to many types of images in future surveys. The total number of images provided is limited both by the imaging depth of SDSS and the elimination of both uncertain and over-represented morphological categories as a function of colour (primarily red elliptical and blue spiral galaxies). This helped to ensure that colour is not used as a proxy for morphology, and that a high-performing model would be based purely on the images' structural parameters.

The final \textbf{training set} of data consisted of 61,578 JPEG colour images of galaxies, along with probabilities\footnote{These are actually post-processed vote fractions obtained from the Galaxy~Zoo participants' answers, but we treat them as probabilities in this paper.} for each of the 37~answers in the decision tree. An \textbf{evaluation set} of 79,975 images was also provided, but with no morphological data -- the goal of the competition was to predict these values. Each image is 424 by 424 pixels in size. The morphological data provided was a modified version of the weighted vote fractions in the GZ2 catalog; these were transformed into ``cumulative'' probabilities that gave higher weights to more fundamental morphological categories higher in the decision tree. Images were anonymized from their SDSS IDs, with any use of metadata (such as colour, size, position, or redshift) to train the algorithm explicitly forbidden by the competition guidelines. 

Because the goal was to predict probabilities for each answer, as opposed to determining the most likely answer for each question in the decision tree, the models built by participants were actually solving a \emph{regression} problem, and not a classification problem in the strictest sense. The predictive performance of a model was determined by computing the root-mean-square error (RMSE) between predictions on the evaluation set and the corresponding crowdsourced probabilities. Let $p_k$ be the answer probabilities associated with an image ($k = 1 \ldots 37$), and $\hat{p}_k$ the corresponding predictions. Then the RMSE $e(\hat{p}_k,p_k)$ can be computed as follows:
\begin{equation}\label{eq:rmse}
  e(\hat{p}_k,p_k) = \sqrt{\sum_{k=1}^{37}(\hat{p}_k - p_k)^2} .
\end{equation}

\noindent This metric was chosen because it puts more emphasis on questions with higher answer probabilities, i.e. the topmost questions in the decision tree, which correspond to broader morphological categories.

The provided answer probabilities are derived from crowdsourced classifications, so they are somewhat noisy and biased in certain ways. As a result, the predictive models that were built exhibited some of the same biases. In other words, they are models of how \emph{the crowd} would classify images of galaxies, which may not necessarily correspond to the ``true'' morphology. An example of such a discrepancy is discussed in Section~\ref{sec:analysis}.

The models built by participants were evaluated as follows. The Kaggle platform automatically computed two scores based on a set of model predictions: a public score, computed on about 25\% of the evaluation data, and a private score, computed on the other 75\%. Public scores were immediately revealed during the course of the competition, but private scores were not revealed until after the competition had finished. The private score was used to determine the final ranking. Because the participants did not know which evaluation images belonged to which set, they could not directly optimize the private score, but were instead encouraged to build a predictive model that generalizes well to new images.

\section{Related work}
\label{sec:related-work}

Machine learning techniques, and artificial neural networks in particular, have been a popular tool in astronomy research for more than two decades. Neural networks were initially applied for star-galaxy discrimination \citep{odewahn1992automated,bertin1994classification} and classification of galaxy spectra \citep{folkes1996artificial}. More recently they have also been used for photometric redshift estimation \citep{firth2003estimating,collister2004annz}.

Galaxy morphology classification is one of the most widespread applications of neural networks in astronomy. Most work in this domain proceeds by preprocessing the photometric data and then extracting a limited set of handcrafted features that are known to be discriminative, such as ellipticity, concentration, surface brightness, and radii and log-likelihood values measured from various types of radial profiles
\citep{storrie1992morphological,1995Sci...267..859L,naim1995automated,lahav1996neural,ball2004galaxy,banerji2010galaxy}. Support vector machines (SVMs) have also been applied in this fashion \citep{huertas2010revisiting}.

Earlier work in this domain typically relied on much smaller datasets and used networks with very few trainable parameters (between $10^1$ and $10^3$). Modern network architectures are capable of handling at least $\sim10^7$~parameters, allowing for more precise fits and a larger morphological classification space.
More recent work has profited from the availability of larger training sets using data from surveys such as the SDSS \citep{banerji2010galaxy,huertas2010revisiting}.

Another recent trend is the use of general purpose image features, instead of features that are specific to galaxies: the WND-CHARM feature set \citep{orlov2008wnd}, originally designed for biological image analysis, has been applied to galaxy morphology classification in combination with nearest neighbour classifiers \citep{shamir2009automatic,shamir2013automatic,kuminski2014combining}.

Other approaches to this problem attempt to forgo any form of handcrafted feature extraction by applying principal component analysis (PCA) to preprocessed images in combination with a neural network \citep{de2004machine}, or by applying kernel SVMs directly to raw pixel data \citep{polsterer2012galaxy}.

Our approach differs from prior work in two main areas:
\begin{itemize}
 \item Most prior work uses handcrafted features (e.g., WND-CHARM) that required expert knowledge and many hours of engineering to develop. We work directly with raw pixel data and our use of convolutional neural networks allows for a set of task-specific features to be learned from the data. The networks learn hierarchies of features, which allow them to detect complex patterns in the images. Handcrafted features mostly rely on image statistics and very local pattern detectors, making it harder to recognize complex patterns. Furthermore, it is usually necessary to perform feature selection because the handcrafted representations are highly redundant and many features are irrelevant for the task at hand. Although many other participants in the Galaxy Challenge used convolutional neural networks, there is little discussion of this approach in the astronomical literature.
 \item Apart from the recent work of \citet{kuminski2014combining}, whose algorithms are also trained on Galaxy Zoo data, most research has focused on classifying galaxies into a limited number of classes (typically between 2 and 5), or predicting scalar values that are indicative of galaxy morphology (e.g., Hubble T-types). Since the classifications made by Galaxy~Zoo users are much more fine-grained, the task the networks must solve is more challenging. Since many outstanding astrophysical questions require more detailed morphological data (such as the number and arrangements of clumps into proto-galaxies, the relation between bar strength and central star formation, link between merging activity and active black holes, etc.), development of models that can handle these more difficult tasks is crucial.
\end{itemize}

Our method for classifying galaxy morphology exploits the rotational symmetry of galaxy images; however, there are other invariances and symmetries (besides translational) that may be exploited for convolutional neural networks. \citet{bruna2013spectral} define convolution operations over arbitrary graphs, generalizing from the typical grid of pixels to other locally connected structures. \citet{sifre2013rotation} extract representations that are invariant to affine transformations, based on scattering transforms. However, these representations are fixed (i.e., not learned from data), and not specifically tuned for the task at hand, unlike the representations learned by convolutional neural networks.

\citet{mairal2014convolutional} propose to train convolutional neural networks to approximate kernel feature maps, allowing for the desired invariance properties to be encoded in the choice of kernel, and subsequently learned. \citet{Gens2014} propose deep symmetry networks, a generalization of convolutional neural networks with the ability to form feature maps over any symmetry group, rather than just the translation group. Our approach for exploiting rotational symmetry in the input images, described in Section \ref{sec:rotation-invariance}, is quite similar in spirit to this work. The major advantage to our implementation is a demonstrably effective result at a reasonable computational cost.

\section{Background}
\label{sec:background}

\subsection{Deep learning}
\label{sec:deep-learning}
The idea of \emph{deep learning} is to build models that represent data at multiple levels of abstraction, and can discover accurate representations autonomously from the data itself \citep{citeulike:3196377}. Deep learning models consist of several layers of processing that form a hierarchy: each subsequent layer extracts a progressively more abstract representation of the input data and builds upon the representation from the previous layer, typically by computing a non-linear transformation of its input. The parameters of these transformations are optimized by \emph{training} the model on a dataset.

A \emph{feed-forward neural network} is an example of such a model, where each layer consists of a number of units (or neurons) that compute a weighted linear combination of the layer input, followed by an elementwise non-linearity. These weights constitute the model parameters. Let the vector $\v{x}_{n-1}$ be the input to layer $n$, $\m{W}_n$ be a matrix of weights, and $\v{b}_n$ be a vector of biases. Then the output of layer $n$ can be represented as the vector

\begin{equation} \label{eq:activation}
 \v{x}_{n} = f(\m{W}_n \v{x}_{n - 1} + \v{b}_n) ,
\end{equation}

\noindent where $f$ is the \emph{activation function}, an elementwise non-linear function. Common choices for the activation function are linear rectification [$f(x) = \max(x, 0)$], which gives rise to \emph{rectified linear units} \citep[ReLUs;][]{NAI10}, or a sigmoidal function [$f(x) = (1 + e^{-x})^{-1}$ or $f(x) = \tanh(x)$]. Another possibility is to compute the maximum across several linear combinations of the input, which gives rise to \emph{maxout units} \citep{goodfellow2013maxout}. We will consider a network with $N$ layers. The network input is represented by $\v{x}_0$, and its output by $\v{x}_N$.

\begin{figure}
\centering

\begin{tikzpicture}[>=latex,text height=1.5ex,text depth=0.25ex]
 
  \matrix[row sep=1.0cm,column sep=0.2cm] {
	\node (so1) [spacer] {}; &
	\node (so2) [spacer] {}; &
	\node (so3) [spacer] {}; &
	\node (so4) [spacer] {}; &
	\node (so5) [spacer] {}; &
	\\
	\node (l_n_1) [hidden] {}; &
	\node (l_n_2) [hidden] {}; &
	\node (l_n_3) [hidden] {}; &
	\node (l_n_4) [hidden] {}; &
	\node (l_n_5) [hidden] {}; &
	\\
	\node (l_nm1_1) [hidden] {}; &
	\node (l_nm1_2) [hidden] {}; &
	\node (l_nm1_3) [hidden] {}; &
	\node (l_nm1_4) [hidden] {}; &
	\node (l_nm1_5) [hidden] {}; &
	\\
	\node (s1) [spacer] {}; &
	\node (s2) [spacer] {}; &
	\node (s3) [rectangle] {$\cdots$}; &
	\node (s4) [spacer] {}; &
	\node (s5) [spacer] {}; &
	\\
	\node (l_1_1) [hidden] {}; &
	\node (l_1_2) [hidden] {}; &
	\node (l_1_3) [hidden] {}; &
	\node (l_1_4) [hidden] {}; &
	\node (l_1_5) [hidden] {}; &
	\\
	\node (si1) [spacer] {}; &
	\node (si2) [spacer] {}; &
	\node (si3) [spacer] {}; &
	\node (si4) [spacer] {}; &
	\node (si5) [spacer] {}; &
	\\
    };
    
    \begin{pgfonlayer}{background}
	\node (layero) [rectangle, fit=(so1) (so5), text height=0.5cm] {output};
	\node (layern) [background, fit=(l_n_1) (l_n_5)] {}; \node[right=0.3cm,align=left,anchor=north west] at (layern.east) {layer $N$ (output)\\ $\v{x}_N = f(\m{W}_N \v{x}_{N-1} + \v{b}_N)$};
	\node (layernm1) [background, fit=(l_nm1_1) (l_nm1_5)] {};  \node[right=0.3cm,align=left,anchor=north west] at (layernm1.east) {layer $N-1$ (hidden)\\ $\v{x}_{N-1} = f(\m{W}_{N-1} \v{x}_{N-2} + \v{b}_{N-1})$};
	\node (layerd) [rectangle, fit=(s1) (s5)] {};
	\node (layer1) [background, fit=(l_1_1) (l_1_5)] {}; \node[right=0.3cm,align=left,anchor=north west] at (layer1.east) {layer $1$ (hidden)\\ $\v{x}_1 = f(\m{W}_1 \v{x}_0 + \v{b}_1)$};
	\node (layeri) [rectangle, fit=(si1) (si5)] {input};
    \end{pgfonlayer}
    
    \path[->,thick]
	(layern) edge node[anchor=center, right, midway] {$\v{x}_N$} (layero)
	(layernm1) edge node[anchor=center, right, midway] {$\v{x}_{N-1}$} (layern)
	(layerd) edge node[anchor=center, right, midway] {$\v{x}_{N-2}$} (layernm1)
	(layer1) edge node[anchor=center, right, midway] {$\v{x}_1$} (layerd)
	(layeri) edge node[anchor=center, right, midway] {$\v{x}_0$} (layer1)
        ;
\end{tikzpicture}

\caption{Schematic representation of a feed-forward neural network with $N$ layers.}
\label{fig:neuralnetwork}
\end{figure}
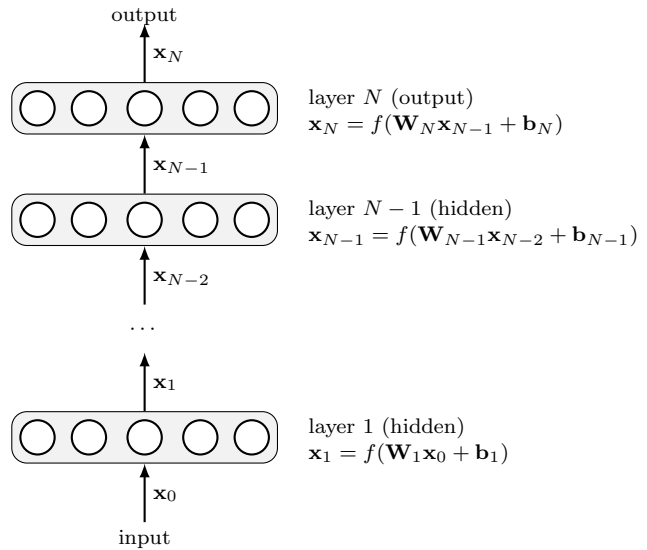

A schematic representation of a feed-forward neural network is shown in Figure~\ref{fig:neuralnetwork}. The network computes a function of the input $\v{x}_0$. The output $\v{x}_N$ of this function is a prediction of one or more quantities of interest. We will use $\v{t}$ to represent the desired output (target) corresponding to the network input $\v{x}_0$. The topmost layer of the network is referred to as the \emph{output layer}. All the other layers below it are \emph{hidden layers}.

During training, the parameters of all layers of the network are jointly optimized to make the output $\v{x}_N$ approximate the desired output $\v{t}$ as closely as possible. We quantify the prediction error using an error measure $e(\v{x}_N, \v{t})$. As a result, the hidden layers will learn to produce representations of the input data that are useful for the task at hand, and the output layer will learn to predict the desired output from these representations.

To determine how the parameters should be changed to reduce the prediction error across the dataset, we use \emph{gradient descent}: the gradient of $e(\v{x}_N, \v{t})$ is computed with respect to the model parameters $\m{W}_n$, $\v{b}_n$ for $n = 1 \ldots N$. The parameter values of each layer are then modified by repeatedly taking small steps in the direction opposite to the gradient:

\begin{align}
 \m{W}_n &\leftarrow \m{W}_n - \eta \dfrac{\partial e(\v{x}_N, \v{t})}{\partial \m{W}_n} , \\
 \v{b}_n &\leftarrow \v{b}_n - \eta \dfrac{\partial e(\v{x}_N, \v{t})}{\partial \v{b}_n} .
\end{align}

\noindent Here, $\eta$ is the \emph{learning rate}, a hyperparameter controlling the step size.

Traditionally, models with many non-linear layers of processing have not been commonly used because they were difficult to train: gradient information would vanish as it propagated through the layers, making it difficult to learn the parameters of lower layers \citep{hochreiter2001gradient}. Practical applications of neural networks were limited to models with one or two hidden layers. Since 2006, the invention of several new techniques, along with a significant increase in available computing power, have made this task much more feasible.

Initially \emph{unsupervised pre-training} was proposed as a method to facilitate training deeper networks \citep{1161605}. Single-layer unsupervised models (such as restricted Boltzmann machines or auto-encoders \citep{citeulike:3196377}) are stacked on top of each other and trained, and the learned parameters of these models are then used to initialize the parameters of a deep neural network. These are then fine-tuned using standard gradient descent. This initialization scheme makes it possible to largely avoid the vanishing gradient problem. \citet{NAI10} and \citet{glorot2011deep} proposed the use of rectified linear units (ReLUs) in deep neural networks. By replacing traditional activation functions with linear rectification, the vanishing gradient problem was significantly reduced. This also makes pre-training unnecessary in most cases.

The introduction of dropout regularization \citep{Hinton2012,JMLR:v15:srivastava14a} has made it possible to train larger networks with many more parameters. Dropout is a regularization method that can be applied to a layer $n$ by randomly removing the output values of the previous layer $n-1$ (setting them to zero) with probability $p$. Typically $p$ is chosen to be $0.5$. The remaining values are rescaled by a factor of $(1 - p)^{-1}$ to preserve the scale of the total input to each unit in layer $n$. For each training example that is presented to the network, a different subset of values is removed. During evaluation, no values are removed and no rescaling is performed.

Dropout is an effective regularizer because it prevents \emph{coadaptation} between units: each unit is forced to learn to be useful by itself, because its utility cannot depend on the presence of other units in the same layer (as they can be removed at random).

\subsection{Convolutional neural networks}
\label{sec:convnets}

Convolutional neural networks or \emph{convnets} \citep{fukushima1980neocognitron,lecun1998gradient} are a subclass of neural networks with constrained connectivity patterns between some of the layers. They can be used when the input data exhibits some kind of topological structure, like the ordering of image pixels in a grid, or the temporal structure of an audio signal.

Convolutional neural networks contain two types of layers with restricted connectivity: \emph{convolutional layers} and \emph{pooling layers}. A convolutional layer takes a stack of \emph{feature maps} (e.g. the colour channels of an image) as input and convolves each of these with a set of learnable filters to produce a stack of output feature maps. This is efficiently implemented by replacing the matrix-vector product $\m{W}_n \v{x}_{n - 1}$ in Equation~\ref{eq:activation} with a sum of convolutions. We represent the input of layer $n$ as a set of $K$ matrices $\m{X}_{n-1}^{(k)}$, with $k = 1 \ldots K$. Each of these matrices represents a different input feature map. The output feature maps $\m{X}_n^{(l)}$, $l = 1 \ldots L$ are represented as follows:

\begin{equation} \label{eq:activation-conv}
 \m{X}_n^{(l)} = f\left( \sum_{k=1}^{K} \m{W}_n^{(k,l)} * \m{X}_{n - 1}^{(k)} + b_n^{(l)}\right) .
\end{equation}

\noindent Here, $*$ represents the two-dimensional convolution operation, the matrices $\m{W}_n^{(k,l)}$ represent the \emph{filters} of layer $n$, and $b_n^{(l)}$ represents the bias for feature map $l$. Note that a feature map $\m{X}_n^{(l)}$ is obtained by computing a sum of $K$ convolutions with the feature maps of the previous layer. The bias $b_n^{(l)}$ can optionally be replaced by a matrix $\m{B}_n^{(l)}$, so that each spatial position in the feature map has its own bias (`untied' biases). This allows the sensitivity of the filters to vary across the input.

\begin{figure*}
\centering

\begin{tikzpicture}
    \node[anchor=south west,inner sep=0] (image) at (0,0) {\includegraphics[width=0.75\textwidth]{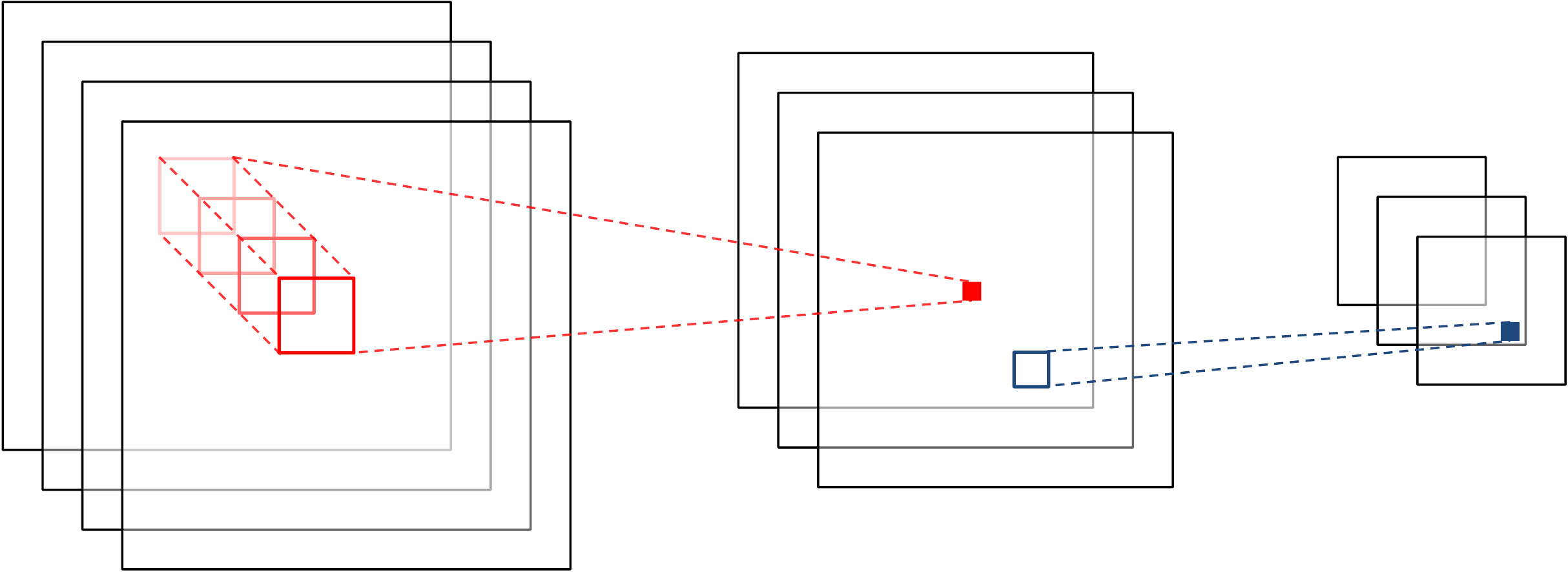}};
    \begin{scope}[x={(image.south east)},y={(image.north west)}]

        \node (idxinput) at (0.20, -0.10) {\textbf{layer $n-1$}};
        \node (idxconv) at (0.61, -0.10) {\textbf{layer $n$}};
	\node (idxpool) at (0.93, -0.10) {\textbf{layer $n+1$}};

        \node (labelinput) at (0.20, -0.17) {input};
        \node (labelconv) at (0.61, -0.17) {convolutions};
	\node (labelpool) at (0.93, -0.17) {pooling};
	
	\node (txt1) at (0.405, 0.07) {$\m{X}_{n-1}^{(k)}$};
	\node (txt2) at (0.78, 0.24) {$\m{X}_{n}^{(l)}$};
	\node (txt3) at (1.04, 0.40) {$\m{X}_{n+1}^{(l)}$};
	
  	\node[red] (txt4) at (0.20, 0.30) {$\m{W}_{n}^{(k,l)}$};
    \end{scope}
\end{tikzpicture}

\caption{A schematic overview of a convolutional layer followed by a pooling layer: each unit in the convolutional layer is connected to a local neighborhood in all feature maps of the previous layer. The pooling layer aggregates groups of neighboring units from the layer below.}
\label{fig:convpool}
\end{figure*}
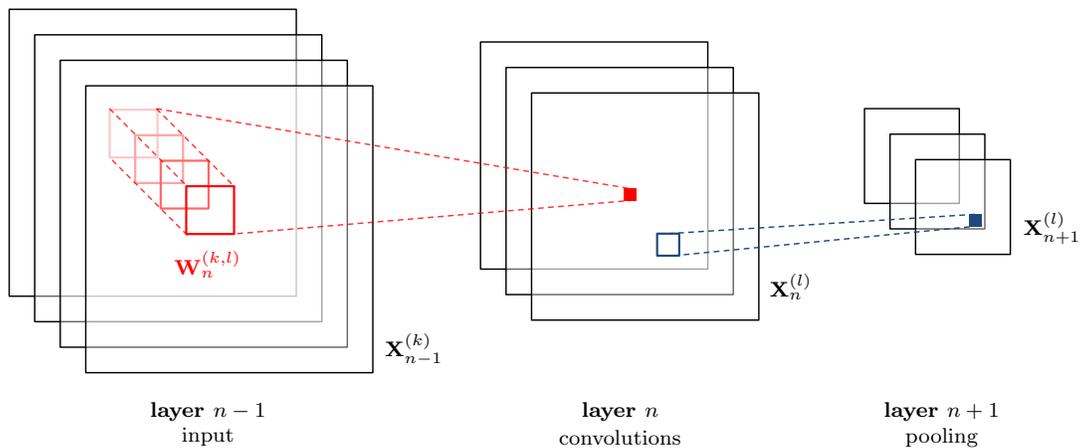

By replacing the matrix product with a sum of convolutions, the connectivity of the layer is effectively restricted to take advantage of the input structure and to reduce the number of parameters. Each unit is only connected to a local subset of the units in the layer below, and each unit is replicated across the entire input. This is shown in the left side of Figure~\ref{fig:convpool}. This means that each unit can be seen as detecting a particular feature across the input (for example, an oriented edge in an image). Applying feature detectors across the entire input enables the exploitation of translational symmetry in images. 

As a consequence of this restricted connectivity pattern, convolutional layers typically have far fewer parameters than traditional \emph{dense} (or \emph{fully-connected}) layers that compute a transformation of their input according to Equation~\ref{eq:activation}. This reduction in parameters can drastically improve generalization performance (i.e., predictive performance on unseen examples) and make the model scale to larger input dimensionalities.

Because convolutional layers are only able to model local correlations in the input, the dimensionality of the feature maps is often reduced between convolutional layers by inserting pooling layers. This allows higher layers to model correlations across a larger part of the input, albeit with a lower resolution. A pooling layer reduces the dimensionality of a feature map by computing some aggregation function (typically the maximum or the mean) across small local regions of the input \citep{Boureau10atheoretical}, as shown in the right side of Figure~\ref{fig:convpool}. This also makes the model invariant to small translations of the input, which is a desirable property for modelling images and many other types of data. Unlike convolutional layers, pooling layers typically do not have any trainable parameters.

By alternating convolutional and pooling layers, higher layers in the network see a progressively more coarse representation of the input. As a result, these layers are able to model higher-level abstractions more easily because each unit is able to see a larger part of the input.

Convolutional neural networks constitute the state of the art in many computer vision problems. Since their effectiveness for large-scale image classification was demonstrated, they have been ubiquitous in computer vision research \citep{Krizhevsky2012,razavian2014cnn,szegedy2014going,simonyan2014very}.

\section{Exploiting rotational symmetry}
\label{sec:rotation-invariance}

The restricted connectivity patterns used in convolutional neural networks drastically reduce the number of parameters required to model large images, by exploiting translational symmetry. However, there are many other types of symmetries that occur in images. For images of galaxies, rotating an image should not affect its morphological classification. This rotational symmetry is exploited by applying the same set of feature detectors to various rotated versions of the input. This further increases parameter sharing, which has a positive effect on generalization performance.

Whereas convolutions provide an efficient way to exploit translational symmetry, applying the same filter to multiple rotated versions of the input requires explicitly instantiating these versions. Additionally, rotating an image by an angle that is not a multiple of 90\degr~requires interpolation and results in an image whose edges are not aligned with the rows and columns of the pixel grid. These complications make exploiting rotational symmetry more challenging.

We note that the original Galaxy~Zoo project experimented with crowdsourced classifications of galaxies in which the images were both vertically and diagonally mirrored. \citet{lan08} showed that the raw votes had an excess of 2.5\% for S-wise (anticlockwise) spiral galaxies over Z-wise (clockwise) galaxies. Since this effect was seen in both the raw and mirrored images, it was interpreted as a bias due to preferences in the human brain, rather than as a true excess in the number of apparent S-wise spirals in the Universe. The existence of such a directional bias in the brain was demonstrated by \citet{gori2006reversal}. The \gztwo~probabilities do not contain any structures related to handedness or rotation-variant quantities, and no rotational or translational biases have yet been discovered in the data. If such biases do exist, however, this would presumably reduce the predictive power of the model since the assumption of rotational invariance to the output probabilities would no longer apply. 

\begin{figure*}
\centering

\begin{tikzpicture}
  \matrix[row sep=0.1cm, column sep=0.5cm] {
	&
	\node (rotate1) [minimum width=2.2cm, minimum height=2.2cm] {\pgftext{\includegraphics[width=2cm]{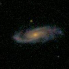}}}; \draw[red,thick] (-1, -0.3) rectangle (0.3, 1); &
	\node (crop1) [minimum width=1.4cm, minimum height=1.4cm] {\pgftext{\includegraphics[width=1.2cm]{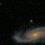}}}; &
	\node (conv1) [minimum width=2.7cm] {\pgftext{\includegraphics[width=2.5cm]{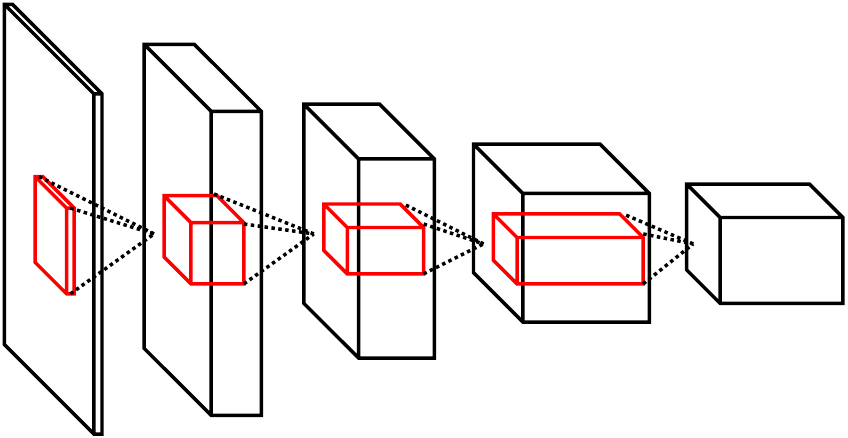}}}; &
	&
	& \\

	\node (inputimage) [minimum width=2.2cm, minimum height=2.2cm] {\pgftext{\includegraphics[width=2cm]{images/view_0.png}}}; &
	\node (rotate2) [minimum width=2.2cm, minimum height=2.2cm] {\pgftext{\includegraphics[width=2cm]{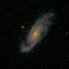}}}; \draw[red,thick] (-1, -0.3) rectangle (0.3, 1);  &
	\node (crop2) [minimum width=1.4cm, minimum height=1.4cm] {\pgftext{\includegraphics[width=1.2cm]{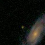}}}; &
	\node (conv2) [minimum width=2.7cm] {\pgftext{\includegraphics[width=2.5cm]{images/convpart_cropped.pdf}}}; &
	\node (dense) [minimum width=1.7cm, minimum height=1.7cm] {\pgftext{\includegraphics[width=1.5cm]{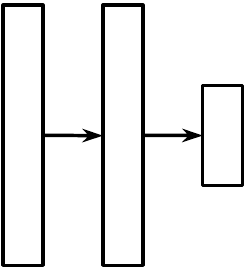}}}; &
	\node (output) [minimum width=1.4cm] {\pgftext{\includegraphics[width=1.2cm,angle=270]{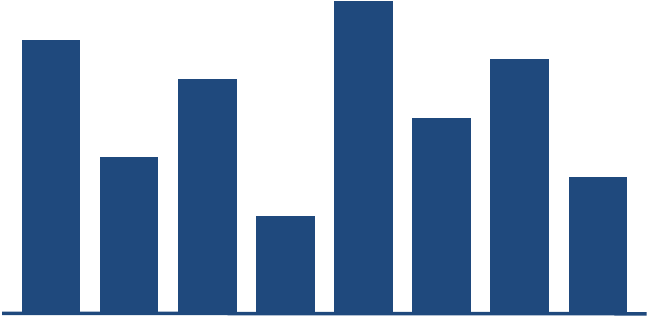}}}; & \\
	
	&
	\node (rotateetc) [] {$\vdots$}; &
	\node (cropetc) [] {$\vdots$}; &
	\node (convetc) [] {$\vdots$}; &
	&
	& \\
	
	&
	\node (rotate3) [minimum width=2.2cm, minimum height=2.2cm] {\pgftext{\includegraphics[width=2cm,angle=90]{images/view_0.png}}}; \draw[red,thick] (-1, -0.3) rectangle (0.3, 1);  &
	\node (crop3) [minimum width=1.4cm, minimum height=1.4cm] {\pgftext{\includegraphics[width=1.2cm]{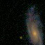}}}; &
	\node (conv3) [minimum width=2.7cm] {\pgftext{\includegraphics[width=2.5cm]{images/convpart_cropped.pdf}}}; &
	&
	& \\
	
  	\node (inputtext) [font=\scriptsize, anchor=base] {1. input}; &
	\node (rotatetext) [font=\scriptsize, anchor=base] {2. rotate}; &
	\node (croptext) [font=\scriptsize, anchor=base] {3. crop}; &
	\node (convtext) [align=center, font=\scriptsize, anchor=base] {4. convolutions}; &
	\node (densetext) [align=center, font=\scriptsize, anchor=base] {5. dense}; &
	\node (outputtext) [font=\scriptsize, anchor=base] {6. predictions}; & \\
};

    \path[-latex]
	(inputimage.north east) edge (rotate1.west)
	(inputimage.east) edge (rotate2.west)
	(inputimage.south east) edge (rotate3.west)
	
	(rotate1) edge (crop1)
	(rotate2) edge (crop2)
	(rotate3) edge (crop3)
	
	(crop1) edge (conv1)
	(crop2) edge (conv2)
	(crop3) edge (conv3)
	
	(conv1.east) edge (dense.north west)
	(conv2.east) edge (dense.west)
	(conv3.east) edge (dense.south west)
	
	(dense) edge (output)
        ;
\end{tikzpicture}

\caption{Schematic overview of a neural network architecture for exploiting rotational symmetry. The input image (1) is first rotated to various angles and optionally flipped to yield different viewpoints (2), and the viewpoints are subsequently cropped to reduce redundancy (3). Each of the cropped viewpoints is processed by the same stack of convolutional layers and pooling layers (4), and their output representations are concatenated and processed by a stack of dense layers (5) to obtain predictions (6).}
\label{fig:rotations}
\end{figure*}

Our approach for exploiting symmetry is visualized in Figure~\ref{fig:rotations}. We compute rotated and flipped versions of the input images, which are referred to as \emph{viewpoints}, and process each of these separately with the same convolutional network architecture, consisting of alternating convolutional layers and pooling layers. The output feature maps of this network for the different viewpoints are then concatenated, and one or more dense layers are stacked on top. This arrangement allows the dense layers to aggregate high-level features extracted from different viewpoints.

In practice, we also crop the top left part of each viewpoint image both to reduce redundancy between the viewpoints and to reduce the size of the input images (and hence computation time). Images are cropped in such a way that each one contains the centre of the galaxy, because this part of the image tends to be very informative. The practical implementation of viewpoint extraction is discussed in Section~\ref{sec:viewpoint-extraction}, and the modified network architecture is described in Section~\ref{sec:architecture}.

\section{Approach}
\label{sec:approach}

\begin{figure*}
\centering

\begin{tikzpicture}
  \tikzstyle{p}=[rectangle,align=center]
  \tikzstyle{pdata}=[p]
  \tikzstyle{pproc}=[p,draw=black, minimum width=4cm, minimum height=1.2cm]
  
  \matrix[row sep=0.5cm,column sep=1.0cm] {
	\node (input) [pdata] {input\\[0.3cm] \includegraphics[width=2.5cm]{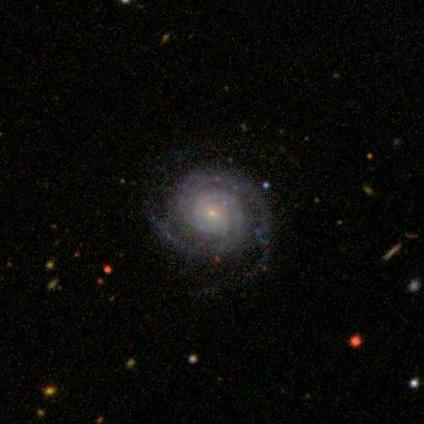}}; &
	\node (augmented) [pdata] {augmented input\\[0.3cm] \includegraphics[width=2.5cm]{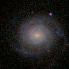}}; &
	\node (predictions) [pdata] {predictions\\[0.3cm] \includegraphics[width=2cm]{images/prob_bars1_cropped.pdf} };
	\\
	\node (preprocessing) [pproc] {preprocessing \\ {\scriptsize Section~\ref{sec:preprocessing}}}; &
	\node (viewpointextraction) [pproc] {viewpoint extraction \\ {\scriptsize Section~\ref{sec:viewpoint-extraction}}}; &
	\node (modelaveraging) [pproc] {model averaging \\ {\scriptsize Section~\ref{sec:model-averaging}}};
	\\
	\node (preprocessed) [pdata] {preprocessed input\\[0.3cm] \includegraphics[width=2.5cm]{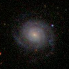}}; &
	\node (viewpoints) [pdata] {viewpoints\\[0.3cm] \includegraphics[width=1.2cm]{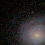} \includegraphics[width=1.2cm]{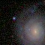}\\[0.1cm]  \includegraphics[width=1.2cm]{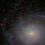}  \includegraphics[width=1.2cm]{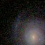} }; &
	\node (averagedpredictions) [pdata] {averaged predictions\\[0.3cm] \includegraphics[width=2cm]{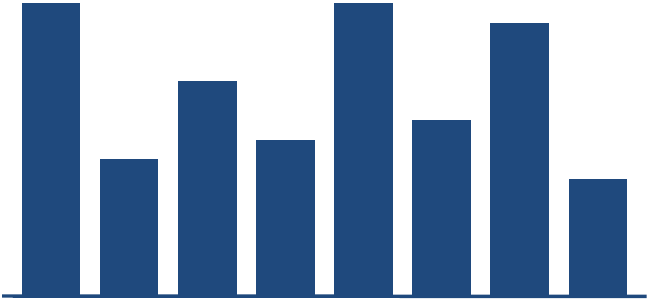}};
	\\
	\node (augmentation) [pproc] {augmentation \\ {\scriptsize Section~\ref{sec:augmentation}}}; &
	\node (convnet) [pproc] {convnet \\ {\scriptsize Section~\ref{sec:architecture}}}; &
	\\
    };

    \path[-latex]
	(input) edge (preprocessing)
	(preprocessing) edge (preprocessed)
	(preprocessed) edge (augmentation)
	(augmented) edge (viewpointextraction)
	(viewpointextraction) edge (viewpoints)
	(viewpoints) edge (convnet)
	(predictions) edge (modelaveraging)
	(modelaveraging) edge (averagedpredictions)
        ;
        
      \draw[-latex,rounded corners=5pt] (augmentation.east) -- +(0.5,0) |- (augmented.west);   
      \draw[-latex,rounded corners=5pt] (convnet.east) -- +(0.5,0) |- (predictions.west);
\end{tikzpicture}

\caption{Schematic overview of the processing pipeline.}
\label{fig:pipeline}
\end{figure*}

In this section, we describe our practical approach to developing and training a model for galaxy morphology prediction.
We first discuss our experimental setup and the problem of overfitting, which was the main driver behind our design decisions. We then describe the successive steps in our processing pipeline to obtain a set of answer probabilities from an image. This pipeline consists of five steps (Figure~\ref{fig:pipeline}): input preprocessing, augmentation, viewpoint extraction, a convolutional neural network and model averaging. We also briefly discuss the practical implementation of the pipeline from a software perspective.

\subsection{Experimental setup}
\label{sec:experimental-setup}

As described in Section~\ref{sec:galaxy-challenge}, the provided dataset consists of a training set with 61,578 images with associated answer probabilities, and an evaluation set of 79,975 images. Feedback could be obtained during the competition by submitting predictions for the images in the evaluation set. During the competition, submitted predictions were scored by computing the RMSE on a subset of approximately 25\% of the evaluation images. It was not revealed which images were part of this subset. The scores used to determine the final ranking were obtained by computing the RMSE on the remaining 75\% of images. This arrangement is typical for competitions hosted on the Kaggle platform. We split off a further 10\% of the training set images for real-time evaluation during model training, and trained our models only on the remaining 90\%.

\subsection{Avoiding overfitting}
\label{sec:overfitting}

Modern neural networks typically have a large number of learnable parameters -- several million in the case of our model. This is in stark contrast with the limited size of the training set, which had only $5\times10^4$ images. As a result, there is a high risk of \emph{overfitting}: a network will tend to memorize the training examples because it has enough capacity to do so, and will not generalize well to new data. We used several strategies to avoid overfitting:

\begin{itemize}
 \item \textbf{data augmentation}: extending the training set by randomly perturbing images in a way that leaves their associated answer probabilities unchanged;
 \item \textbf{regularization}: penalizing model complexity through use of dropout \citep{Hinton2012};
 \item \textbf{parameter sharing}: reducing the number of model parameters by exploiting translational and rotational symmetry in the input images;
 \item \textbf{model averaging}: averaging the predictions of several models.
\end{itemize}

\subsection{Preprocessing}
\label{sec:preprocessing}

Images are first cropped and rescaled to reduce the dimensionality of the input. It was useful to crop the images because the object of interest is in the middle of the image with a large amount of sky background, and typically fits within a square with a side of approximately half the image height. We then rescaled the images to speed up training, with little to no effect on predictive performance. Images were cropped from $424\times424$ pixels to $207\times207$, and then downscaled 3 times to $69\times69$ pixels.

For a small subset of the images, the cropping operation removed part of the object of interest, either because it had an unusually large angular size or because it was not perfectly centred. We looked into recentering and rescaling the images by detecting and measuring the objects in the images using \emph{SExtractor} \citep{bertin1996sextractor}. This allowed us to independently estimate both the position and Petrosian radii of the objects.
This information is then used to centre and rescale all images to standardize the sizes of the objects before further processing. 

This normalization step had no significant effect on the predictive performance of our models. Nevertheless, we did train a few models using this approach, because even though they achieved the same performance in terms of RMSE compared to models trained without it, the models make different mistakes. This is useful in the context of model averaging (Section~\ref{sec:model-averaging}), where high variance among a set of comparably performing models is desirable \citep{bishop2006pattern}.

The images for the competition were provided in the same format that is used on the Galaxy~Zoo website ($424\times424$ JPEG colour images). We found that keeping the colour information (instead of converting the images to grayscale) improved the predictive performance considerably, despite the fact that the colours are artificial and intended for human eyes. These artificial colours are nevertheless correlated with morphology, and our models are able to exploit this correlation.

\subsection{Data augmentation}
\label{sec:augmentation}

\begin{figure*}
        \centering
        \begin{subfigure}[b]{0.12\textwidth}
                \includegraphics[width=\textwidth]{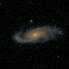}
                \caption{none}
                \label{fig:augmentations-none}
        \end{subfigure}%
        \quad\quad 
        \begin{subfigure}[b]{0.12\textwidth}
                \includegraphics[width=\textwidth]{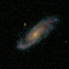}
                \caption{rotation}
                \label{fig:augmentations-rotate}
        \end{subfigure}%
        \quad\quad 
        \begin{subfigure}[b]{0.12\textwidth}
                \includegraphics[width=\textwidth]{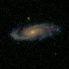}
                \caption{translation}
                \label{fig:augmentations-translate}
        \end{subfigure}%
        \quad\quad 
        \begin{subfigure}[b]{0.12\textwidth}
                \includegraphics[width=\textwidth]{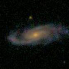}
                \caption{scaling}
                \label{fig:augmentations-zoom}
        \end{subfigure}%
        \quad\quad 
        \begin{subfigure}[b]{0.12\textwidth}
                \includegraphics[width=\textwidth]{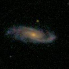}
                \caption{flipping}
                \label{fig:augmentations-flip}
        \end{subfigure}%
        \quad\quad 
        \begin{subfigure}[b]{0.12\textwidth}
                \includegraphics[width=\textwidth]{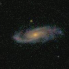}
                \caption{brightness}
                \label{fig:augmentations-brightness}
        \end{subfigure}
        \caption{The five types of random data augmentation used in this model. Note that the effect of translation and brightness adjustment is fairly subtle.}\label{fig:augmentations}
\end{figure*}

Due to the limited size of the training set, performing data augmentation to artificially increase the number of training examples is instrumental. Each training example was randomly perturbed in five ways, which are shown in Figure \ref{fig:augmentations}:

\begin{itemize}
 \item \textbf{rotation}: random rotation with an angle sampled uniformly between 0\degr~and 360\degr, to exploit rotational symmetry in the images.
 \item \textbf{translation}: random shift sampled uniformly between $-4$ and $4$ pixels (relative to the original image size of 424 by 424) in the x and y direction. The size of the shift is limited to ensure that the object of interest remains in the centre of the image.
 \item \textbf{scaling}: random rescaling with a scale factor sampled log-uniformly between $1.3^{-1}$ and $1.3$.
 \item \textbf{flipping}: the image is flipped with a probability of $0.5$.
 \item \textbf{brightness adjustment}: the colour of the image is adjusted as described by \citet{Krizhevsky2012}, with two differences: the first eigenvector has a much larger eigenvalue than the other two, so only this one is used, and the standard deviation for the scale factor is set to $\alpha=0.5$. In practice, this amounts to a brightness adjustment.
\end{itemize}

The first four of these are affine transformations, which can be collapsed into a single transformation together with the one used for preprocessing. This means that the data augmentation step has no noticeable computational cost. To maximize the effect of data augmentation, we randomly perturbed the images on demand during training, so the models were never presented with the exact same training example more than once.

\subsection{Viewpoint extraction}
\label{sec:viewpoint-extraction}

\begin{figure}
        \centering
        \begin{subfigure}[b]{0.45\textwidth}
                \includegraphics[width=\textwidth]{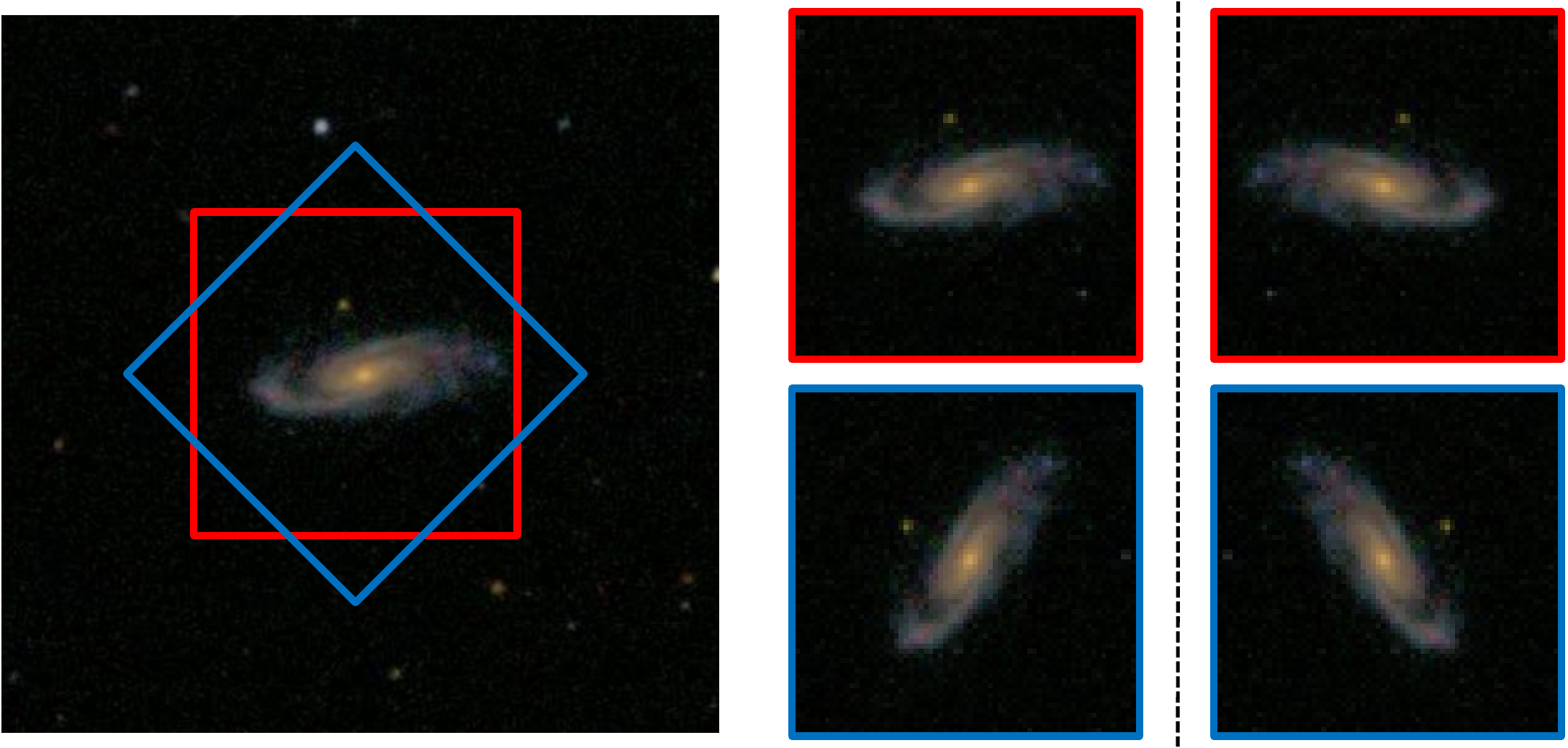}
                \caption{4 crops from an image}
                \label{fig:viewpoints-crops}
        \end{subfigure}%
        \vspace{1em}
        \begin{subfigure}[b]{0.45\textwidth}
                \includegraphics[width=\textwidth]{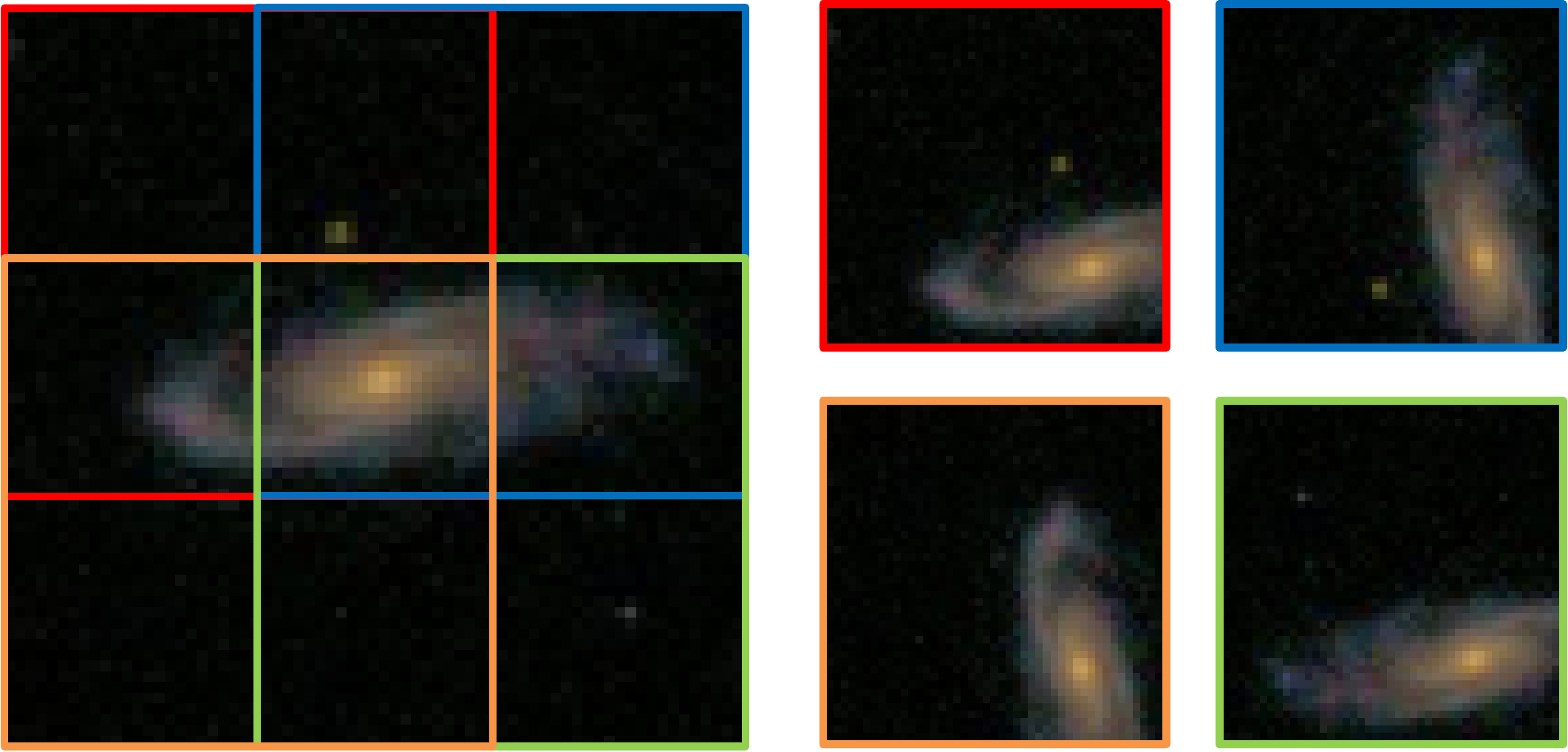}
                \caption{4 viewpoints from each crop}
                \label{fig:viewpoints-corner-patches}
        \end{subfigure}
        \caption{Obtaining 16 viewpoints from an input image. (\subref{fig:viewpoints-crops}) First, two square-shaped crops are extracted from the image, one at 0\degr~(red outline) and one at 45\degr~(blue outline). Both are also flipped horizontally to obtain 4 crops in total. (\subref{fig:viewpoints-corner-patches}) Then, four overlapping corner patches are extracted from each crop, and they are rotated so that the galaxy centre is in the bottom right corner of each patch. These 16 rotated patches constitute the viewpoints. This figure is best viewed in colour.}\label{fig:viewpoints}
\end{figure}

After preprocessing and augmentation, we extracted viewpoints by rotating, flipping and cropping the input images. We extracted 16 different viewpoints for each image: first, two square-shaped crops were extracted from an input image, one at 0\degr~and one at 45\degr. Both were also flipped horizontally to obtain 4 crops in total. Each of these crops is $69\times69$ pixels in size. Then, four overlapping corner patches of $45\times45$ pixels were extracted from each crop, and rotated so that the centre of the galaxy is in the bottom right corner of each patch. These 16 rotated patches constitute the viewpoints (Figure~\ref{fig:viewpoints}).

This approach allowed us to obtain 16 different viewpoints with just two affine transformation operations, thus avoiding additional computation. All viewpoints can be obtained from the two original crops without interpolation (which in practice are array indexing operations).
This also means that image edges and padding have no effect on the input, and that the loss of image fidelity after preprocessing, augmentation and viewpoint extraction is minimal.

\subsection{Network architecture}
\label{sec:architecture}

All viewpoints were presented to the network as 45 by 45 by 3 arrays of RGB values, scaled to the interval $[0,1]$, and processed by the same convolutional architecture. The resulting feature maps were then concatenated and processed by a stack of three fully connected layers to map them to the 37~answer probabilities.

\begin{figure*}
\centering
\includegraphics[width=0.9\textwidth]{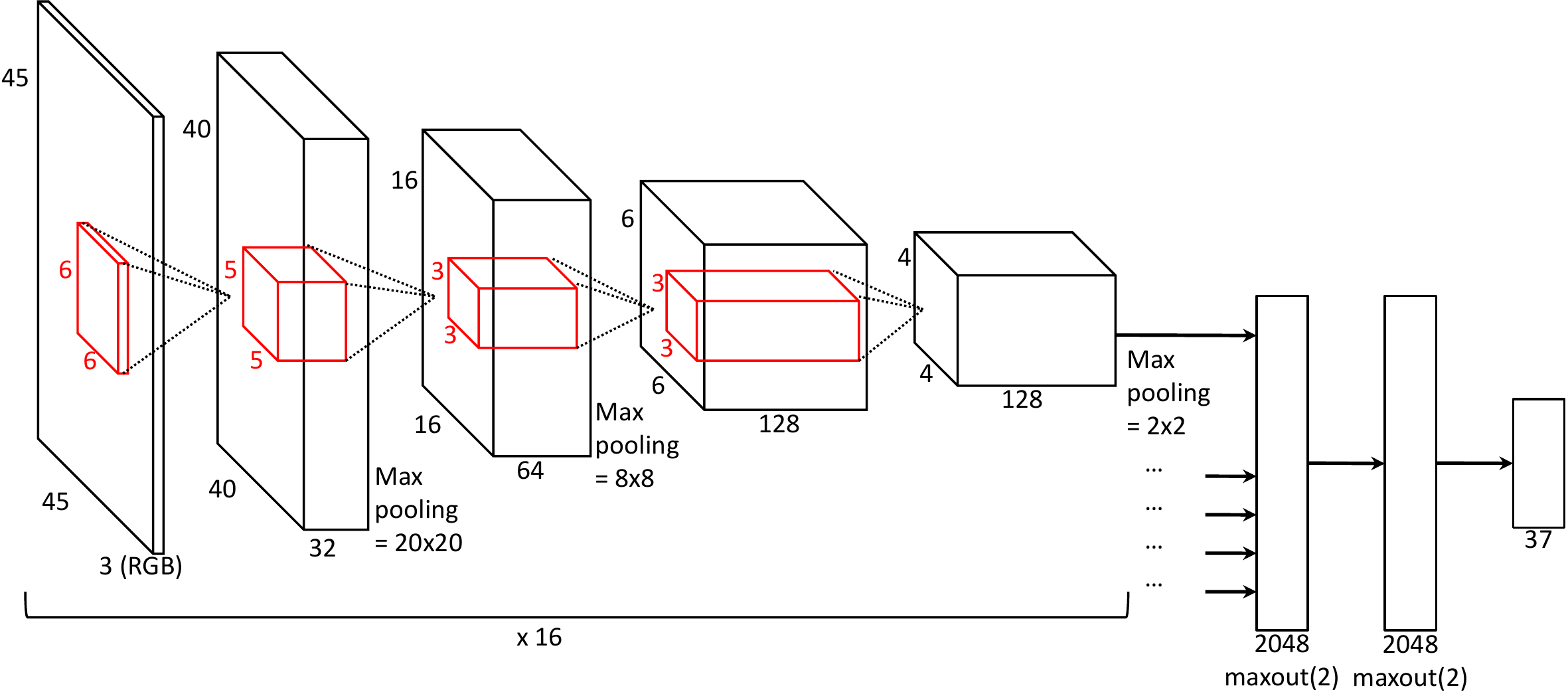}
\caption{Schematic overview of the architecture of the best performing network that we trained. The sizes of the filters and feature maps are indicated for each layer.}
\label{fig:architecture}
\end{figure*}

The architecture for the best performing network is visualized in Figure~\ref{fig:architecture}. There are four convolutional layers, all with square filters, with filter sizes 6, 5, 3 and 3 respectively, and with untied biases (i.e. each spatial position in each feature map has a separate bias, see Section~\ref{sec:convnets}). The rectification non-linearity is applied after each layer \citep{NAI10}. 2 by 2 max-pooling follows the first, second and fourth convolutional layers. The concatenated feature maps from all viewpoints are processed by a stack of three fully connected layers, consisting of two maxout layers \citep{goodfellow2013maxout} with 2048 units with two linear filters each, and a linear layer that outputs 37 real numbers. Maxout layers were used instead of ReLU layers to reduce the number of connections to the next layer (and thus the number of parameters). We did not use maxout in the convolutional layers because it proved too computationally intensive.

We arrived at this particular architecture after a manual parameter search: more than 100 architectures were evaluated over the course of the competition, and this one was found to yield the best predictive performance. The network has roughly 42~million trainable parameters in total. Table~\ref{tab:hyperparams} lists the hyperparameter settings for the trainable layers.

\begin{table*}
 \begin{tabular}{c|ccccccc}
    \noalign{\hrule height 1pt}
    & \textbf{type} & \textbf{\# features} & \textbf{filter size} & \textbf{non-linearity} & \textbf{initial biases} & \textbf{initial weights} \\
    \hline
    \textbf{1} & convolutional & 32 & $6 \times 6$ & ReLU & $0.1$ & $\mathcal{N}(0, 0.01)$ \\
    \textbf{2} & convolutional & 64 & $5 \times 5$ & ReLU & $0.1$ & $\mathcal{N}(0, 0.01)$ \\
    \textbf{3} & convolutional & 128 & $3 \times 3$ & ReLU & $0.1$ & $\mathcal{N}(0, 0.01)$ \\
    \textbf{4} & convolutional & 128 & $3 \times 3$ & ReLU & $0.1$ & $\mathcal{N}(0, 0.1)$ \\
    \textbf{5} & dense & 2048 & --  & maxout (2) & $0.01$ & $\mathcal{N}(0, 0.001)$ \\
    \textbf{6} & dense & 2048 & --  & maxout (2) & $0.01$ & $\mathcal{N}(0, 0.001)$ \\
    \textbf{7} & dense & 37 & -- & constraints & $0.1$ & $\mathcal{N}(0, 0.01)$ \\
    \noalign{\hrule height 1pt}
 \end{tabular} 
\caption{The hyperparameters of the trainable layers of the best performing network that we trained, also depicted in Figure~\ref{fig:architecture}. The last two columns describe the initialization distributions of the weights and biases of each layer. See Section~\ref{sec:architecture} for a description of the incorporation of the output constraints into the last layer of the network.}
 \label{tab:hyperparams}
\end{table*}

The 37~values that the network produces for an input image are converted into a set of probabilities. First, the values are passed through a rectification non-linearity, and then normalized per question to obtain a valid categorical probability distribution for each question. Valid probability distributions could also be obtained by using a softmax function per question, instead of rectification followed by normalization. However, this decreased the overall performance since it was harder for the network to predict a probability of exactly 0 or 1.

The distributions still need to be rescaled, however; they give the probability of an answer conditional on its associated question being asked, but each user is only asked a subset of the questions. This implies that some questions have a lower probability of being asked, so the probabilities of the answers to these questions should be scaled down to obtain unconditional probabilities. In practice we scale them by the probabilities of the answers that preceded them in the decision tree (see Figure~\ref{fig:decision-tree}).

This post-processing operation is incorporated into the network. Because it consists only of differentiable operations\footnote{Although the rectification operation is not technically differentiable everywhere, it is subdifferentiable so this does not pose a problem in practice.}, the gradient of the objective function can be backpropagated through it. This guarantees that the output of the network will not violate the constraints that the answer probabilities must adhere to (for example, $p_\textrm{bar}$ must be greater to or equal to $p_\textrm{spiral}$ in the cumulative probabilities, since it is a higher-level question in the decision tree). This resulted in a small but significant performance improvement.

In addition to the best performing network, we also trained variants for the purpose of model averaging (see Section~\ref{sec:model-averaging}). These networks differ slightly from the best performing network, and make slightly different predictions as a result.
Variants included:

\begin{itemize}
 \item a network with only two dense layers instead of three;
 \item a network with a different filter size configuration (filter sizes 8, 4, 3, 3 respectively instead of 6, 5, 3, 3);
 \item a network with ReLUs in the dense layers instead of maxout units;
 \item a network with 256 filters instead of 128 in the topmost convolutional layer.
\end{itemize}

In total, 17 different networks were trained on this data set.

\subsection{Training}
\label{sec:training}

To train the models we used minibatch gradient descent with a batch size\footnote{The batch size chosen is small because the convolutional part of the network is applied 16 times to different viewpoints of the input images, yielding an effective batch size of 256.} of 16 and \emph{Nesterov momentum} \citep{bengio2013advances} with coefficient $\mu = 0.9$. Nesterov momentum is a method for accelerating gradient descent by accumulating gradients over time in directions that consistently decrease the objective function value. This and similar methods have are commonly used neural network training because they speed up the training process and often lead to improved predictive performance \citep{sutskever2013importance}.

We performed approximately 1.5~million gradient updates, corresponding to 25~million training examples. Following \citet{Krizhevsky2012}, we used a discrete learning rate schedule to improve convergence. We began with a constant learning rate $\eta = 0.04$ and decreased it tenfold twice: it was decreased to $0.004$ after 18~million examples, and to $0.0004$ after 23~million examples. For the first 10,000 examples, the output constraints were ignored, and the linear output of the top layer of the network was simply clipped between 0 and 1. This was necessary to ensure convergence.

Weights in the model were initialized by sampling from zero-mean normal distributions \citep{Bengio2012practical}. The variances of these distributions were fixed at each layer, and were manually chosen to ensure proper flow of the gradient through the network. All biases were initialized to positive values to decrease the risk of units getting stuck in the saturation region. Although this is not necessary for maxout units, the same strategy was used for the dense layers. The initialization strategy for all layers is shown in the last two columns of Table~\ref{tab:hyperparams}.

During training, we used dropout \citep{Hinton2012} in all three dense layers. Using dropout was essential to reduce overfitting to manageable levels.

\subsection{Model averaging}
\label{sec:model-averaging}

To further improve the prediction accuracy, we averaged the predictions of several different models, and across several transformations of the input images. Two requirements for model averaging to be effective is that each individual model must have roughly the same prediction accuracy, and the prediction errors should be as uncorrelated as possible.

For each model, we computed predictions for 60~affine transformations of the input images: a combination of 10~rotations, spaced by 36\degr, 3~rescalings (with scale factors $1.2^{-1}$, $1$ and $1.2$) and optional horizontal flipping. An unweighted average of the predictions was computed. Even though the model is trained to be robust to these types of deformations (see Section~\ref{sec:augmentation}), computing averaged predictions in this fashion still helped to increase prediction accuracy (see Table~\ref{tab:results}).

In total, 17 variants of the model were trained with predictions computed from the mean across 60 transformations. This resulted in 1020 sets of predictions averaged in total.

\subsection{Implementation}
\label{sec:implementation}

All aspects of the model were implemented using Python and the Theano library \citep{bergstra+al:2010-scipy,bastien2012theano}. This allowed the use of GPU acceleration without any additional effort. Theano is also able to perform automatic differentiation, which simplifies the implementation of gradient-based optimization techniques. Networks were trained on NVIDIA GeForce GTX~680 cards. Data augmentation was performed on the CPU using the {\tt scikit-image} package \citep{van2014scikit} in parallel with model training on the GPU. Training the network described in Section~\ref{sec:architecture} took roughly 67~hours in real time.

The code to reproduce the winning submission for the Galaxy Challenge is available at \texttt{https://github.com/benanne/kaggle-galaxies}.

\section{Results}
\label{sec:results}

Competition results of the models are listed in Table~\ref{tab:results}. We report the performance of our best performing network, with and without averaging across 60~transformations, as well as that of the combination of all 17~variants. The root-mean-square error in Table~\ref{tab:results} is the same metric used to score submissions in the Galaxy Challenge (Equation \ref{eq:rmse}).
Both averaging across transformations and averaging across different models contributed significantly to the final score. It is worth noting that our model performs well even without any model averaging, which is important because fast inference is desirable for practical applications. If predictions are to be generated for millions of images, combining a large number of predictions for each image would require an impractical amount of computation.

\begin{table}
\centering
 \begin{tabular}{c|cc}
    \noalign{\hrule height 1pt}
    \textbf{model} & \multicolumn{2}{c}{\textbf{leaderboard score}} \\
    & public & private \\
    \hline
    best performing network & 0.07671 & 0.07693 \\
    + averaging over 60 transformations & 0.07579 & 0.07603 \\
    + averaging over 17 networks & 0.07467 & 0.07492 \\
    \noalign{\hrule height 1pt}
 \end{tabular} 
\caption{Performance (in RMSE) of the best performing network, as well as the performance after averaging across 60~transformations of the input, and across 17~variants of the network. Please refer to Section~\ref{sec:galaxy-challenge} for details on how the scores were computed.}
 \label{tab:results}
\end{table}

Although morphology prediction was framed as a regression problem in the competition (see Section~\ref{sec:galaxy-challenge}), it is fundamentally a classification task. To demonstrate the capabilities of our model in a more interpretable fashion, we can look at classification accuracies. For each question, we can obtain classifications by selecting the answer with the highest probability for each image. We can do this both for the probabilities obtained from Galaxy~Zoo participants, and for the probabilities predicted by our model. We can then compute the classification accuracy simply by counting the number of images for which the classifications match up. Reducing the probability distributions to classifications in this fashion clearly causes some information to be discarded, but classification accuracy is a metric that is much easier to interpret.

To find out how the level of agreement between the Galaxy~Zoo participants affects the accuracy of the predictions of our model, we can compute the entropy of the probability distribution over the answers for a given question.
The entropy of a discrete probability distribution $p$ over $n$ options $x_1, \ldots, x_n$ is given by:

\begin{equation}\label{eq:entropy}
 H(p) = - \sum_{i=1}^{n} p(x_i) \log p(x_i) .
\end{equation}

If the entropy is minimal, all participants selected the same answer (i.e. everyone agreed). If the entropy is maximal, all answers were equally likely to be selected. The entropy ranges between $0$ and $\log(n)$. We can convert it into a measure of agreement $a(p)$ as follows:

\begin{equation}\label{eq:agreement}
 a(p) = 1 - \dfrac{H(p)}{\log(n)}.
\end{equation}

The quantity $a(p)$ will equal $0$ in case of maximal disagreement, and $1$ in case of maximal agreement.

To assess the conditions under which the predictions of the model can be trusted, we can measure the confidence of a prediction using the same measure $a(p)$ by applying it to the probability distributions predicted by the model, instead of the distributions of the crowdsourced answers. This allows us to relate model confidence and prediction accuracy.

\begin{figure*}
        \centering
        \begin{subfigure}[b]{0.32\textwidth}
                \includegraphics[width=\textwidth]{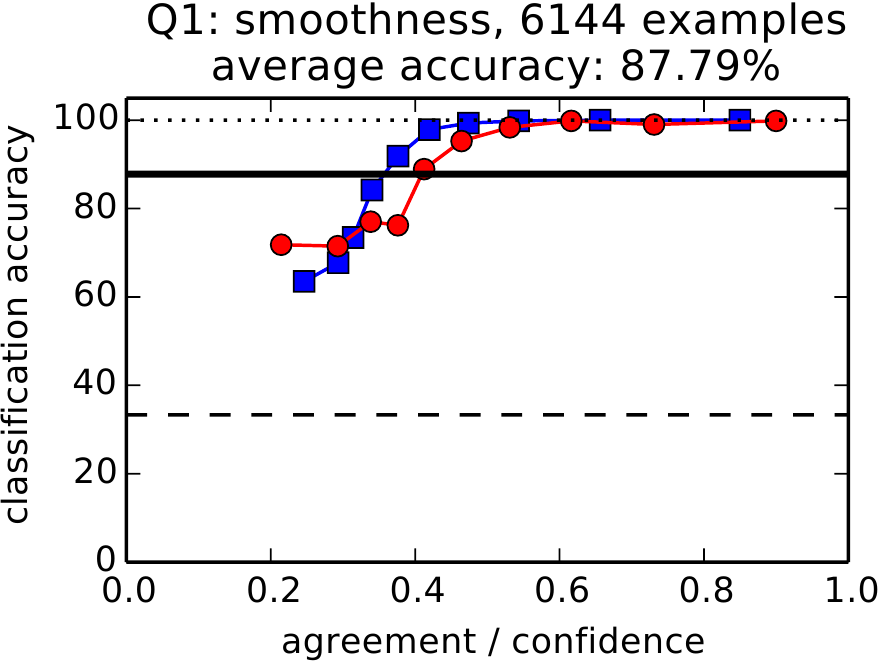}
                \label{fig:disagreement0}
        \end{subfigure}%
        ~ 
        \begin{subfigure}[b]{0.32\textwidth}
                \includegraphics[width=\textwidth]{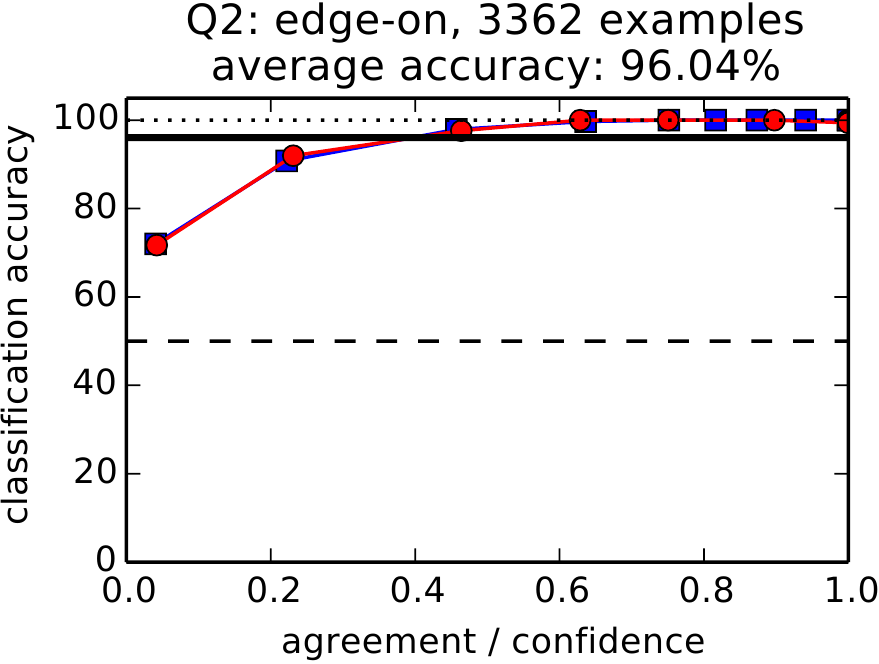}
                \label{fig:disagreement1}
        \end{subfigure}
        ~ 
        \begin{subfigure}[b]{0.32\textwidth}
                \includegraphics[width=\textwidth]{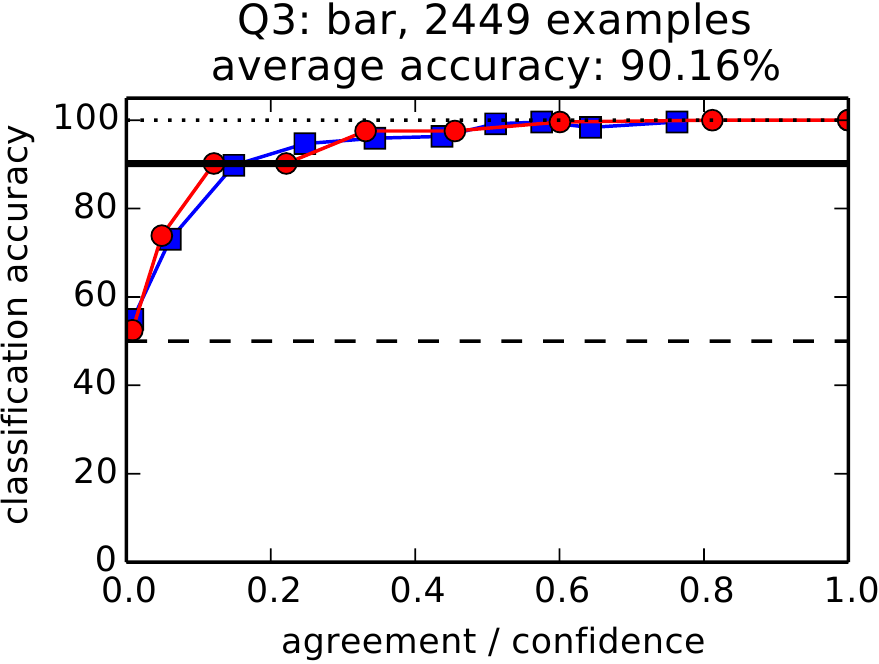}
                \label{fig:disagreement2}
        \end{subfigure}
        
	\begin{subfigure}[b]{0.32\textwidth}
                \includegraphics[width=\textwidth]{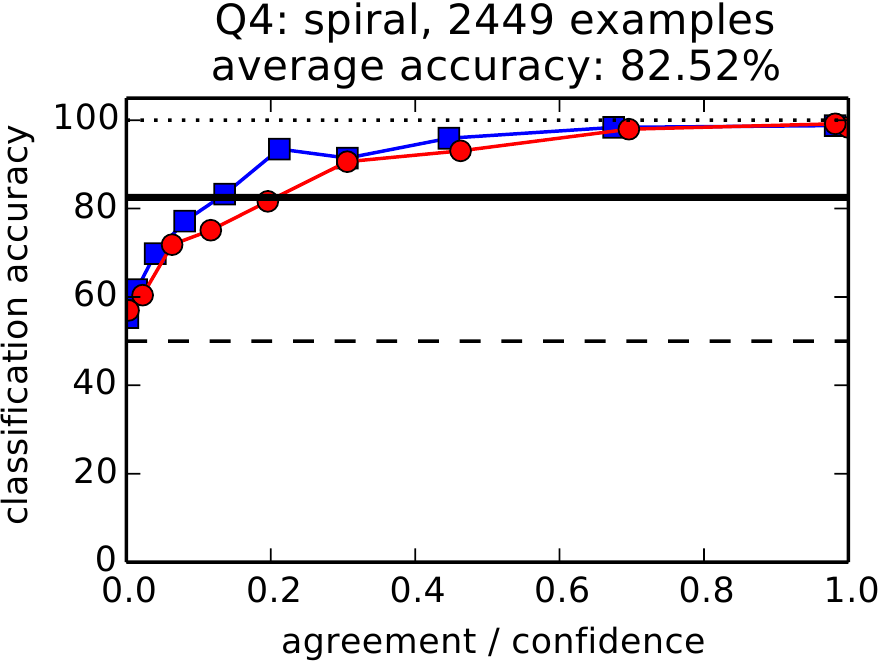}
                \label{fig:disagreement3}
        \end{subfigure}%
        ~ 
        \begin{subfigure}[b]{0.32\textwidth}
                \includegraphics[width=\textwidth]{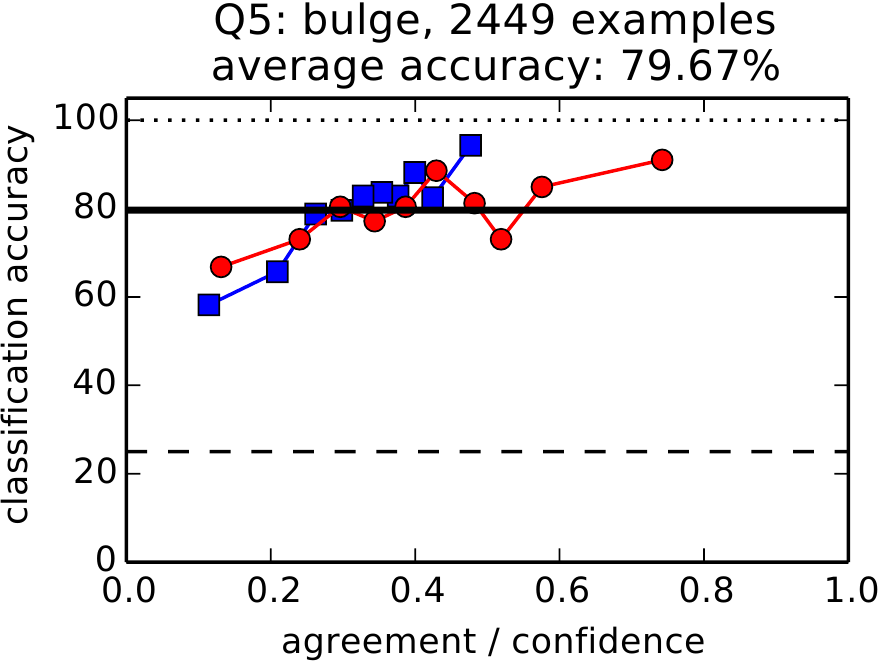}
                \label{fig:disagreement4}
        \end{subfigure}
        ~ 
        \begin{subfigure}[b]{0.32\textwidth}
                \includegraphics[width=\textwidth]{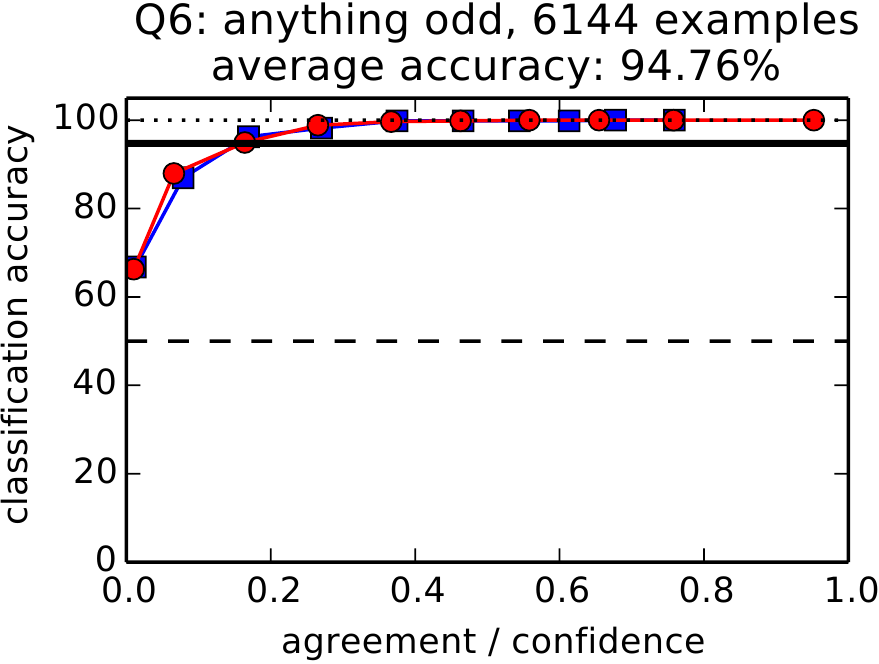}
                \label{fig:disagreement5}
        \end{subfigure}
        
	\begin{subfigure}[b]{0.32\textwidth}
                \includegraphics[width=\textwidth]{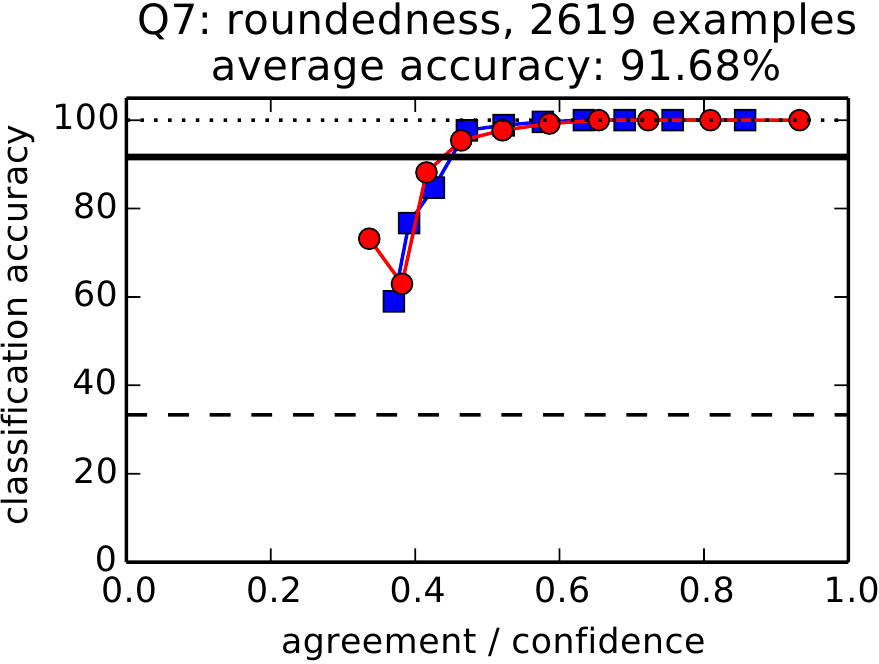}
                \label{fig:disagreement6}
        \end{subfigure}%
        ~ 
        \begin{subfigure}[b]{0.32\textwidth}
                \includegraphics[width=\textwidth]{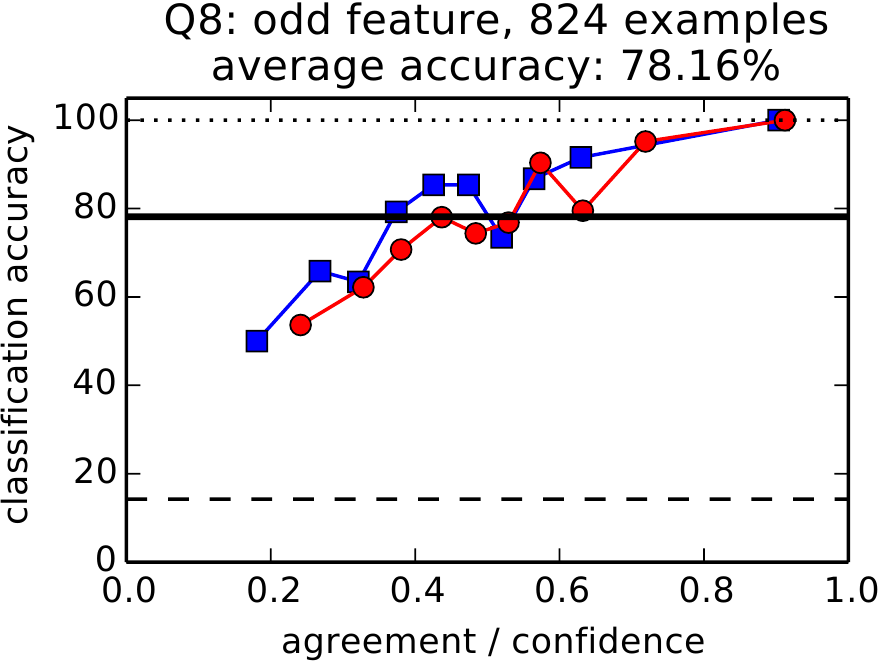}
                \label{fig:disagreement7}
        \end{subfigure}
        ~ 
        \begin{subfigure}[b]{0.32\textwidth}
                \includegraphics[width=\textwidth]{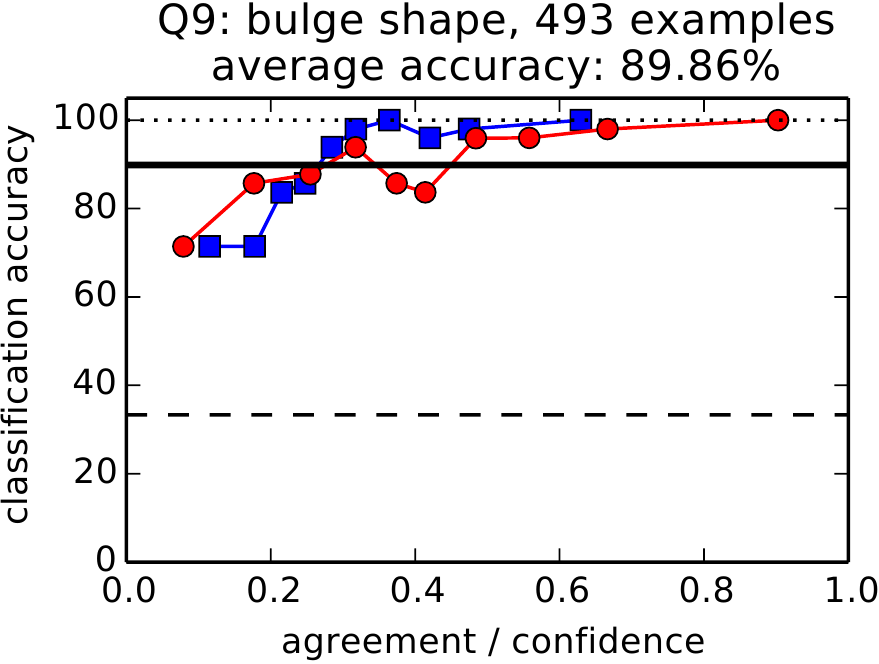}
                \label{fig:disagreement8}
        \end{subfigure}
        
	\begin{subfigure}[b]{0.32\textwidth}
                \includegraphics[width=\textwidth]{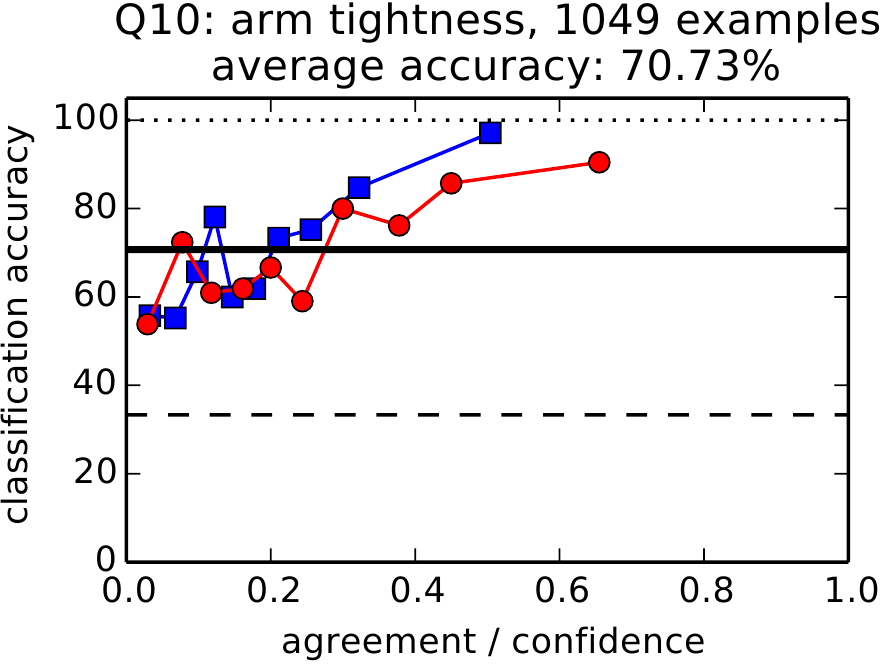}
                \label{fig:disagreement9}
        \end{subfigure}%
        ~ 
        \begin{subfigure}[b]{0.32\textwidth}
                \includegraphics[width=\textwidth]{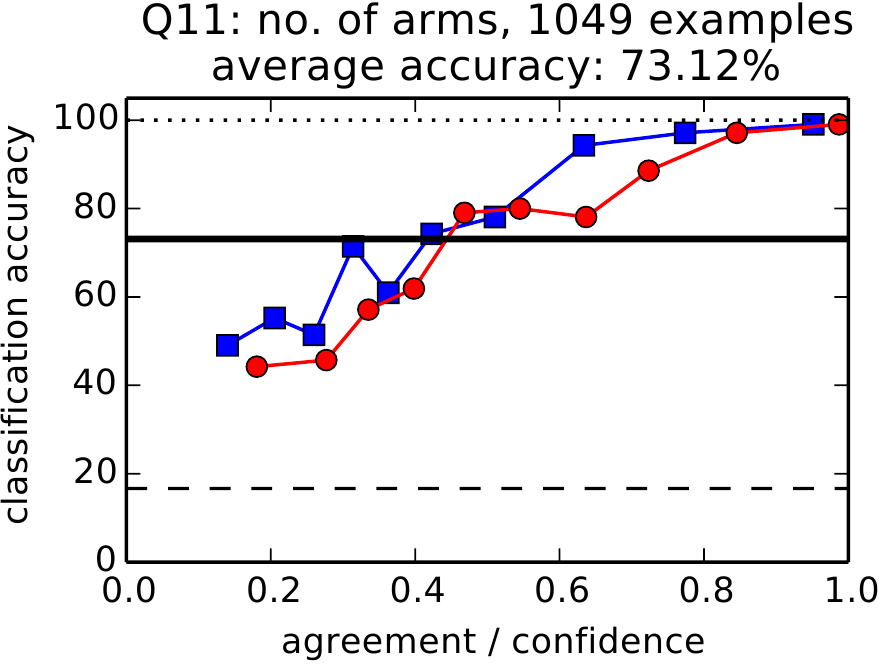}
                \label{fig:disagreement10}
        \end{subfigure}
        \caption{Level of agreement (red circles) and model confidence (blue squares) versus classification accuracy for all questions (see Table~\ref{tab:questions}), computed on the real-time evaluation set. The overall classification accuracy is indicated as a thick horizontal line. The dotted and dashed horizontal lines indicate the maximal accuracy of 100\% and the chance-level accuracy respectively. The number of images that were included in the analysis and the overall classification accuracy for each question are indicated above the graphs.}\label{fig:disagreement}
\end{figure*}

For each question, we selected the subset of images from the real-time evaluation set\footnote{We could also have conducted this analysis on the evaluation set from the competition, but since the true answer probabilities for the real-time evaluation set were readily available and this set contains over 6,000 images, we used this instead.} for which at least 50\% of participants answered the question. This is to ensure that we only consider images for which the question is likely relevant. We ranked all images in this subset according to the measure $a(p)$ and divided them into 10 equal bins. We did this seperately for both the crowdsourced answers and the model predictions. For each bin, we computed the average of $a(p)$ and the classification accuracy using the best performing network (no averaging). These values are visualized in a set of graphs for each question in Figure~\ref{fig:disagreement}. The red circles show classification accuracy versus agreement. The blue squares show classification accuracy versus model confidence. The classification accuracy across the entire subset is also shown as a thick horizontal line. The dashed horizontal lines indicate the maximal accuracy of 100\% and the chance-level accuracy, which depends on the number of options. The number of images in each subset and the overall classification accuracy are indicated above the graphs.

For all questions, the classification accuracy tapers off as the level of agreement between Galaxy~Zoo participants decreases. This makes sense, as those images are harder to classify. \citet{kuminski2014combining} report similar results using the WND-CHARM algorithm, with lowest accuracies for features describing spiral arm and irregular structures. Our model achieves near-perfect accuracy for most of the questions when the level of agreement is high. Classifications for bulge dominance (Q5) and spiral arm tightness (Q10) have low agreement overall, and are also more difficult to answer for the model.

Similarly, the confidence of the model in its predictions is correlated with classification accuracy: we achieve near-perfect accuracy for most questions when the model is highly confident. This is a useful property, because it allows us to determine when we are able to trust the predictions, and when we should defer to an expert instead. As a consequence, the model could be used to filter a large collection of images, in order to obtain a much smaller set that can be annotated manually by experts. Such a two-stage approach would greatly reduce the experts' workload at virtually no cost in terms of accuracy.

For questions 1, 2, 3, 6 and 7 in particular, we are able to make confident, accurate predictions for the majority of examples. This would allow us to largely automate the assessment of e.g. smoothness (Q1) and roundedness (Q7). For questions 5 and 10 on the other hand, confidence is low across the board and the classification accuracy is usually too low to be of practical use. As a result, determining bulge dominance (Q5) and spiral arm tightness (Q10) would still require a lot of manual input. The level to which we are able to automate the annotation process depends on the morphological properties we are interested in, as well as the distribution of morphology types in the dataset we wish to analyse.

To assess how well the model is able to predict various different morphology types, we computed precision and recall scores for all answers individually. The precision ($P$) and recall ($R$) scores are defined in terms of the number of true positive ($TP$), false positive ($FP$) and false negative ($FN$) classifications as follows:

\begin{equation}\label{eq:precision-recall}
 P = \dfrac{TP}{TP + FP}, \quad
 R = \dfrac{TP}{TP + FN}.
\end{equation}

The scores are listed in Table \ref{tab:precision-recall}. We used the same strategy as before to obtain classifications, and only considered those examples for which at least 50\% of the Galaxy Zoo participants answered the question. The numbers of examples that were available for each question and answer are also shown.

From these scores, we can establish that the model has more difficulty with morphology types that occur less frequently in the dataset, e.g., \textit{star or artifact} (A1.3), \textit{no bulge} (A5.1), \textit{dominant bulge} (A5.4) and \textit{dust lane} (A8.7). We note that images in the first category are attempted to be delibrately excluded from the Galaxy Zoo data set via flags in the SDSS pipeline. Both the precision and recall scores are affected, so this effect cannot be attributed entirely to a bias towards more common morphologies. However, recall is generally affected more strongly than precision, which indicates that the model is more conservative in predicting rare morphology types. For a few very rare answers, we were unable to compute precision scores because the model never predicted them for the examples that were considered: \textit{lens or arc} (A8.2), \textit{boxy bulge} (A9.2) and \textit{four spiral arms} (A11.4). While these are all rare morphologies, they have considerable scientific interest and constructing a model that can accurately identify them is still a primary goal.

\begin{table}
\centering
\begin{tabular}{ccccc}
  \noalign{\hrule height 1pt}
  & & \textbf{precision} & \textbf{recall} & \textbf{\# examples} \\
  \hline
  \multicolumn{4}{l}{Q1: smoothness} & 6144 \\
  \hline
  A1.1 & smooth & 0.8459 & 0.8841 & 2700 \\
  A1.2 & features or disk & 0.9051 & 0.8742 & 3435 \\
  A1.3 & star or artifact & 1.0000 & 0.4444 & 9 \\

  \hline
  \multicolumn{4}{l}{Q2: edge-on} & 3362 \\
  \hline
  A2.1 & yes & 0.9065 & 0.8885 & 655 \\
  A2.2 & no & 0.9732 & 0.9778 & 2707 \\

  \hline
  \multicolumn{4}{l}{Q3: bar} & 2449 \\
  \hline
  A3.1 & yes & 0.7725 & 0.7101 & 483 \\
  A3.2 & no & 0.9302 & 0.9486 & 1966 \\

  \hline
  \multicolumn{4}{l}{Q4: spiral} & 2449 \\
  \hline
  A4.1 & yes & 0.8715 & 0.8270 & 1451 \\
  A4.2 & no & 0.7659 & 0.8226 & 998 \\

  \hline
  \multicolumn{4}{l}{Q5: bulge} & 2449 \\
  \hline
  A5.1 & no bulge & 0.6697 & 0.5000 & 146 \\
  A5.2 & just noticeable & 0.7828 & 0.8475 & 1174 \\
  A5.3 & obvious & 0.8292 & 0.8049 & 1092 \\
  A5.4 & dominant & 0.4444 & 0.1081 & 37 \\

  \hline
  \multicolumn{4}{l}{Q6: anything odd} & 6144 \\
  \hline
  A6.1 & yes & 0.8438 & 0.7500 & 828 \\
  A6.2 & no & 0.9617 & 0.9784 & 5316 \\

  \hline
  \multicolumn{4}{l}{Q7: roundedness} & 2619 \\
  \hline
  A7.1 & completely round & 0.9228 & 0.9282 & 1197 \\
  A7.2 & in between & 0.9128 & 0.9171 & 1279 \\
  A7.3 & cigar-shaped & 0.9000 & 0.8182 & 143 \\

  \hline
  \multicolumn{4}{l}{Q8: odd feature} & 824 \\
  \hline
  A8.1 & ring & 0.9097 & 0.9161 & 143 \\
  A8.2 & lens or arc & ?     & 0.0000 & 2 \\
  A8.3 & disturbed & 0.8000 & 0.4138 & 29 \\
  A8.4 & irregular & 0.8579 & 0.8674 & 181 \\
  A8.5 & other & 0.6842 & 0.6810 & 210 \\
  A8.6 & merger & 0.7398 & 0.7773 & 256 \\
  A8.7 & dust lane & 0.5000 & 0.6667 & 3 \\

  \hline
  \multicolumn{4}{l}{Q9: bulge shape} & 493 \\
  \hline
  A9.1 & rounded & 0.9143 & 0.9412 & 340 \\
  A9.2 & boxy & ?     & 0.0000 & 8 \\
  A9.3 & no bulge & 0.8601 & 0.8483 & 145 \\

  \hline
  \multicolumn{4}{l}{Q10: arm tightness} & 1049 \\
  \hline
  A10.1 & tight & 0.7500 & 0.7350 & 449 \\
  A10.2 & medium & 0.6619 & 0.7112 & 457 \\
  A10.3 & loose & 0.7373 & 0.6084 & 143 \\

  \hline
  \multicolumn{4}{l}{Q11: no. of arms} & 1049 \\
  \hline
  A11.1 & 1 & 1.0000 & 0.2037 & 54 \\
  A11.2 & 2 & 0.8201 & 0.8691 & 619 \\
  A11.3 & 3 & 0.4912 & 0.3182 & 88 \\
  A11.4 & 4 & ?     & 0.0000 & 21 \\
  A11.5 & more than 4 & 0.4000 & 0.4000 & 20 \\
  A11.6 & can't tell & 0.5967 & 0.7368 & 247 \\

  \noalign{\hrule height 1pt}
\end{tabular}
\caption{Precision and recall scores for each answer. We compute these values only for the subset of examples in the real-time evaluation set where at least 50\% of participants answered the question. We also give the number of examples that are in this subset for each answer. A question mark indicates that we were unable to compute the precision score because the model did not predict this answer for any of the considered examples.}
\label{tab:precision-recall}
\end{table}

\section{Analysis}
\label{sec:analysis}

Traditionally, neural networks are often treated as black boxes that perform some complicated and uninterpretable sequence of computations that yield a good approximation to the desired output. However, analysing the parameters of a trained model can be very informative, and sometimes even leads to new insights about the problem the network is trying to solve \citep{zeiler2014visualizing}. This is especially true for convolutional neural networks trained on images, where the first-layer filters can be interpreted visually.

\begin{figure*}
        \centering
        \begin{subfigure}[b]{0.25\textwidth}
                \includegraphics[width=\textwidth]{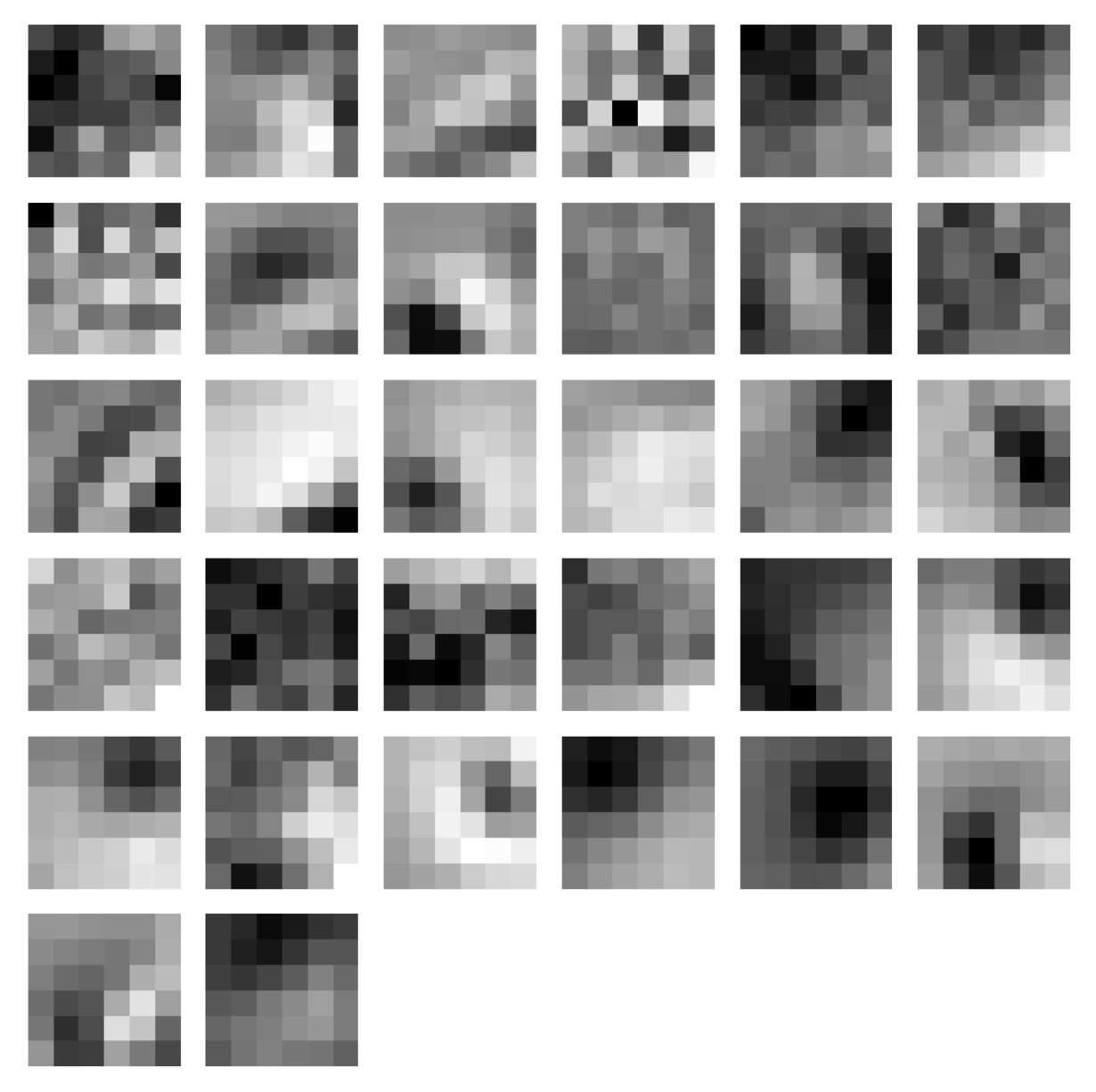}
                \caption{red channel}
                \label{fig:filters-channel0}
        \end{subfigure}%
        \quad\quad 
        \begin{subfigure}[b]{0.25\textwidth}
                \includegraphics[width=\textwidth]{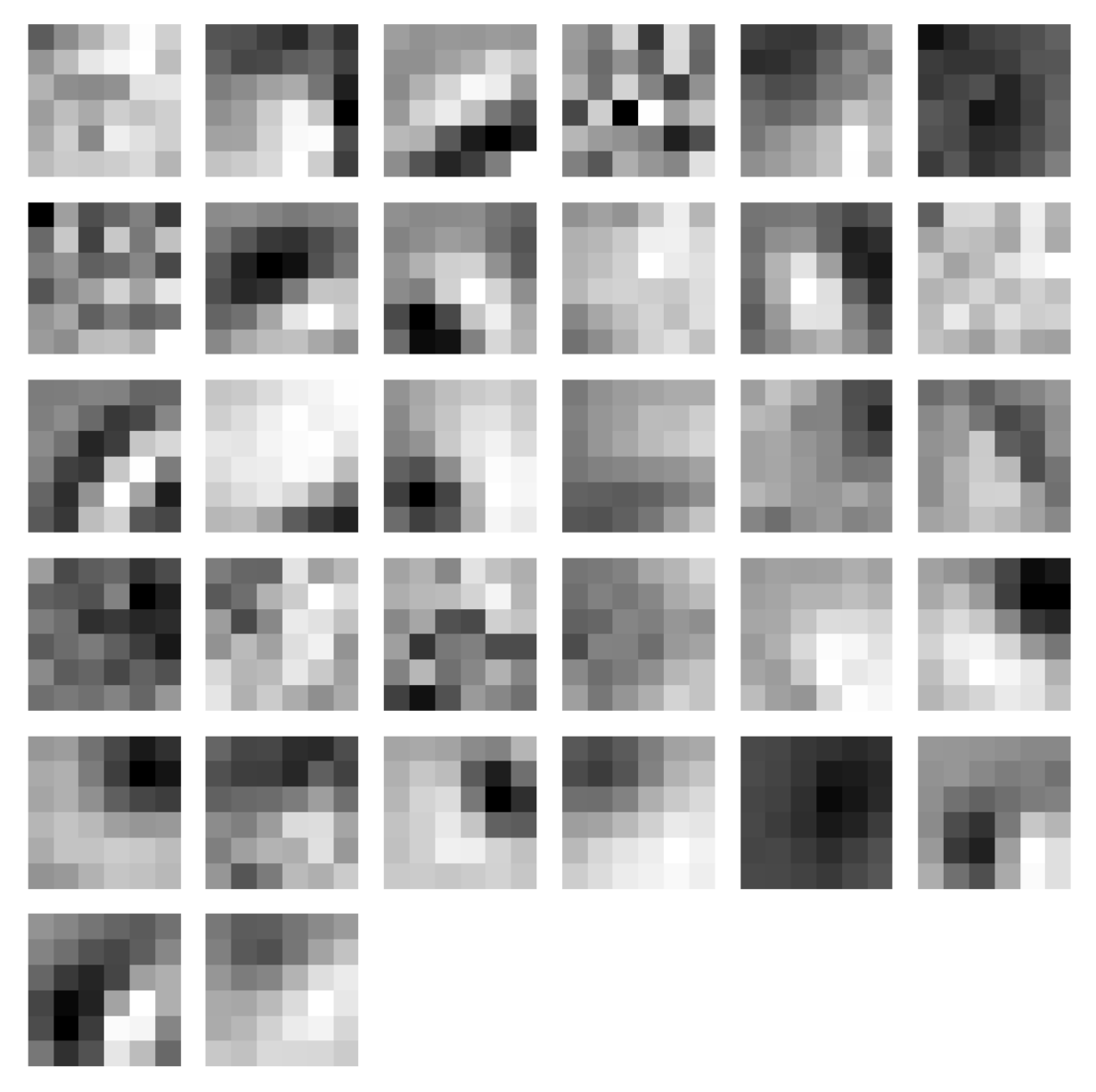}
                \caption{green channel}
                \label{fig:filters-channel1}
        \end{subfigure}
        \quad\quad 
        \begin{subfigure}[b]{0.25\textwidth}
                \includegraphics[width=\textwidth]{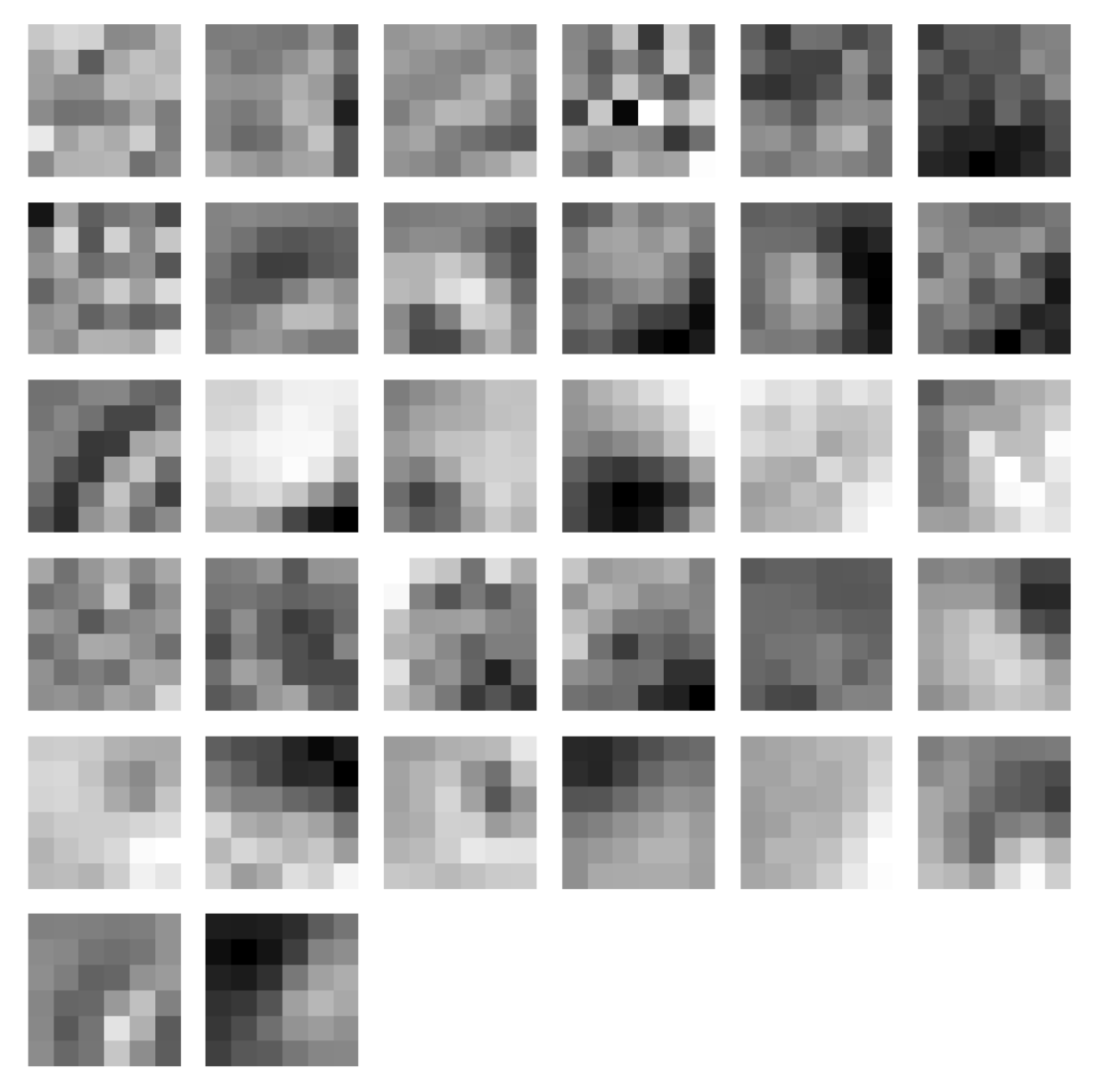}
                \caption{blue channel}
                \label{fig:filters-channel2}
        \end{subfigure}
        \caption{The 32 filters learned in the first convolutional layer of the best-performing network. Each filter was contrast-normalized individually across all channels.}\label{fig:filters}
\end{figure*}

Figure~\ref{fig:filters} shows the 32~filters learned in the first layer of the best performing network described in Section~\ref{sec:architecture}. Each filter was contrast-normalized individually to bring out the details, and the three colour channels are shown separately. Comparing the filter weights across colour channels reveals that some filters are more sensitive to particular colours, while others are sensitive to patterns, edges and textures. The same phenomenon is observed when training convolutional neural networks on more traditional image datasets. The filters for edge detection seem to be looking for curved edges in particular, which is to be expected because of the radial symmetry of the input images.

\begin{figure*}
        \centering
        \begin{subfigure}[b]{0.25\textwidth}
                \includegraphics[width=\textwidth]{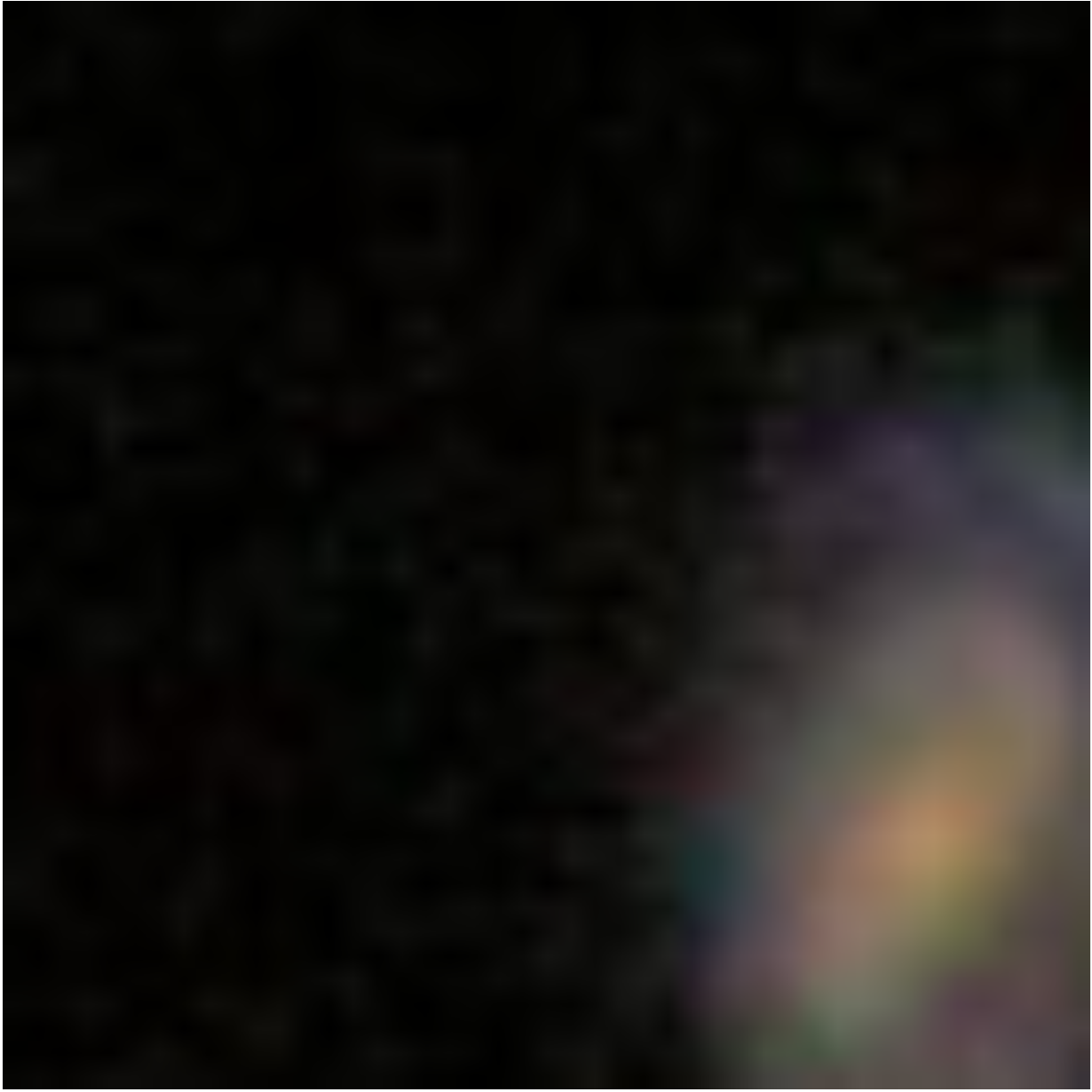}
                \caption*{\footnotesize input ($45 \times 45$)}
                \label{fig:activations-0-input}
        \end{subfigure}%
        ~ 
        \begin{subfigure}[b]{0.30\textwidth}
                \includegraphics[width=\textwidth]{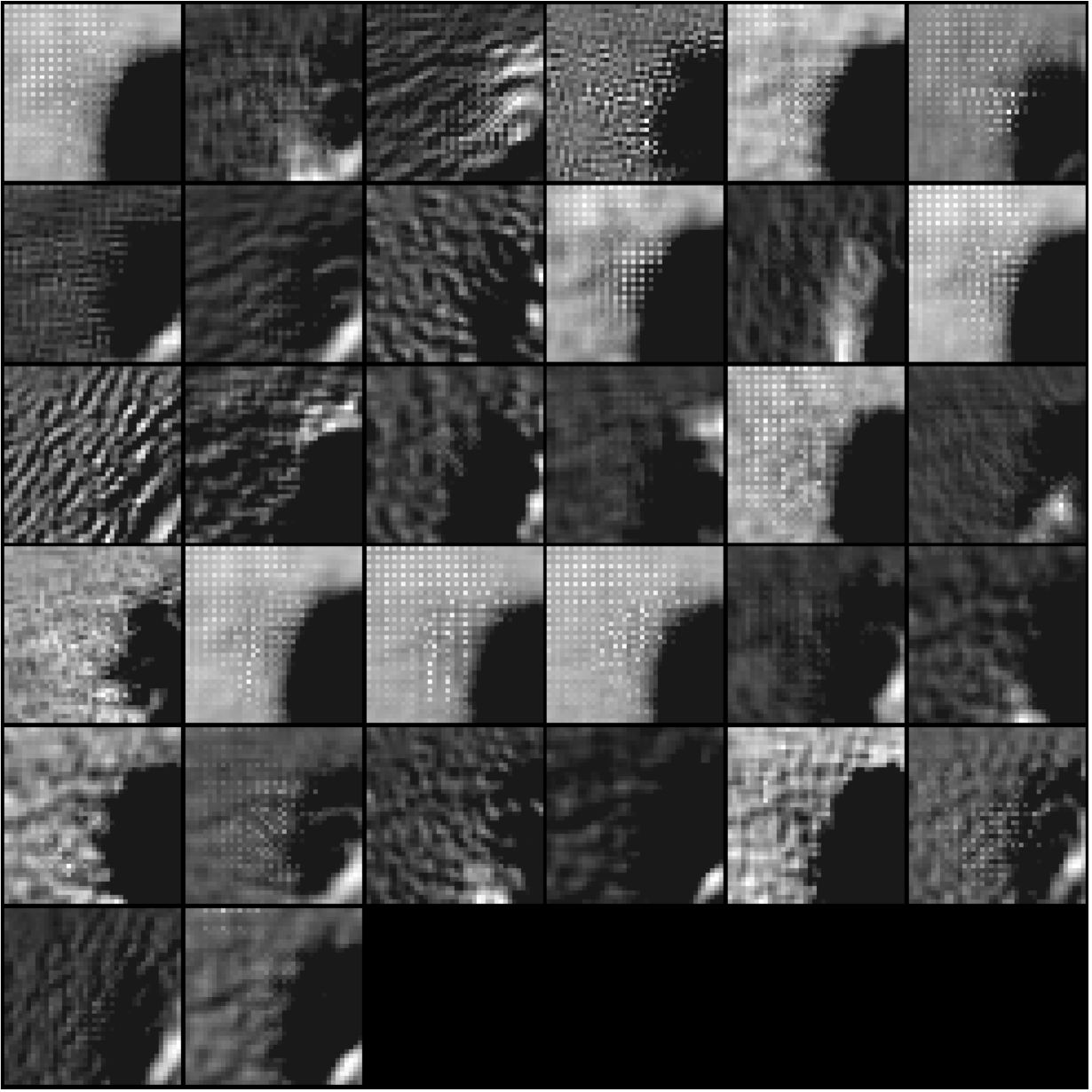}
                \caption*{\footnotesize layer 1 (32 maps, $40 \times 40$)}
                \label{fig:activations-0-l1}
        \end{subfigure}%
        ~ 
        \begin{subfigure}[b]{0.30\textwidth}
                \includegraphics[width=\textwidth]{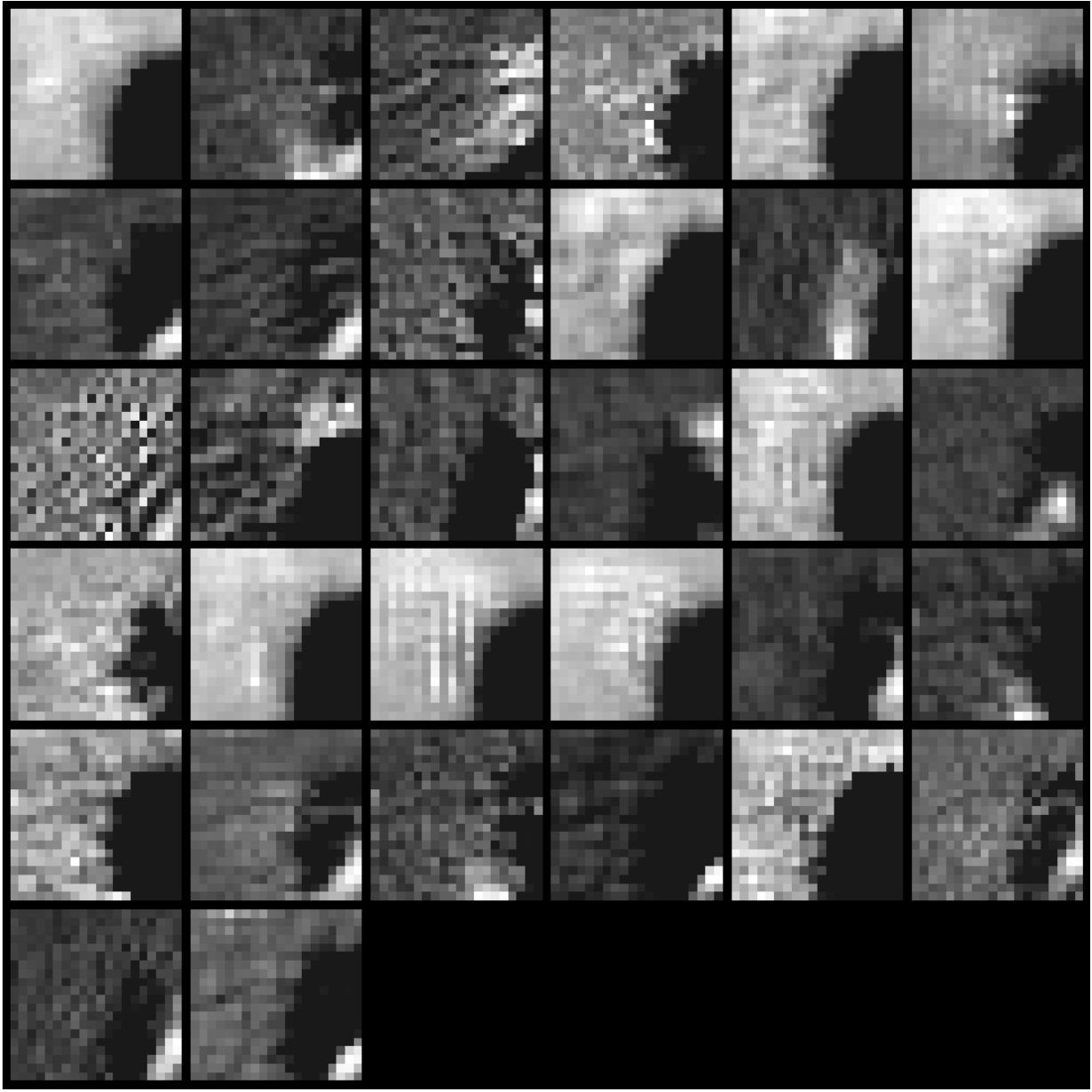}
                \caption*{\footnotesize pooling 1 (32 maps, $20 \times 20$)}
                \label{fig:activations-0-l1p}
        \end{subfigure}%
        \vspace{1em}
        
        \begin{subfigure}[b]{0.17\textwidth}
		\includegraphics[width=\textwidth]{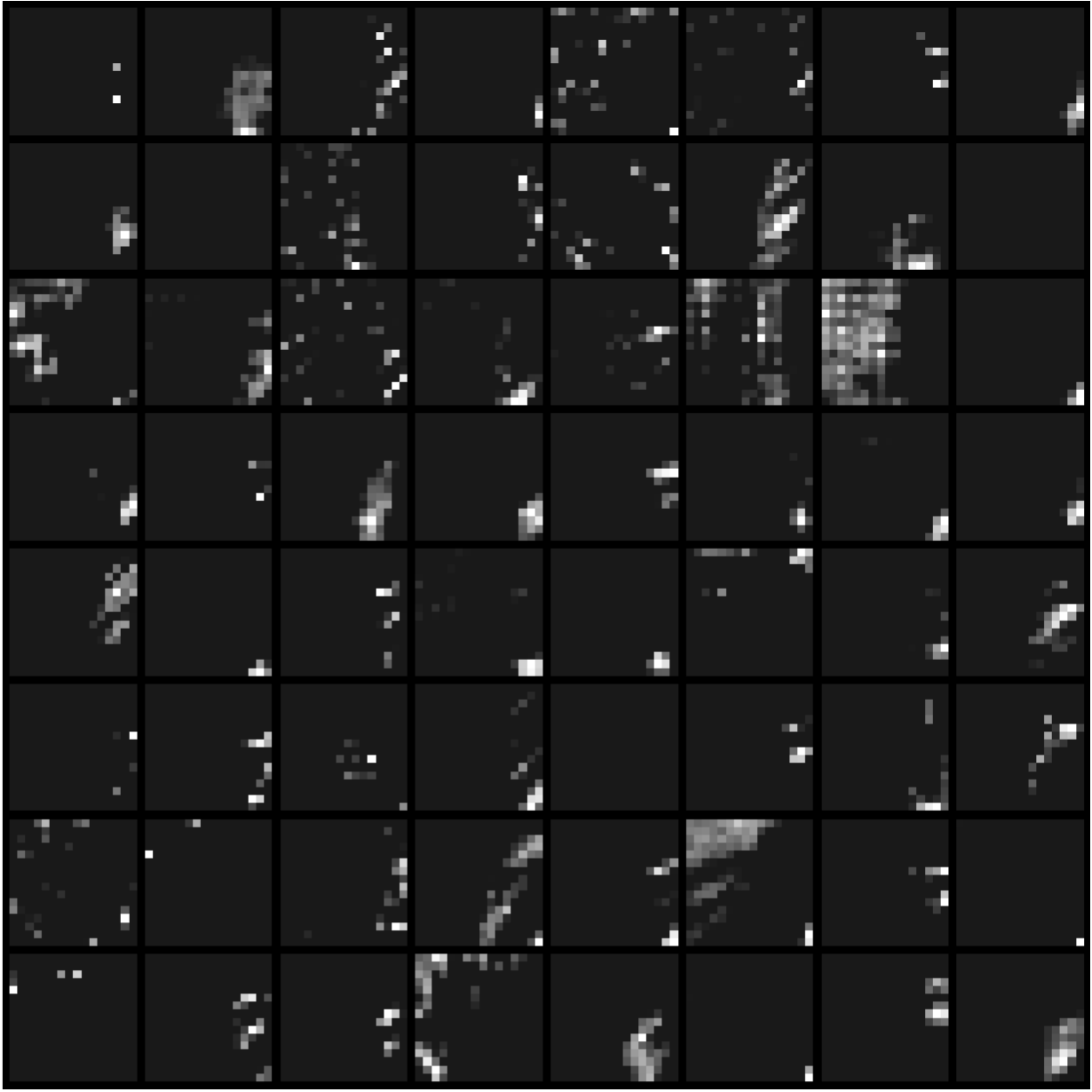}
                \caption*{\footnotesize layer 2\\(64 maps, $16 \times 16$)}
                \label{fig:activations-0-l2}
        \end{subfigure}%
	  ~ 
	\begin{subfigure}[b]{0.17\textwidth}
		\includegraphics[width=\textwidth]{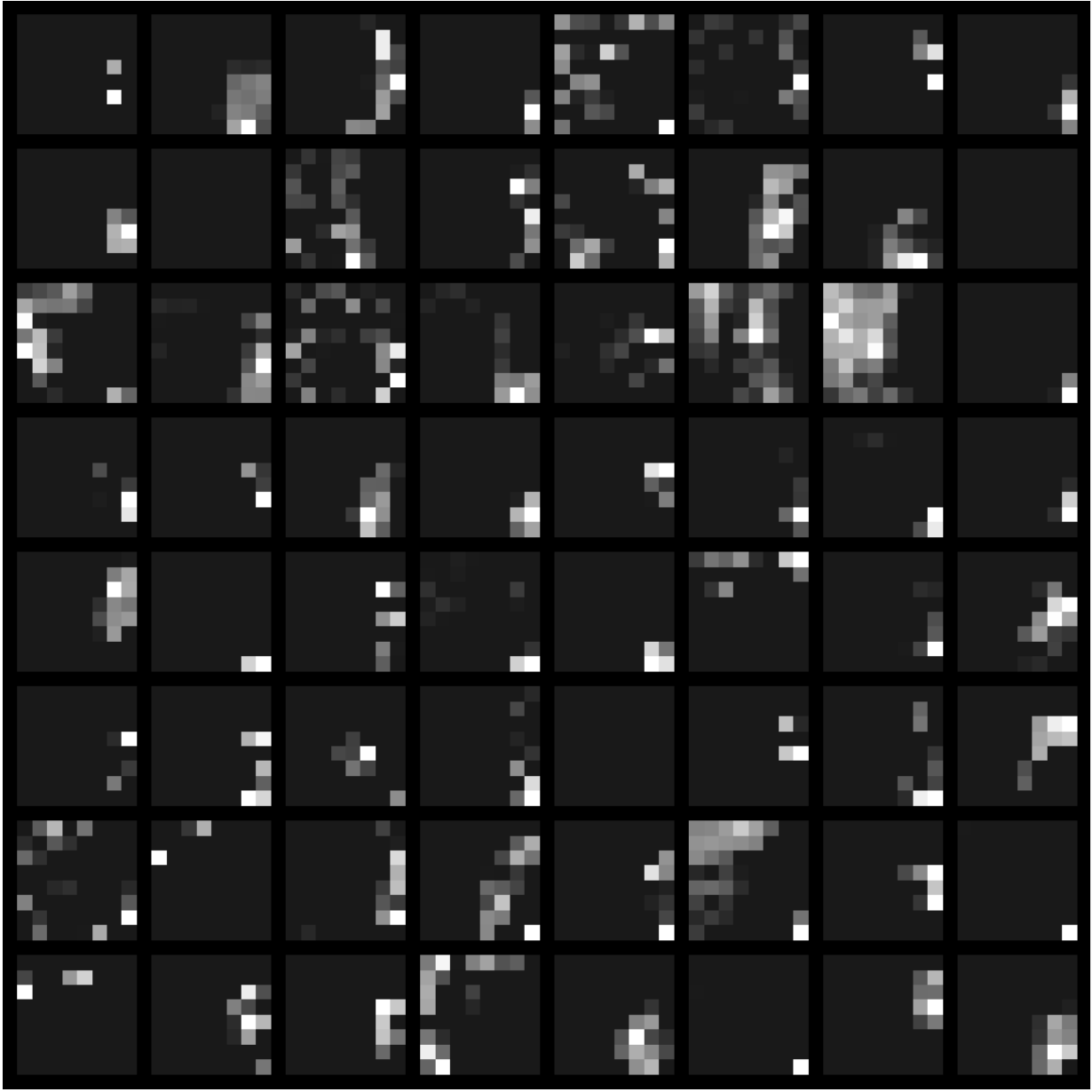}
                \caption*{\footnotesize pooling 2\\(64 maps, $8 \times 8$)}
                \label{fig:activations-0-l2p}
        \end{subfigure}%
	  ~ 
	\begin{subfigure}[b]{0.17\textwidth}
		\includegraphics[width=\textwidth]{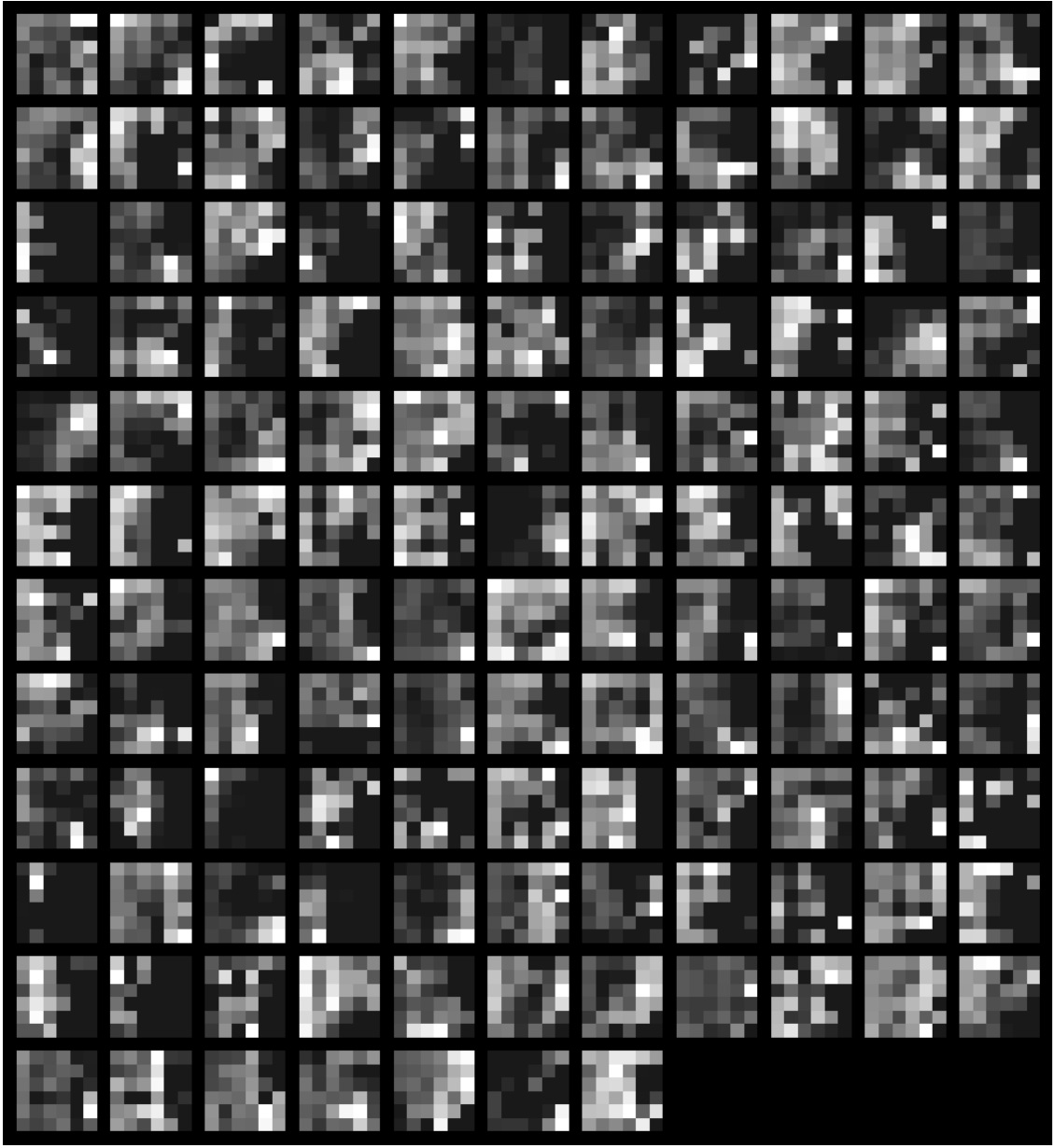}
                \caption*{\footnotesize layer 3\\(128 maps, $6 \times 6$)}
                \label{fig:activations-0-l3}
        \end{subfigure}%
	  ~ 
	\begin{subfigure}[b]{0.17\textwidth}
		\includegraphics[width=\textwidth]{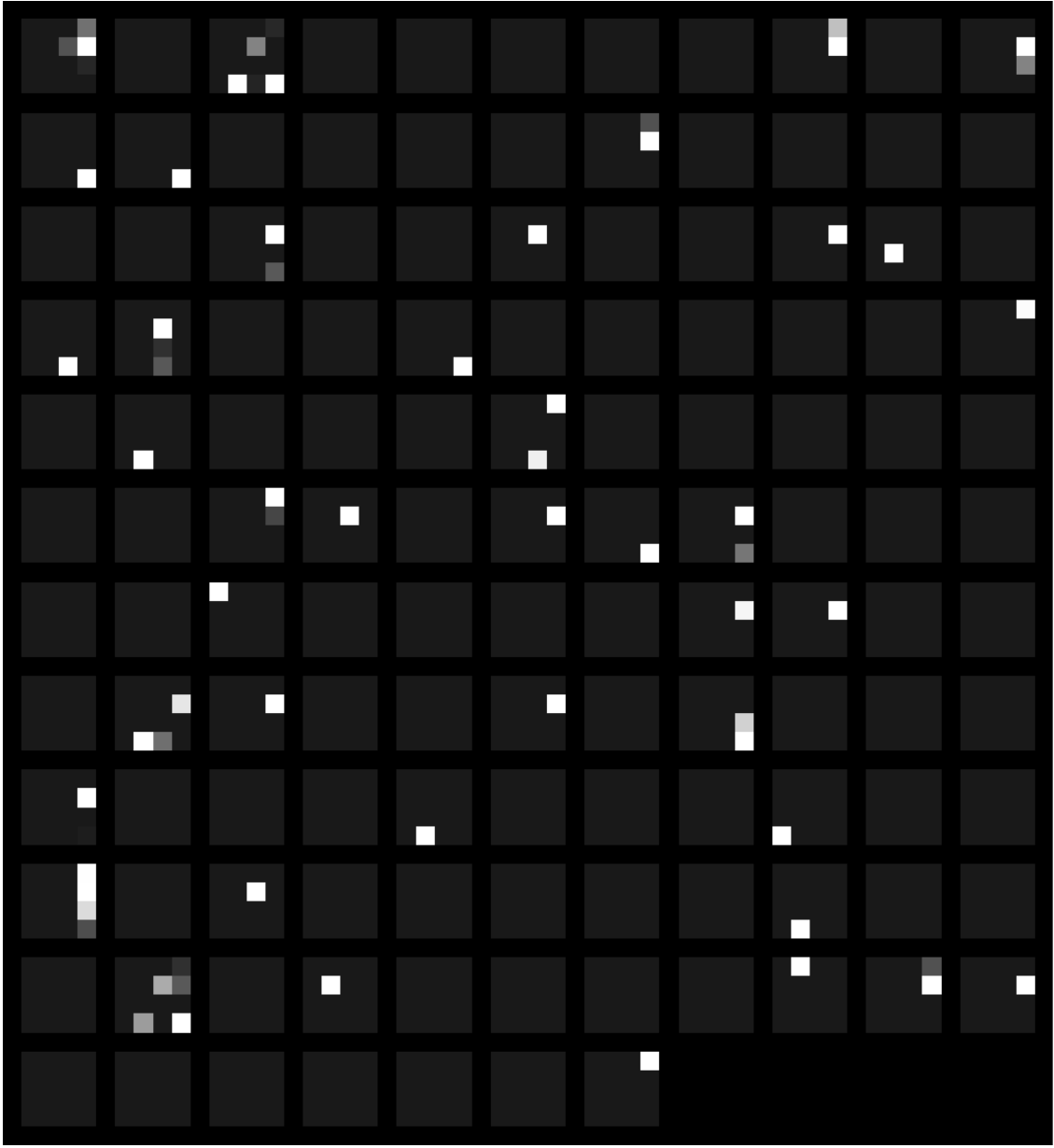}
                \caption*{\footnotesize layer 4\\(128 maps, $4 \times 4$)}
                \label{fig:activations-0-l4}
        \end{subfigure}%
	  ~ 
	\begin{subfigure}[b]{0.17\textwidth}
		\includegraphics[width=\textwidth]{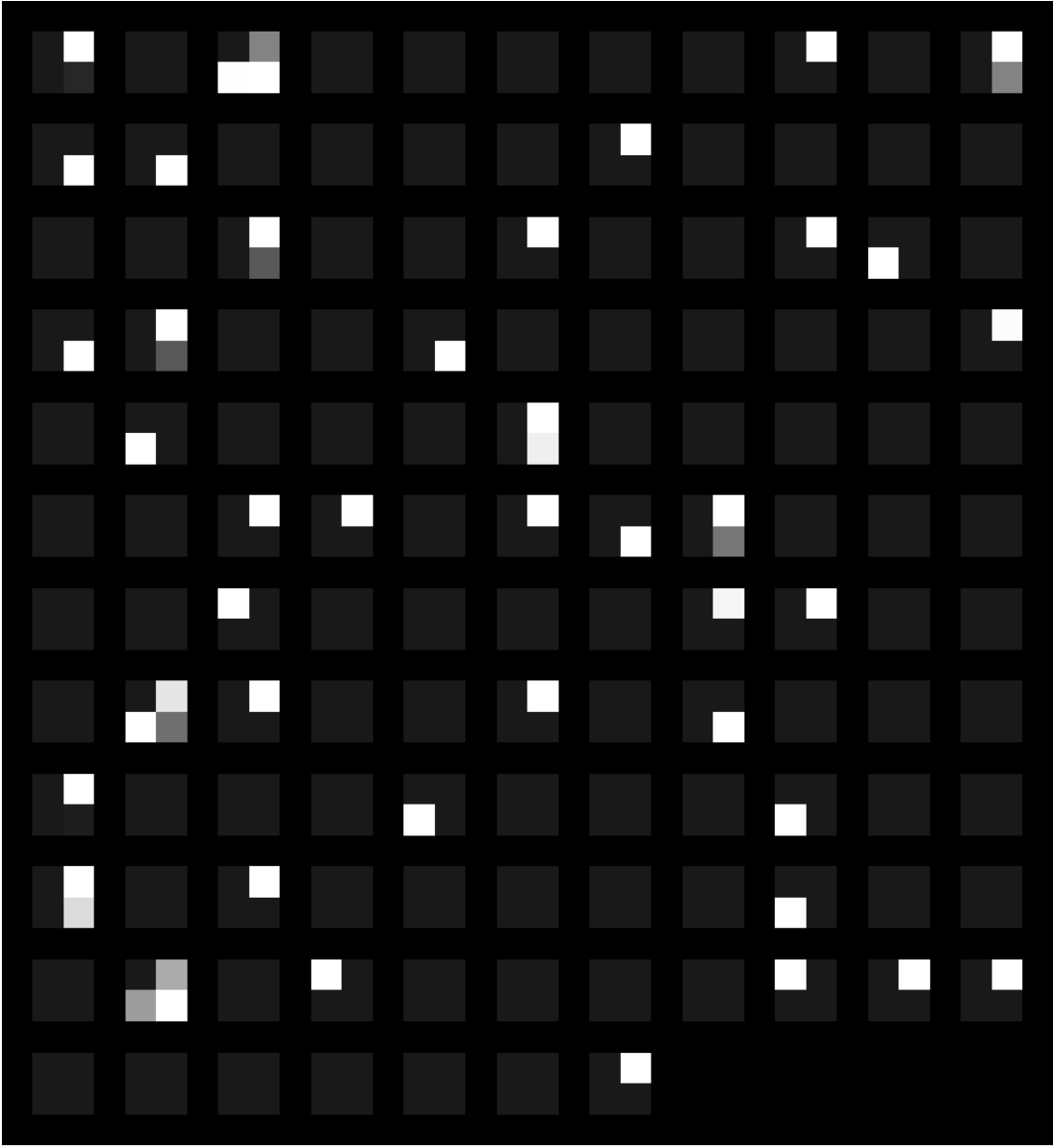}
                \caption*{\footnotesize pooling 4\\(128 maps, $2 \times 2$)}
                \label{fig:activations-0-l4p}
        \end{subfigure}%
        \caption{Activations of each layer in the convolutional part of the best performing network, given the input viewpoint shown in the top left. The number of feature maps and the size of each map is indicated below each figure. The geometry of the input image is still apparent in the activations of higher convolutional layers. The activations of all layers except the third are also quite sparse.}\label{fig:activations-0}

	\vspace{2em}
	
        \begin{subfigure}[b]{0.25\textwidth}
                \includegraphics[width=\textwidth]{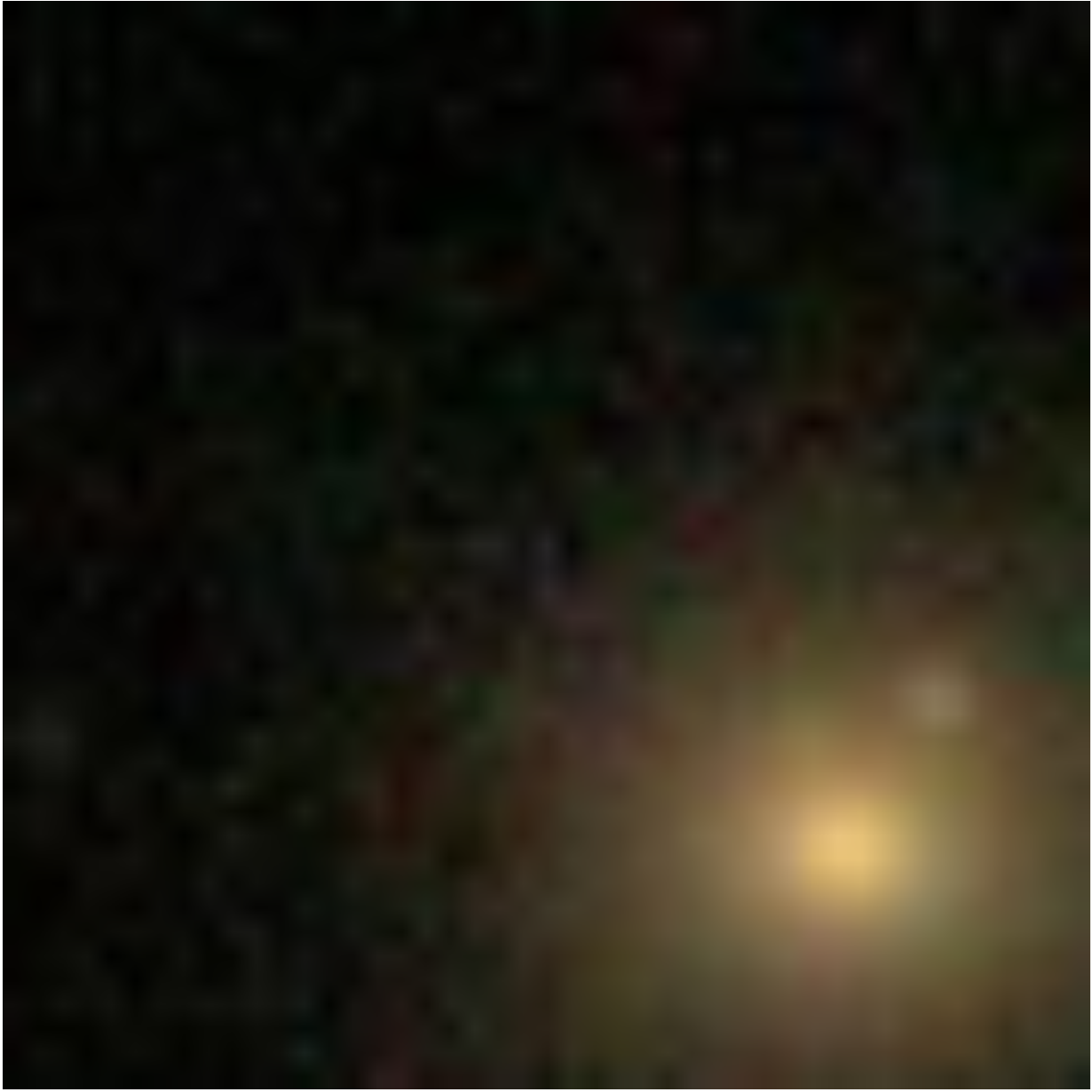}
                \caption*{\footnotesize input ($45 \times 45$)}
                \label{fig:activations-1-input}
        \end{subfigure}%
        ~ 
        \begin{subfigure}[b]{0.30\textwidth}
                \includegraphics[width=\textwidth]{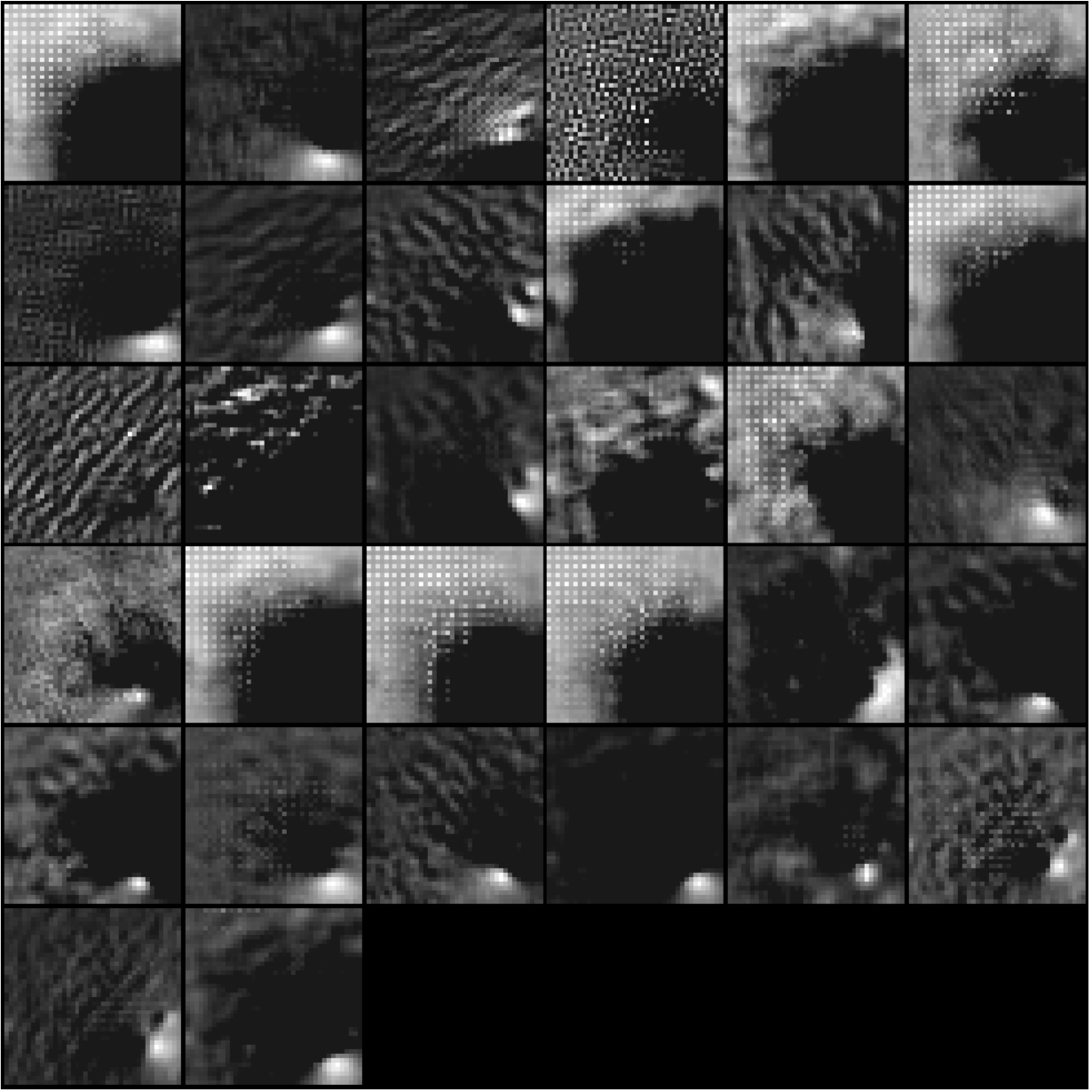}
                \caption*{\footnotesize layer 1 (32 maps, $40 \times 40$)}
                \label{fig:activations-1-l1}
        \end{subfigure}%
        ~ 
        \begin{subfigure}[b]{0.30\textwidth}
                \includegraphics[width=\textwidth]{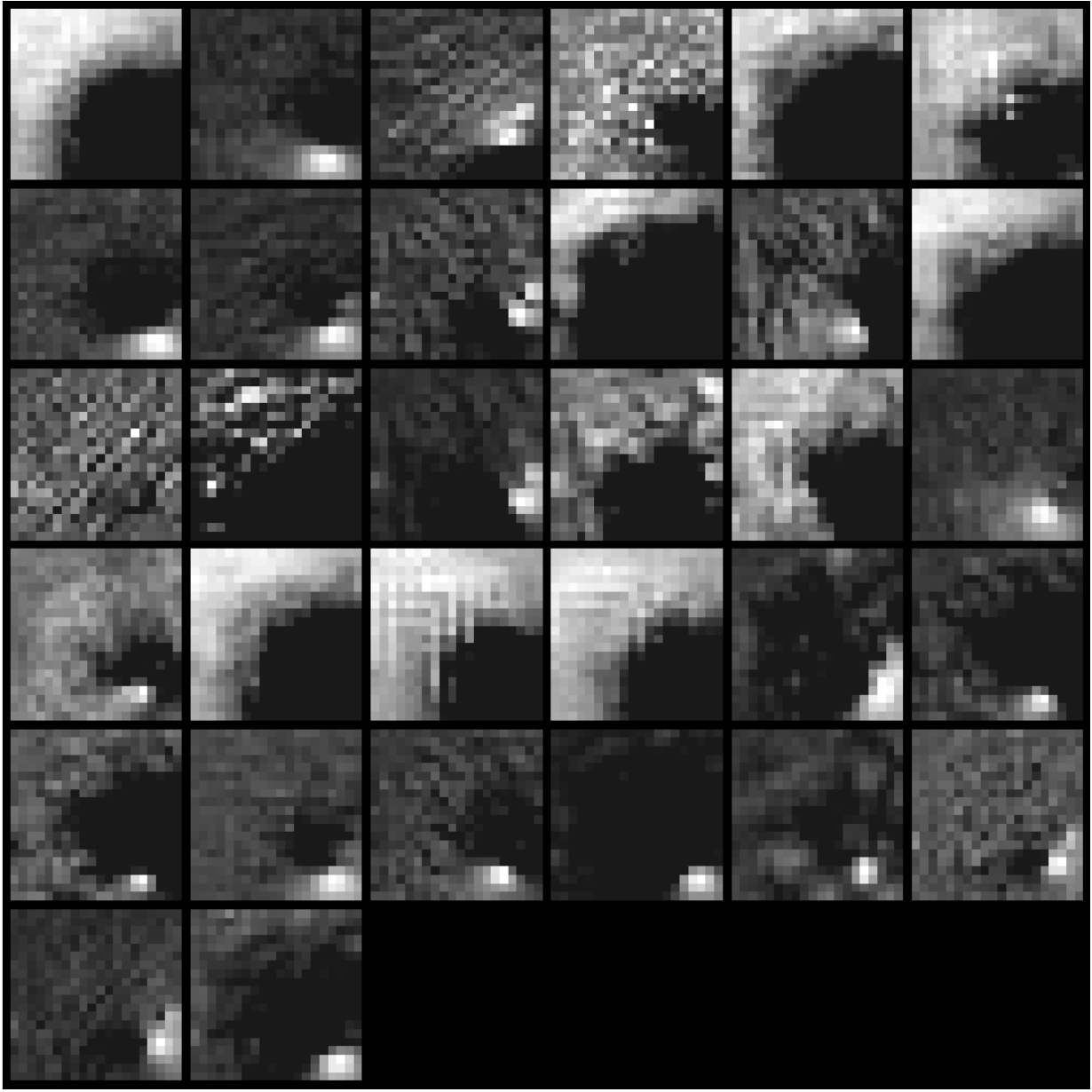}
                \caption*{\footnotesize pooling 1 (32 maps, $20 \times 20$)}
                \label{fig:activations-1-l1p}
        \end{subfigure}%
        \vspace{1em}
        
        \begin{subfigure}[b]{0.17\textwidth}
		\includegraphics[width=\textwidth]{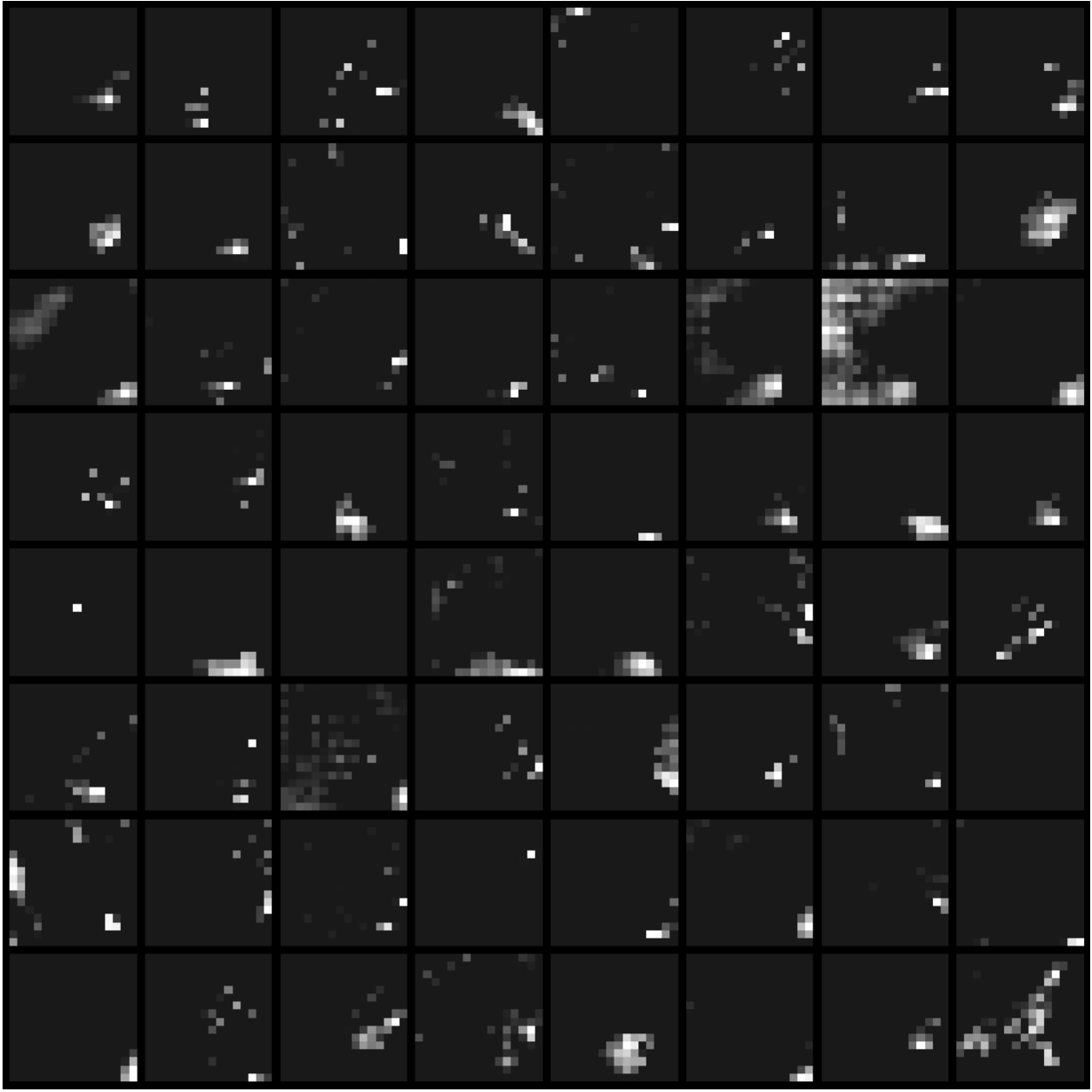}
                \caption*{\footnotesize layer 2\\(64 maps, $16 \times 16$)}
                \label{fig:activations-1-l2}
        \end{subfigure}%
	  ~ 
	\begin{subfigure}[b]{0.17\textwidth}
		\includegraphics[width=\textwidth]{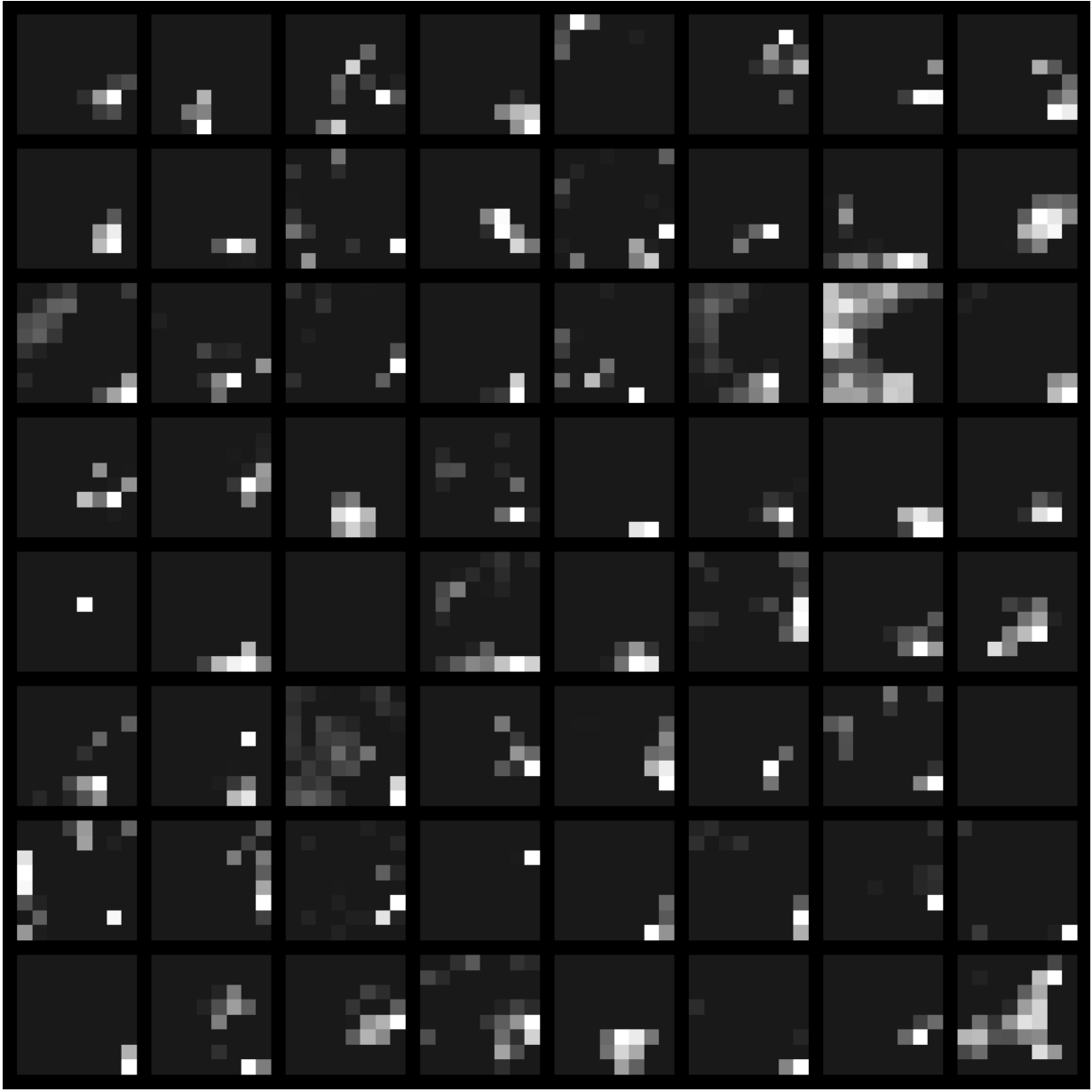}
                \caption*{\footnotesize pooling 2\\(64 maps, $8 \times 8$)}
                \label{fig:activations-1-l2p}
        \end{subfigure}%
	  ~ 
	\begin{subfigure}[b]{0.17\textwidth}
		\includegraphics[width=\textwidth]{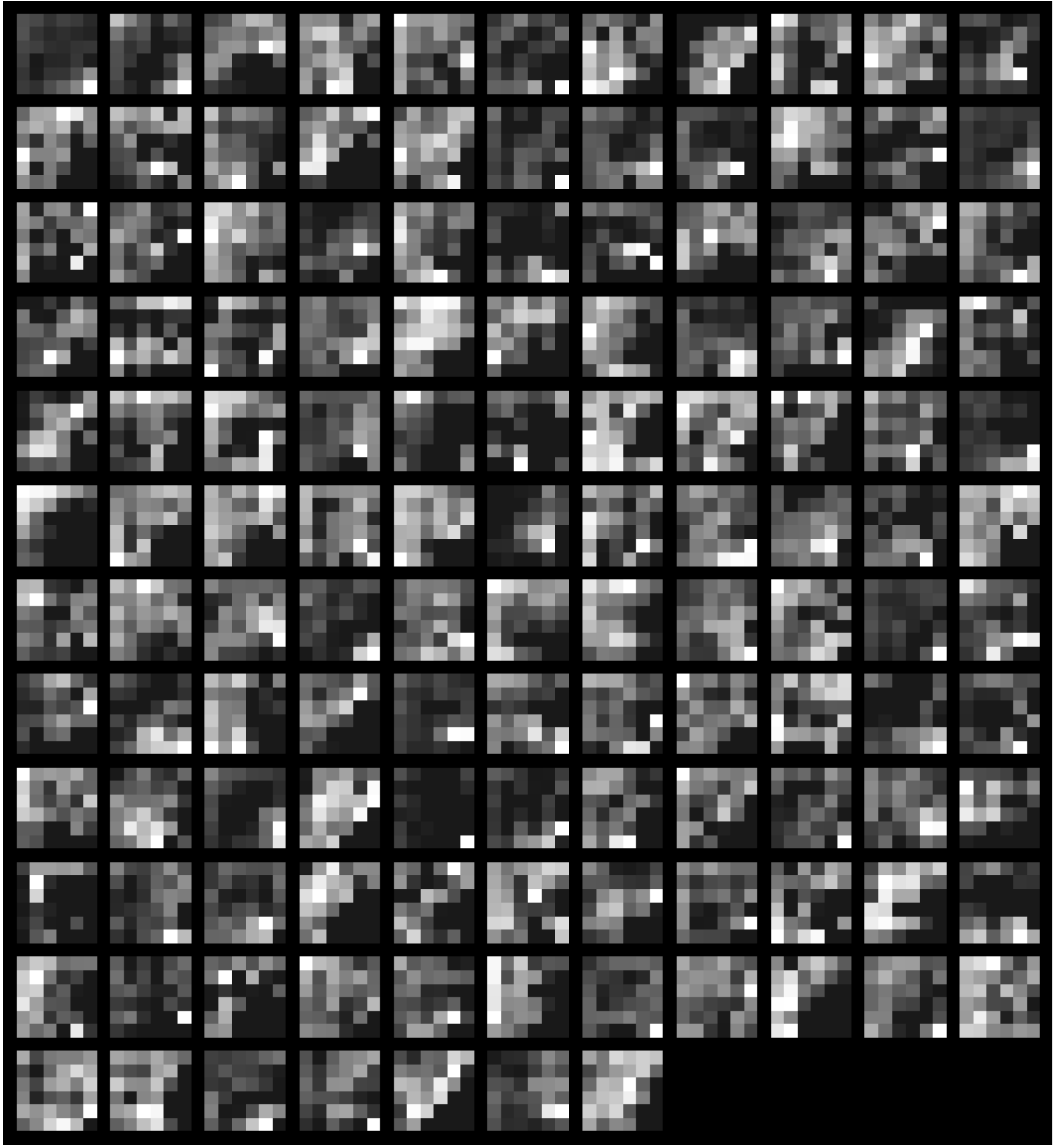}
                \caption*{\footnotesize layer 3\\(128 maps, $6 \times 6$)}
                \label{fig:activations-1-l3}
        \end{subfigure}%
	  ~ 
	\begin{subfigure}[b]{0.17\textwidth}
		\includegraphics[width=\textwidth]{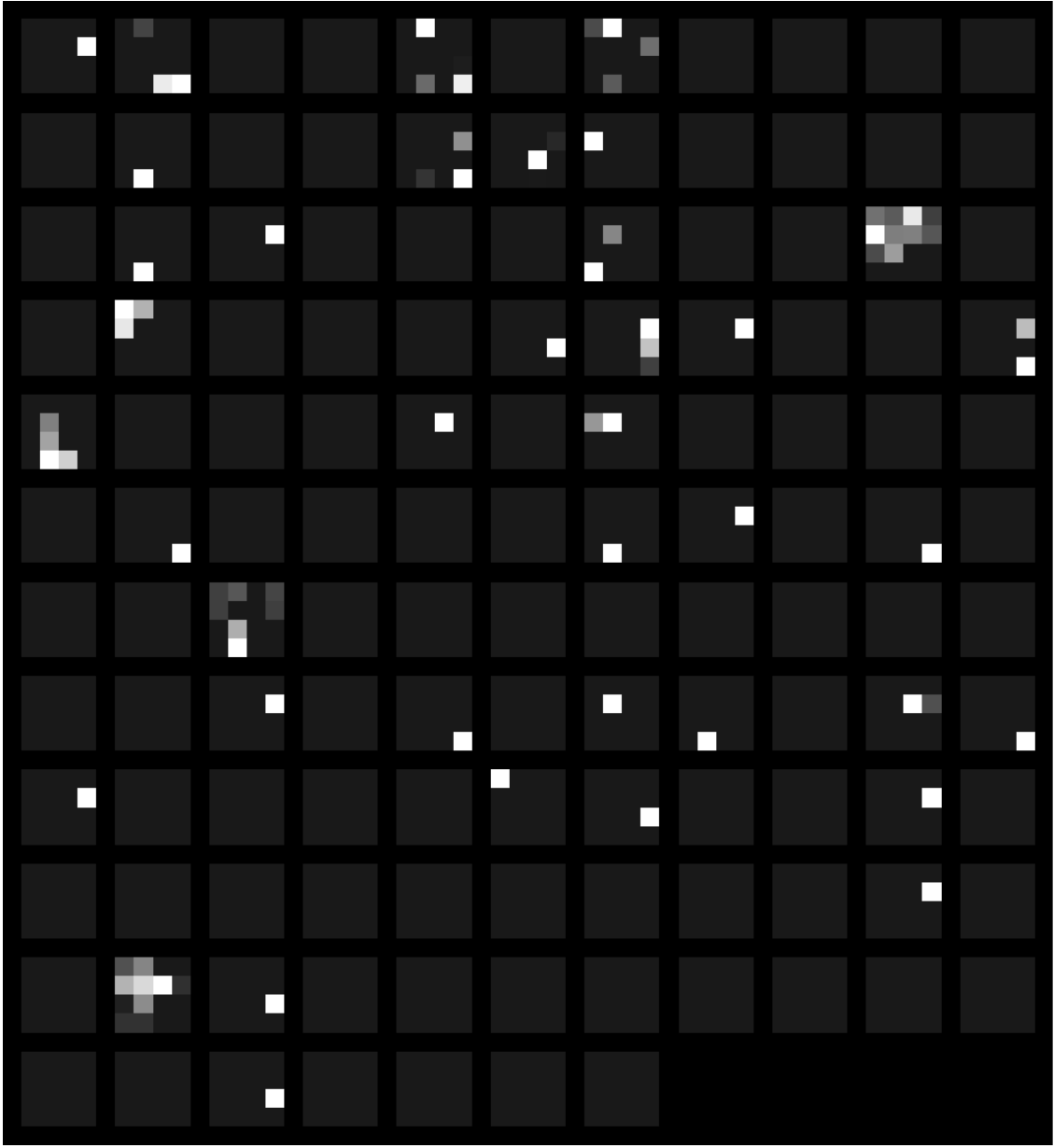}
                \caption*{\footnotesize layer 4\\(128 maps, $4 \times 4$)}
                \label{fig:activations-1-l4}
        \end{subfigure}%
	  ~ 
	\begin{subfigure}[b]{0.17\textwidth}
		\includegraphics[width=\textwidth]{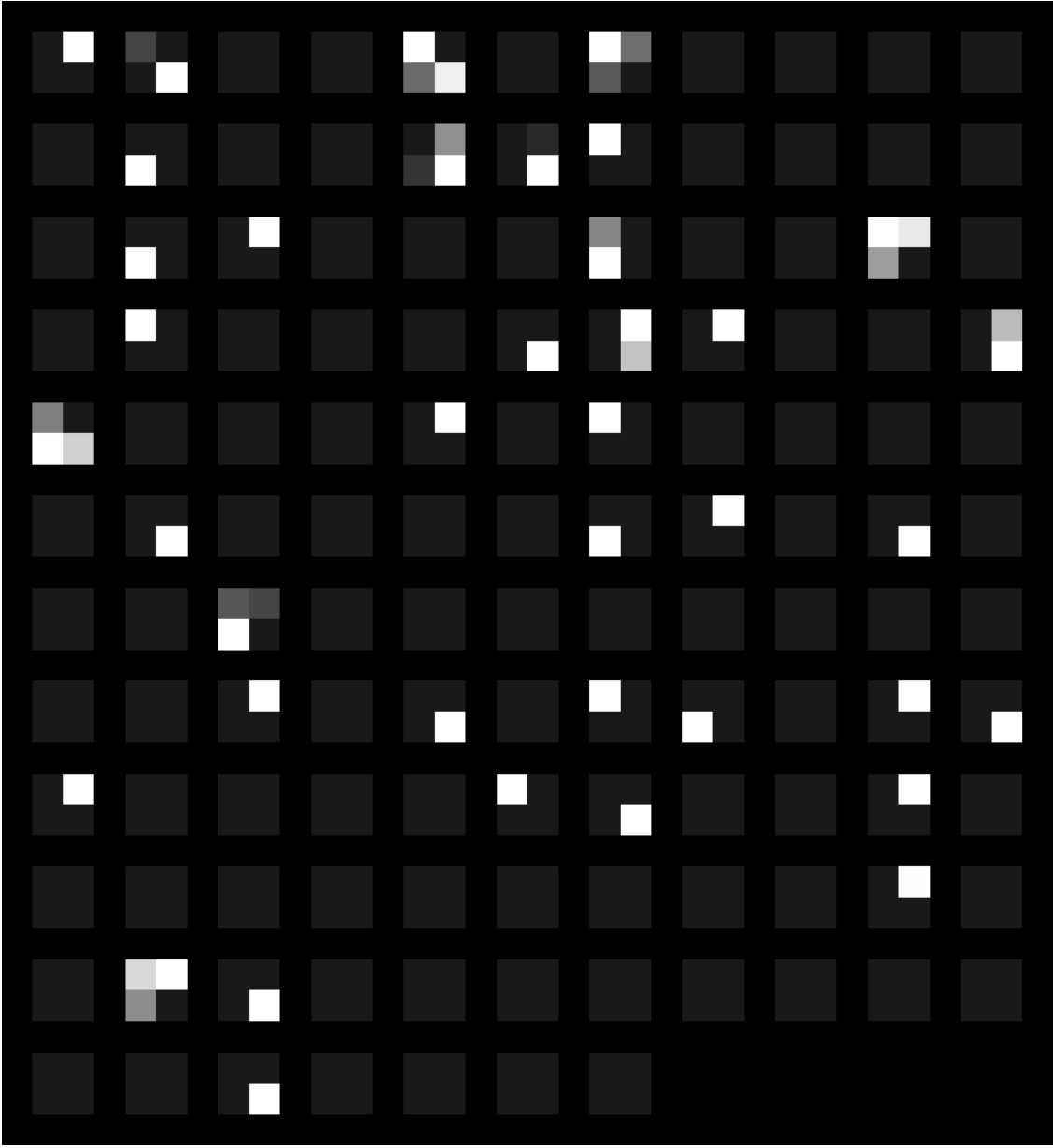}
                \caption*{\footnotesize pooling 4\\(128 maps, $2 \times 2$)}
                \label{fig:activations-1-l4p}
        \end{subfigure}%
        \caption{Same as Figure~\ref{fig:activations-0}, but for a different input viewpoint in the upper left.}\label{fig:activations-1}
\end{figure*}

Figures~\ref{fig:activations-0} and~\ref{fig:activations-1} show how an input viewpoint (i.e. a $45\times45$ part of an input image, see Section~\ref{sec:viewpoint-extraction}) activates the units in the convolutional part of the network. Note that the geometry of the input image is still apparent in the activations of higher convolutional layers. The activations of all layers except the third are also quite sparse, especially those of the fourth layer. One possible reason why the third layer activations are not as sparse is because there is no pooling layer directly following it.

It is also possible to visualize what neurons in the topmost hidden layer of the network (i.e. just before the output layer) have learned about the data, by selecting representative examples from the test set that maximize their activations. This reveals what type of inputs the unit is sensitive to, and what kind of invariances it has learned. Because we used maxout units in this layer, we can also select examples that minimally activate the units, allowing us to determine which types of inputs each unit discriminates between.

\begin{figure*}
        \centering
        \begin{subfigure}[b]{0.90\textwidth}
                \includegraphics[width=\textwidth]{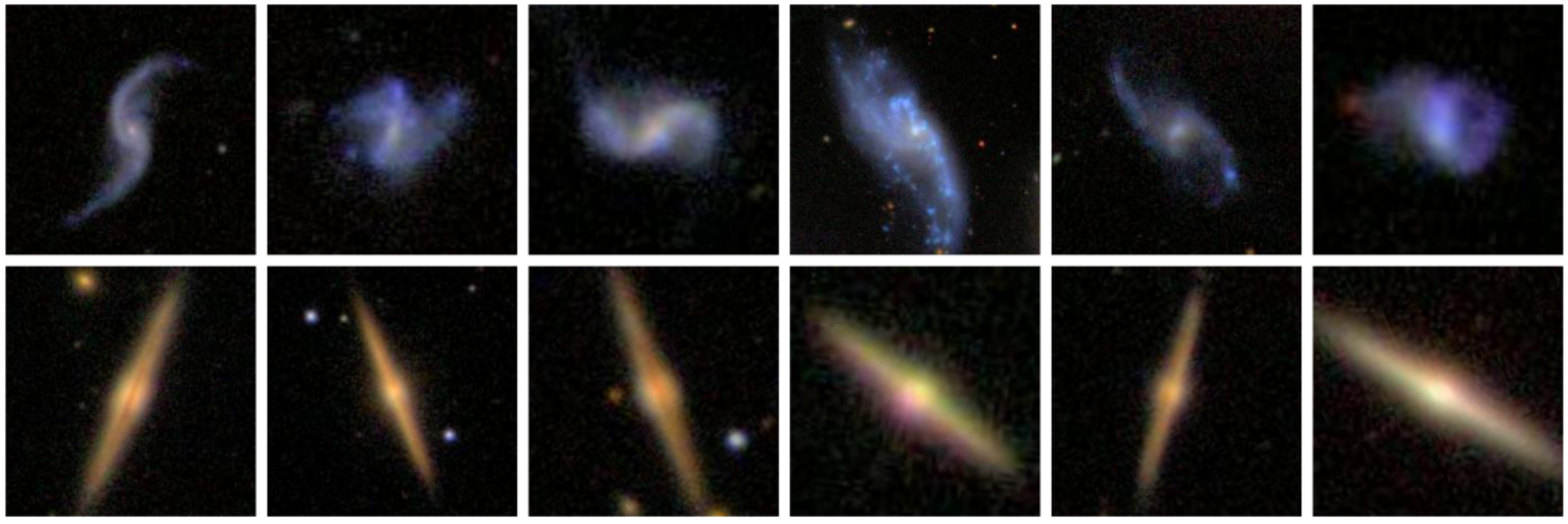}
                \caption{}
                \label{fig:exemplars-0}
        \end{subfigure}%
        \vspace{1em}

        \begin{subfigure}[b]{0.90\textwidth}
                \includegraphics[width=\textwidth]{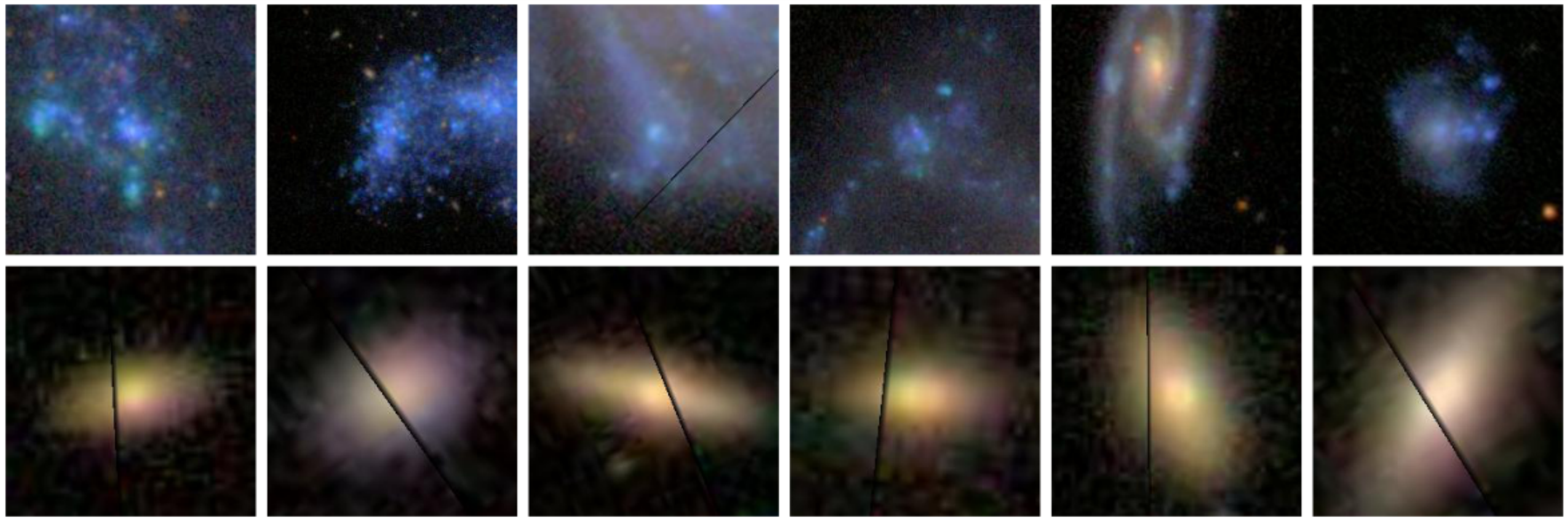}
                \caption{}
                \label{fig:exemplars-15}
        \end{subfigure}
        \vspace{1em}

        \begin{subfigure}[b]{0.90\textwidth}
                \includegraphics[width=\textwidth]{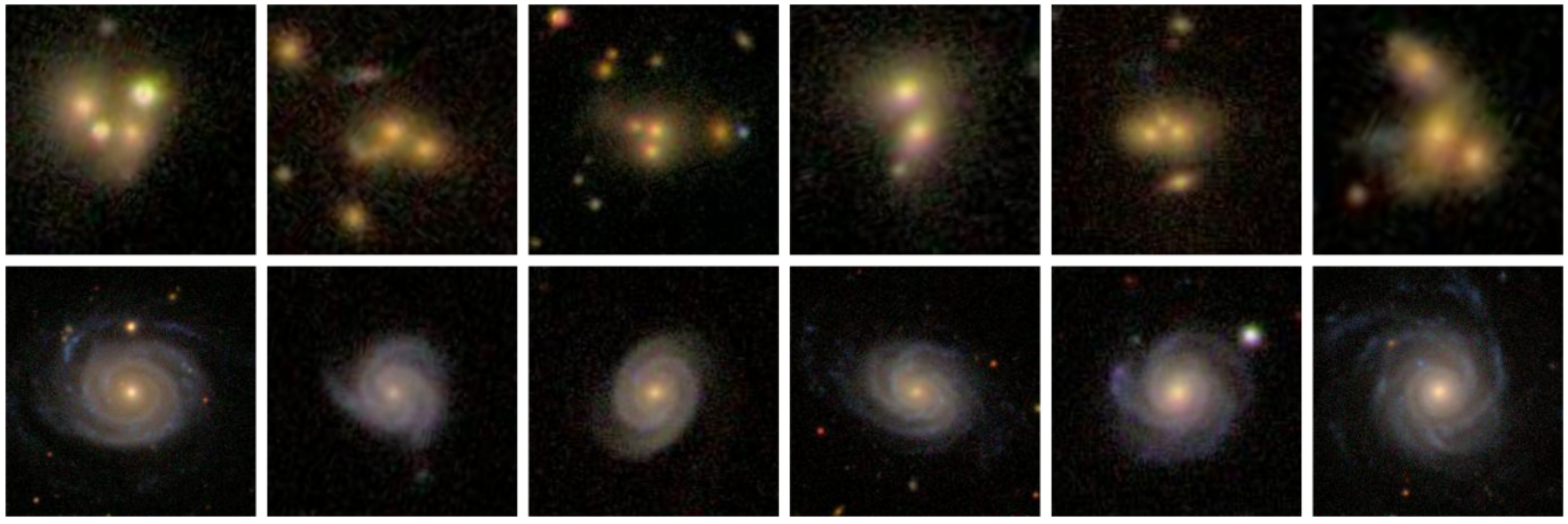}
                \caption{}
                \label{fig:exemplars-19}
        \end{subfigure}
        \caption{Example images from the test set that maximally and minimally activate units in the topmost hidden layer of the best performing network. Each group of 12 images represents a unit. The top row of images in each group maximally activate the unit, and bottom row of images minimally activate it. From top to bottom, these galaxies primarily correspond to the \gztwo~labels of: loose winding arms, edge-on disks, irregulars, disturbed, other, and tight winding arms.}\label{fig:exemplars}
\end{figure*}

Figure~\ref{fig:exemplars} shows such a visualization for three different units. Clearly each unit is able to discriminate between two distinct types of galaxies. The units also exhibit rotation invariance, as well as some scale invariance. For some units, we observed selectivity only in the positive or in the negative direction (not shown). A minority of units seem to be multimodal, activating in the same direction for two or more distinct types of galaxies. Presumably the activation value of these units is disambiguated in the context of all other unit values.

The unit visualized in Figure~\ref{fig:exemplars-15} detects imaging artifacts: black lines running across the centre of the images, which are the result of dead pixels in the SDSS camera. This is interesting because such (known) artifacts are not morphological features of the depicted galaxies. It turns out that the network is trying to replicate the behaviour of the Galaxy~Zoo participants, who tend to classify images featuring such artifacts as \emph{disturbed} galaxies (answer A8.3 in Table~\ref{tab:questions}), even though this is not the intended meaning of this answer. Most likely this is because the button for this answer in the \gztwo~web interface seems to feature such a black line.

\begin{figure*}
        \centering
        \begin{subfigure}[b]{0.22\textwidth}
                \includegraphics[width=\textwidth]{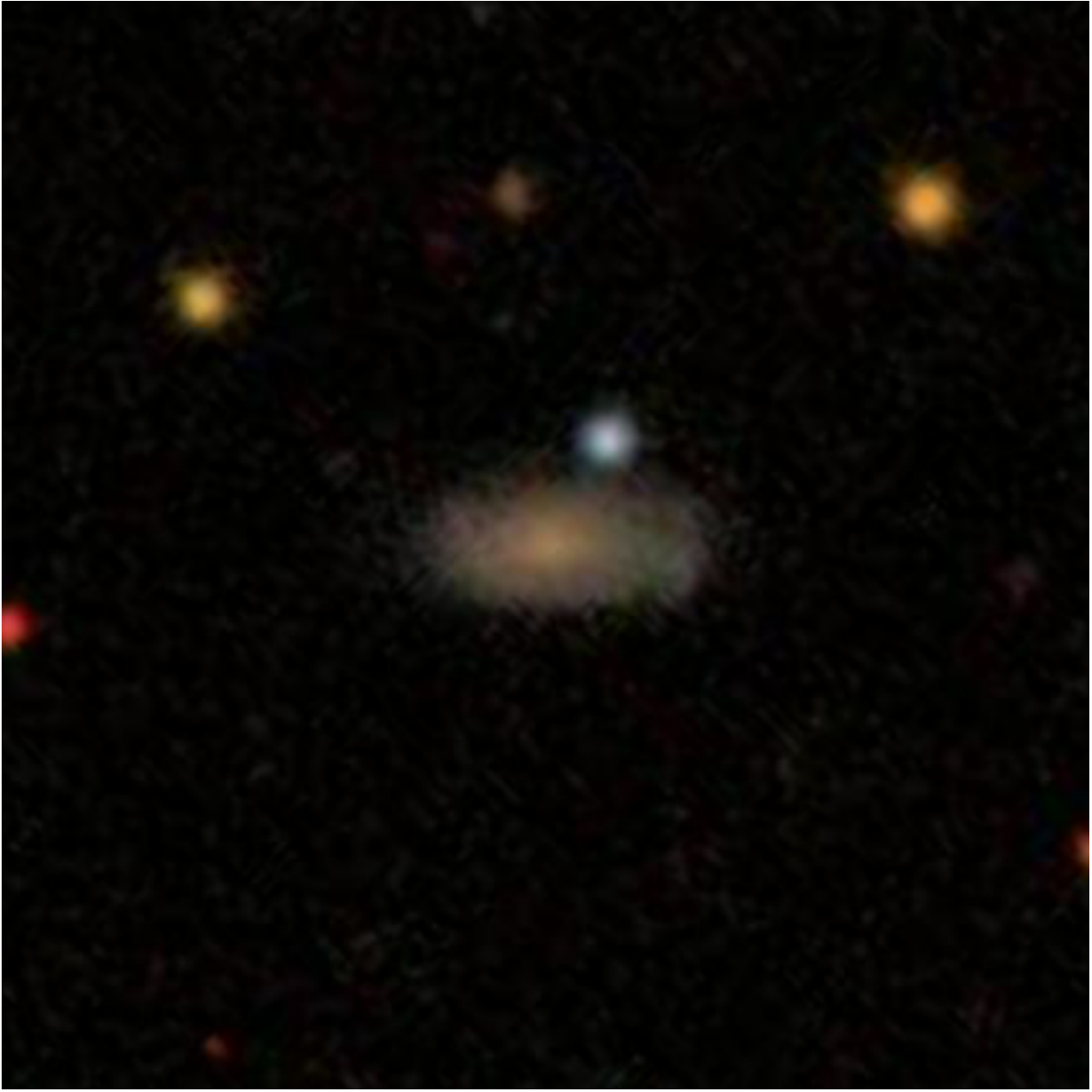}
                \caption{0.24610}
                \label{fig:example-952388}
        \end{subfigure}%
        ~ 
        \begin{subfigure}[b]{0.22\textwidth}
                \includegraphics[width=\textwidth]{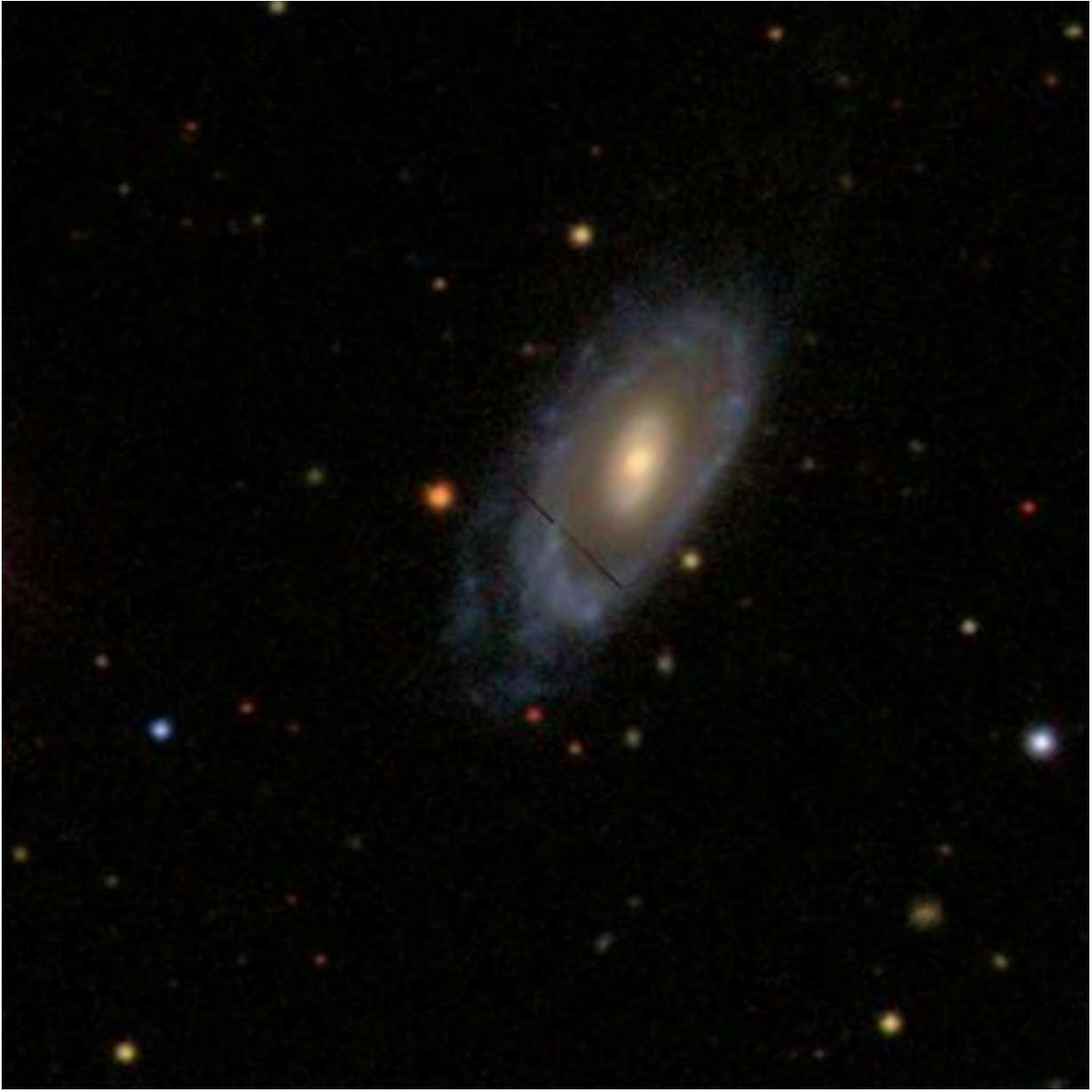}
                \caption{0.21877}
                \label{fig:example-941557}
        \end{subfigure}%
        ~ 
	\begin{subfigure}[b]{0.22\textwidth}
                \includegraphics[width=\textwidth]{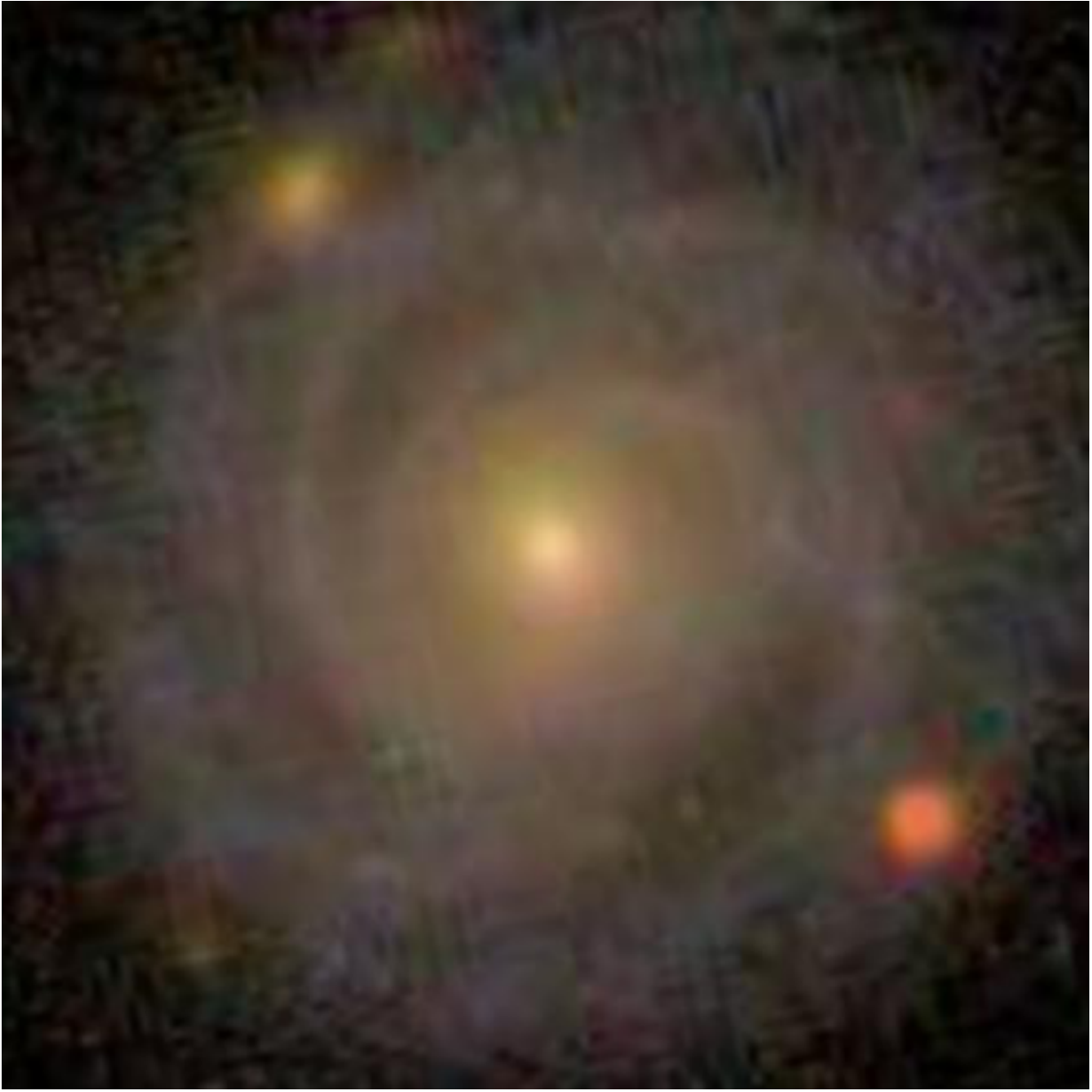}
                \caption{0.21145}
                \label{fig:example-915482}
        \end{subfigure}%
        ~ 
	\begin{subfigure}[b]{0.22\textwidth}
                \includegraphics[width=\textwidth]{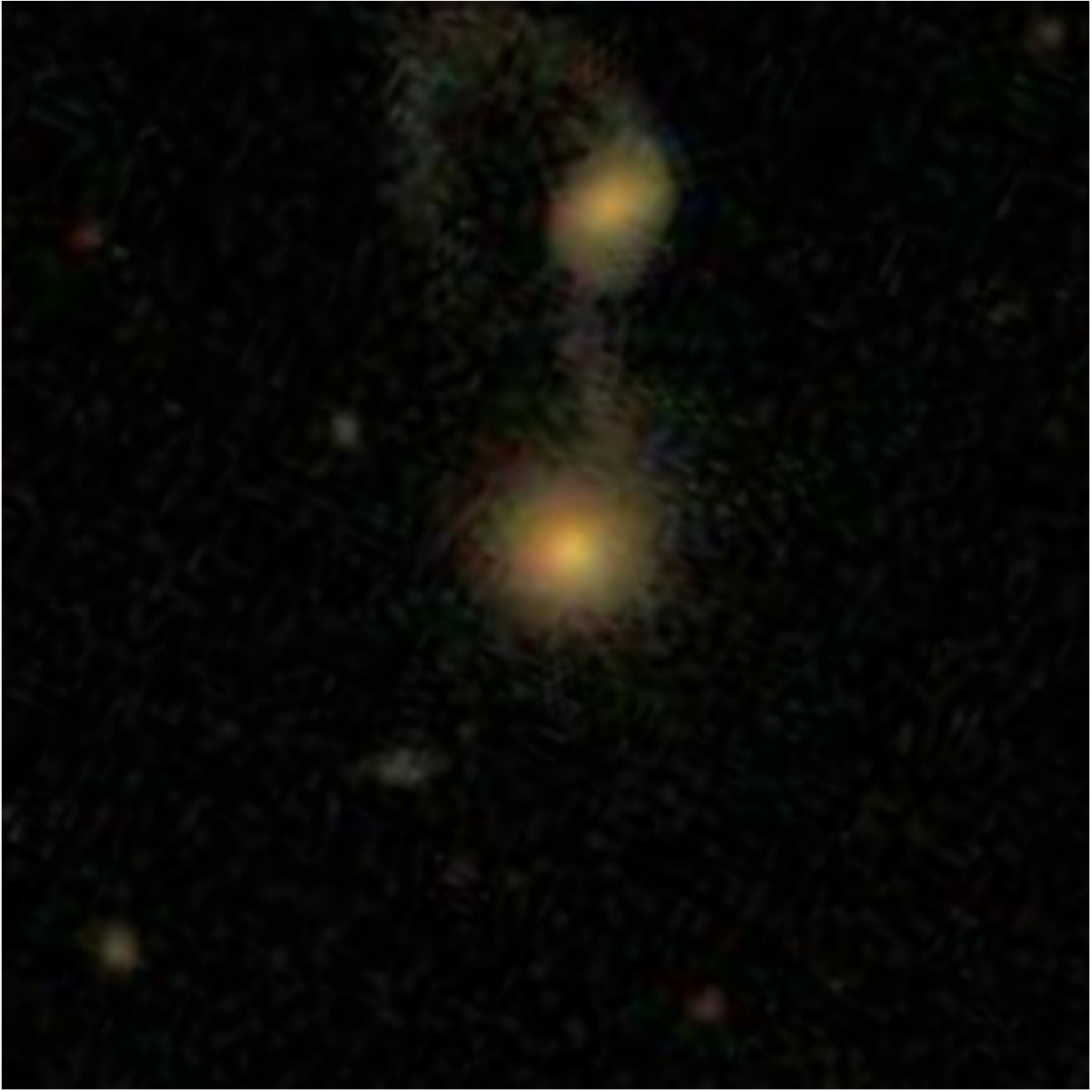}
                \caption{0.20088}
                \label{fig:example-938173}
        \end{subfigure}%
        \vspace{1em}
        
	\begin{subfigure}[b]{0.22\textwidth}
                \includegraphics[width=\textwidth]{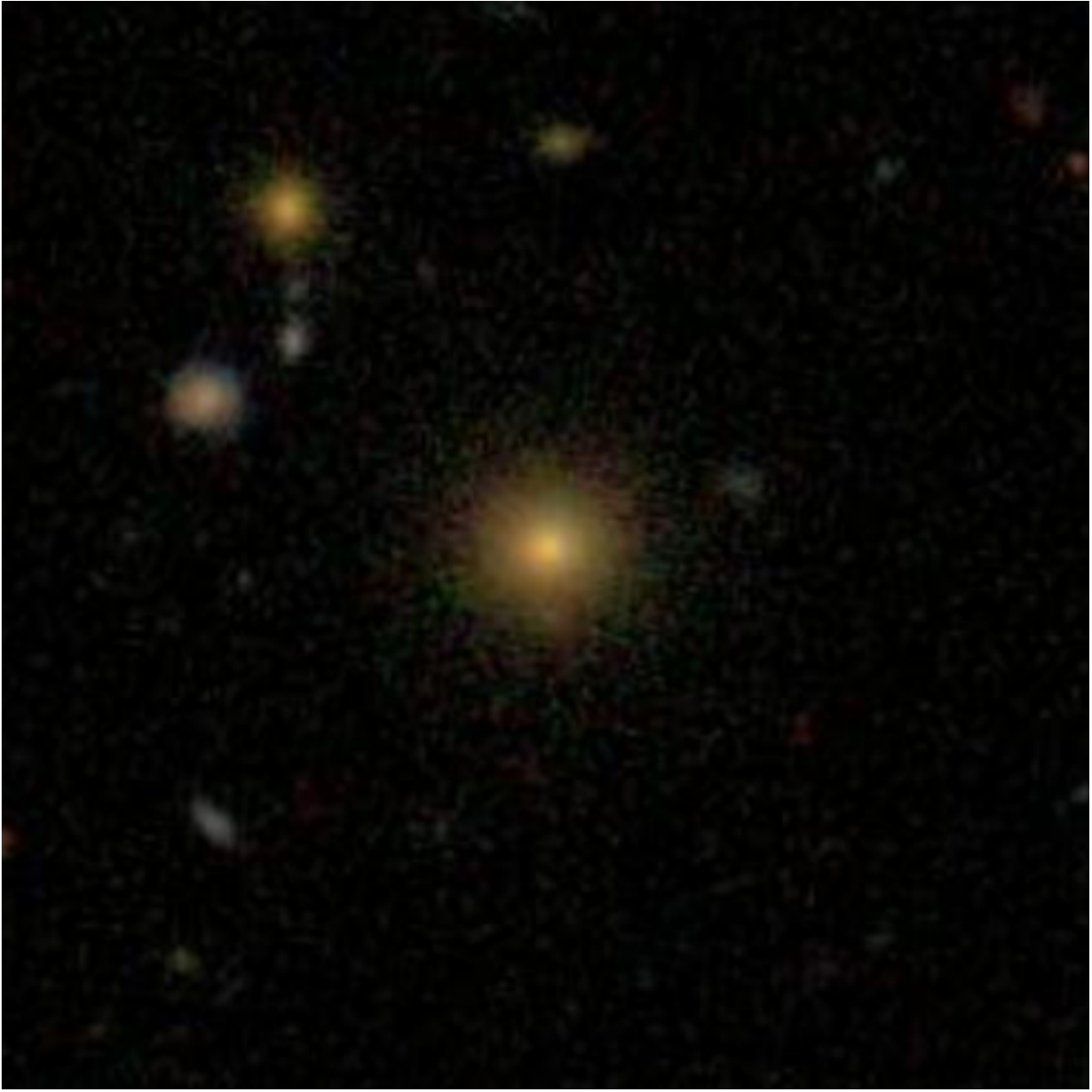}
                \caption{0.01112}
                \label{fig:example-974365}
        \end{subfigure}%
        ~ 
        \begin{subfigure}[b]{0.22\textwidth}
                \includegraphics[width=\textwidth]{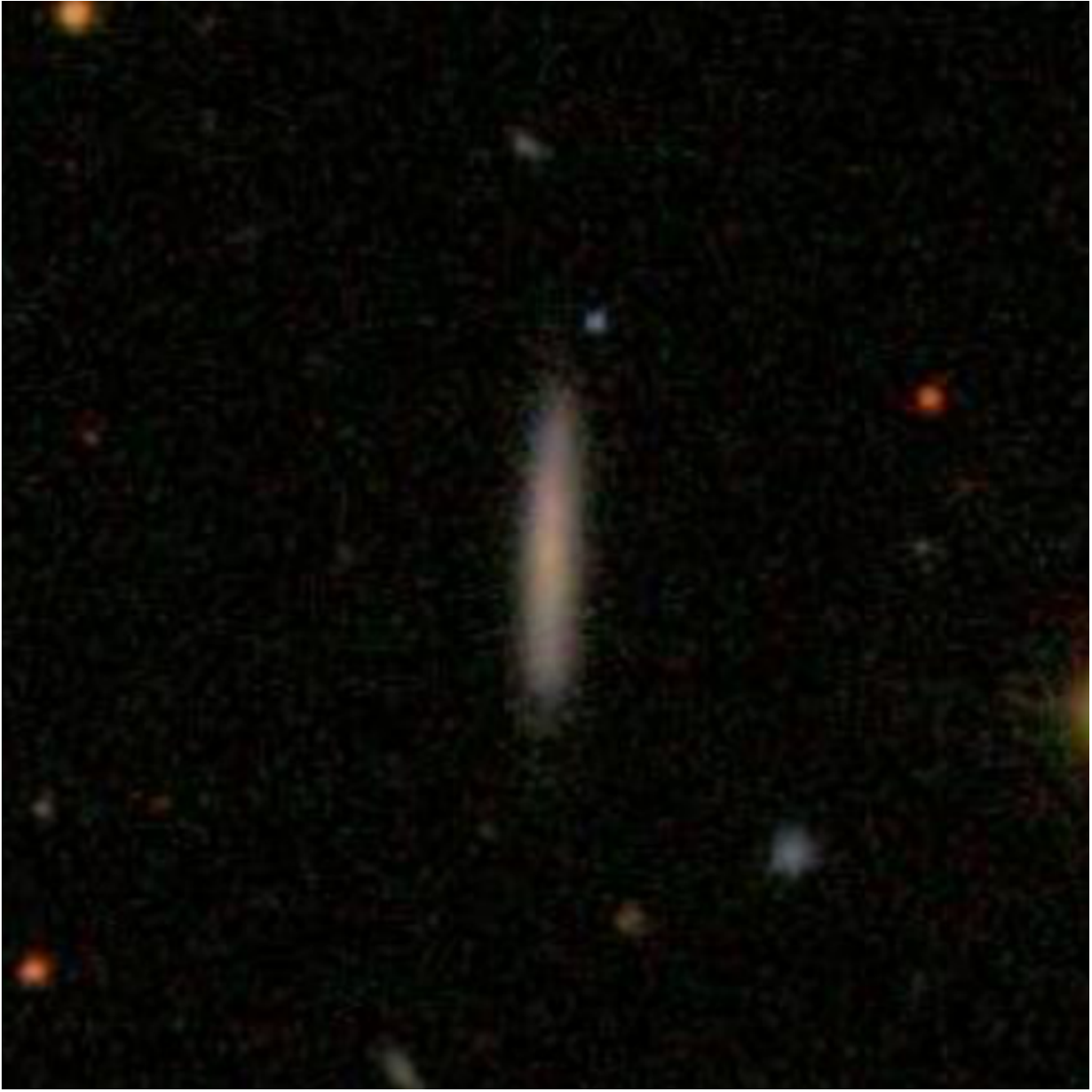}
                \caption{0.01174}
                \label{fig:example-946708}
        \end{subfigure}%
        ~ 
	\begin{subfigure}[b]{0.22\textwidth}
                \includegraphics[width=\textwidth]{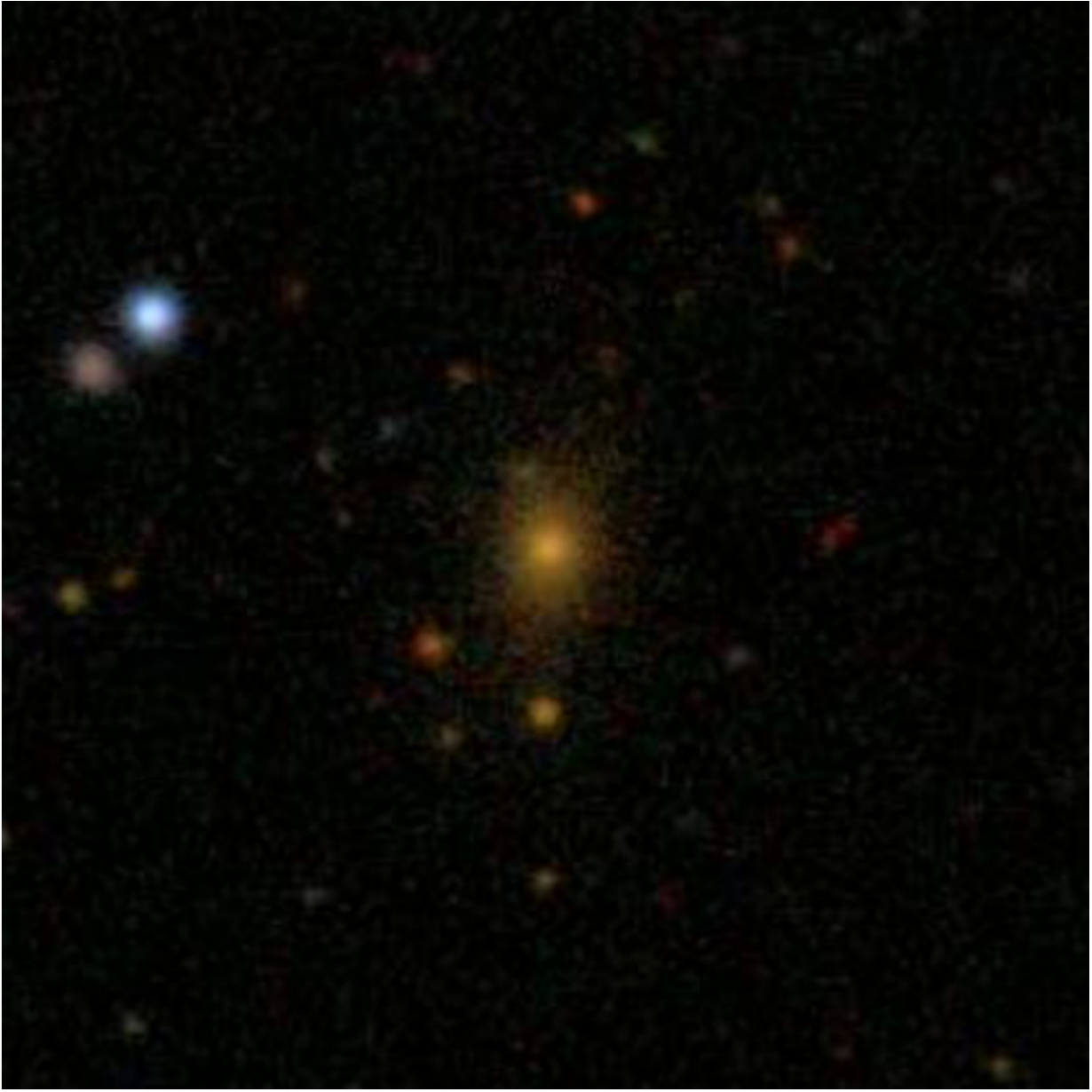}
                \caption{0.01187}
                \label{fig:example-914806}
        \end{subfigure}%
        ~ 
        \begin{subfigure}[b]{0.22\textwidth}
                \includegraphics[width=\textwidth]{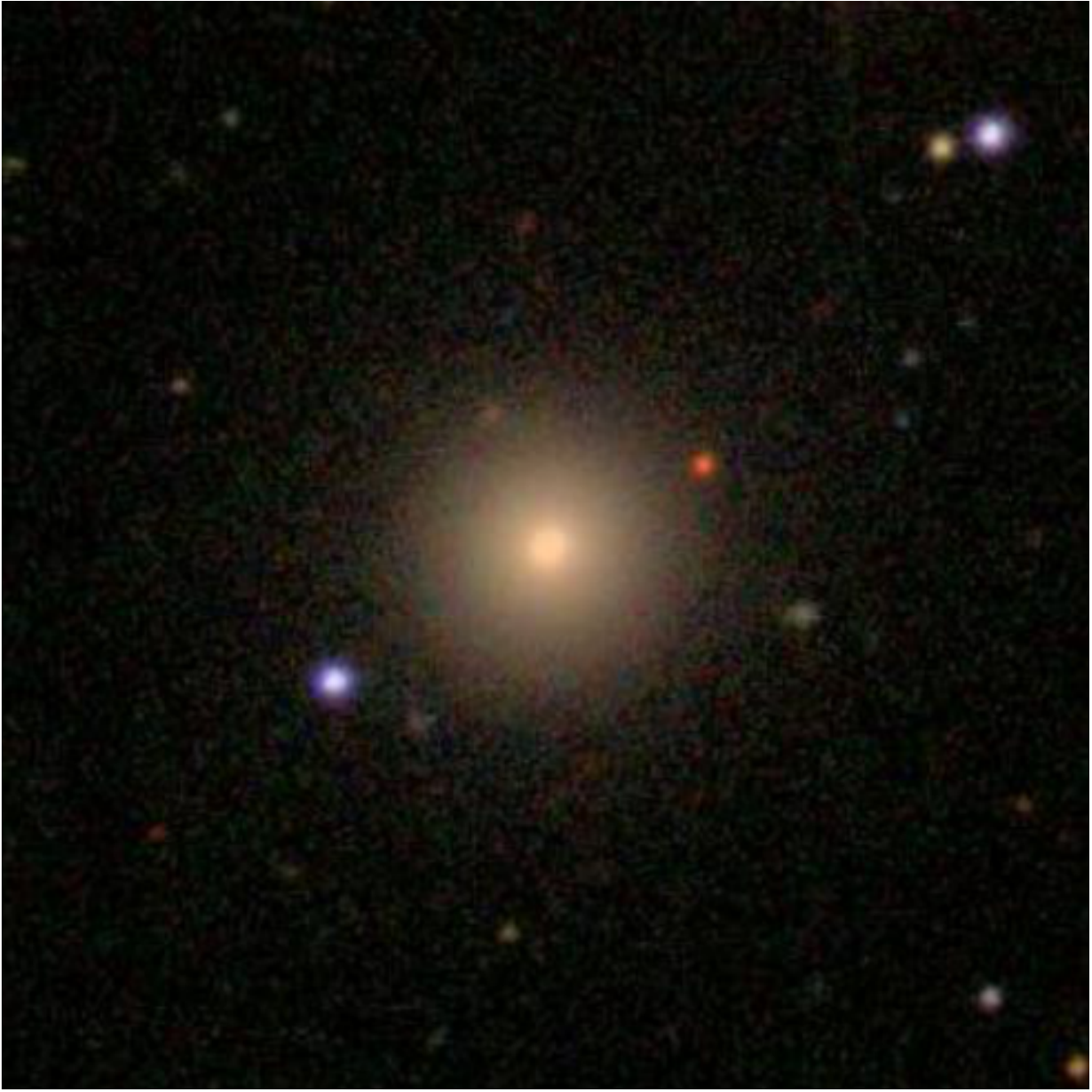}
                \caption{0.01223}
                \label{fig:example-966618}
        \end{subfigure}%
        ~ 
        \caption{Example images from the real-time evaluation set, along with their prediction RMSEs for the best-performing network. The images on the top row were the most difficult for the model to classify; the images on the bottom row were the easiest. Larger angular size and non-radially symmetric morphology are the most challenging targets for the model.}\label{fig:examples}
\end{figure*}

Finally, we can look at some examples from the real-time evaluation set (see Section~\ref{sec:experimental-setup}) with low and high prediction errors, to get an idea of the strengths and weaknesses of the model (Figure~\ref{fig:examples}). The reported RMSE values were obtained with the best performing network and without any averaging, and without centering or rescaling.

The images that are difficult to classify are quite varied. Some are faint, but look fairly typical otherwise, such as Figure~\ref{fig:example-952388}. Most are negatively affected by the cropping operation in various ways: either because they are not properly centred, or because they are very large (Figures~\ref{fig:example-941557} and~\ref{fig:example-915482} respectively). This was the original motivation for introducing an additional rescaling and centering step during preprocessing, but did not end up improving the overall prediction accuracy. The easiest galaxies to classify are mostly smooth, round ellipticals.

\section{Conclusion and future work}
\label{sec:conclusion}

We present a convolutional neural network for fine-grained galaxy morphology prediction, with a novel architecture that allows us to exploit rotational symmetry in the input images. The network was trained on data from the \gztwo~project and is able to reliably predict various aspects of galaxy morphology directly from raw pixel data, without requiring any form of handcrafted feature extraction. It can automatically annotate large collections of images, enabling quantitative studies of galaxy morphology on an unprecedented scale.

Our novel approach to exploiting rotational symmetry was essential to achieve state-of-the-art performance, winning the Galaxy Challenge hosted on Kaggle. Although our winning solution required averaging many sets of predictions from different networks for each image, using a single network also yields competitive results.

Our model can be adapted to work with any collection of centered galaxy images and arbitrary morphological decision trees. Our implementation was developed using open source tools and the source code is publicly available. The model can be trained and used on consumer hardware. Its predictions are highly reliable when they are confident, making our approach applicable for fine-grained morphological analysis of large-scale survey data. Performing such large-scale analyses is an important direction for future research.

For future work, we would like to train networks on larger collections of annotated images. From previous applications in the domain of computer vision, it has become clear that the performance of convolutional neural networks scales very well with the size of the dataset. The $\sim55,000$ galaxy images used in this paper (90\% of the provided training set) is quite a small dataset by modern standards. Even though we combined several techniques to avoid overfitting, which allowed us to train very large models on this dataset effectively, a clear opportunity to improve predictive performance is to train the same model on a larger dataset, since Galaxy~Zoo has already collected annotations for a much larger number of images. More recent iterations of the Galaxy~Zoo project have concentrated on higher redshift samples, so care will have to be taken to ensure that the model is able to generalize across different redshift slices.

The use of larger datasets may also allow for a further increase in model capacity (i.e. the number of trainable parameters) without the risk of excessive overfitting. These high-capacity models could be used as the basis for much larger surveys such as the LSST. The integration of model predictions into existing annotation workflows, both by experts and through crowdsourcing platforms, will also require further study.

Another possibility is the application of our approach to raw photometric data which have not been preprocessed for visual inspection by humans. The networks should be able to learn useful features from this representation, including structural changes from multiple wavebands \citep[eg, ][]{hau13}. Automated classification of other data modalities that exhibit radial symmetry (a commonly occurring property in nature, e.g. in flowers, animals) also presents an interesting opportunity.

From a machine learning point of view, we would like to investigate improved network architectures based on recent developments, such as the trend towards deeper networks with in excess of 20 layers of processing and the use of smaller receptive fields \citep{szegedy2014going,simonyan2014very}.

\section*{Acknowledgements}

We would like to thank Pieter-Jan Kindermans, Francis wyffels, 
A\"aron van den Oord, Pieter Buteneers, Chris Lintott, Philip Marshall and the anonymous reviewer for their valuable feedback. We would like to acknowledge Joyce Noah-Vanhoucke, Chris Lintott, David Harvey, Thomas Kitching and Philip Marshall for their help in designing the Kaggle Galaxy Challenge. We thank Winton Capital for their financial support of the competition, and the Galaxy Zoo volunteers for providing the original morphology classifications. Their efforts are individually acknowledged at \url{http://authors.galaxyzoo.org}. KWW is supported in part by a UMN Grant-in-Aid.

\bibliographystyle{mn2e}
\bibliography{phd_clean}

\bsp

\label{lastpage}

\end{document}